\documentclass[12pt]{article}
\pdfoutput=1
\usepackage{amsmath,amsfonts,color,braket,cite,graphicx,physics,here,mathtools,multirow,tikz,mathtools,array,ulem,bm,hyperref}
\usetikzlibrary{intersections, arrows, calc, angles}
\allowdisplaybreaks[4]

\makeatletter
	\@addtoreset{equation}{section}
	\@addtoreset{figure}{section}
	\renewcommand{\thetable}{\thesection.\arabic{table}}\@addtoreset{table}{section}
\makeatother

\usepackage{color}

\begin{document}

\begin{titlepage}
\begin{flushright}
TIT/HEP-680 \\
NORDITA 2020-034\\
April,  2020
\end{flushright}
\vspace{0.5cm}
\begin{center}
{\Large \bf 
ODE/IM correspondence for affine Lie algebras:\\
A numerical approach
}

\lineskip .75em
\vskip 2.5cm

{\large  Katsushi Ito$^{a,}$\footnote{ito@th.phys.titech.ac.jp}, Takayasu Kondo$^{a,}$\footnote{t.kondo@th.phys.titech.ac.jp}, Kohei Kuroda$^{a,}$\footnote{k.kuroda@th.phys.titech.ac.jp} and  Hongfei Shu$^{b,}$\footnote{hongfei.shu@su.se } 
}
\vskip 2.5em
 {\normalsize\it 
 $^{a}$Department of Physics, Tokyo Institute of Technology,
Tokyo, 152-8551, Japan\\
$^{b}$Nordita, KTH Royal Institute of Technology and Stockholm University\\
Roslagstullsbacken 23, SE-106 91 Stockholm, Sweden
}

\vskip 3.0em
\end{center}
\begin{abstract}
We study numerically the ODE/IM correspondence for untwisted affine Lie algebras associated with simple Lie algebras including exceptional type. We consider the linear problem obtained from the massless limit of that of the modified affine Toda field equation. 
We found that the Q-functions in integrable models are expressed as the inner product of the solution of the dual linear problem and the subdominant solution of the linear problem.
Using Cheng's algorithm to obtain the solution of the linear problem, we can determine efficiently the zeros of the Q-function, which is known to provide the solutions of the Bethe ansatz equations.
We calculate the zeros numerically, which are shown to agree with the results from the Non-Linear Integral Equations for simply-laced affine Lie algebras including the exceptional type.
By the folding procedure of the Dynkin diagrams of simply-laced Lie algebras, we also find the correspondence for the linear problem of the non-simply-laced affine Lie algebras.

\end{abstract}

\end{titlepage}

\baselineskip=0.7cm

\newpage

%%%%%%%%%%%%%%%%%%%%%%%%%%%%%%%%%%%%%%%%%%%%%%%%%%%%%%%%%%%%%%%%%%%%%%%%%%%%%%%%%%%%%%%%%%%%%%%%%%%%%%%%%%%%%%%%%%%%%%%%%%%%%%%%%%%%%%%%

\section{Introduction}

The ODE/IM correspondence \cite{Dorey:1998pt} is a relation between ordinary differential equations (ODE) and two-dimensional quantum integrable models (IM).
This correspondence has been proposed from the spectral analysis of the Schr\"odinger equation, where the functional relations of the integrable model are obtained from the asymptotic solutions to the ODE.
In particular, one can obtain Baxter's T-Q relations and the Bethe ansatz equations (BAEs), which correspond to the ground state, from the coefficients connecting the different bases of the solutions of the ODE\cite{Bazhanov:1998wj,Dorey:1999uk}.
The BAEs, which are satisfied by the zeros of the Q-functions, determine the spectrum of the integrable model.
Although the BAEs are highly non-trivial equations to solve, they are written in the form of the Non-Linear Integral Equations (NLIEs) \cite{Destri:1992qk}, which can be solved numerically.
The spectrum can be recovered from the solutions of the ODE with appropriate boundary conditions.

In \cite{Dorey:1999pv,Suzuki:1999hu,Dorey:2000ma,Bazhanov:2001xm}, the Schr\"odinger type ODE has been generalized to higher-order ODE, which corresponds to $A_r$-type integrable models.
The Q-functions for the fundamental representations has been represented by the Wronskians of the solutions of the ODE.
In \cite{Dorey:2006an}, the correspondence between the ODE and the BAEs has been generalized to classical simple Lie algebras.
The $\psi$-system satisfied by the Wronskians of the solutions plays an important role to write down the BAEs for simple Lie algebras.
The Bethe roots were obtained numerically from the solution to the ODE around the origin\footnote{See \cite{Dorey:2000ma,Dunning-thesis} for the numerical results for the higher order ODE with a linear potential.}, where Cheng's algorithm \cite{PhysRev_127_647} provides an efficient approach to generate the solutions.\cite{Dorey:2001uw,Dorey:2006an}
Numerically, it is hard to find the zeros of the Q-functions represented by the Wronskians.
However, if one uses the adjoint ODE, which is satisfied by the sub-determinants of the Wronskians, one can obtain the zeros from the adjoint ODE efficiently.
For the $D_r$-type Lie algebra, the corresponding ODE is the pseudo-differential equation. 
For this type of ODE, it is difficult to find the Wronskian formula of the Q-functions associated to spinor representations and the adjoint ODE.

In \cite{Sun:2012xw,Masoero:2015lga,Masoero:2015rcz}, it has been shown that the ODE for classical simple Lie algebra
$\mathfrak{g}$ can be obtained from the linear problem associate with the affine Lie algebra $\hat{\mathfrak{g}}^{\vee}$, where the check symbol implies the Langlands dual of $\hat{\mathfrak{g}}$ \cite{Feigin:2007mr} and the $\psi$-system is written in terms of the solutions of the linear system, from which one finds the BAEs of the ground state for the affine Lie algebra $\hat{\mathfrak{g}}$.
In \cite{Ito:2013aea}, motivated by a massive ODE/IM correspondence between the modified sinh-Gordon equation and the quantum sinh-Gordon model \cite{Lukyanov:2010rn} and its generalization \cite{Dorey:2012bx}, the linear differential equation was obtained as the linear problem associated with the modified affine Toda field equation based on $\hat{\mathfrak{g}}$. 
From the linear problem associated with affine Lie algebra, one can write down the ODE for any simple Lie algebra by taking the massless limit.
The $\psi$-system for $\hat{\mathfrak{g}}^{\vee}$ \cite{Masoero:2015lga,Masoero:2015rcz} leads to the BAEs for the Langlands dual $\hat{\mathfrak{g}}$ \cite{Reshetikhin:1986vd,Reshetikhin:1987bz}.\footnote{See \cite{Ito:2015nla,Ito:2018wgj} for massive case.}
This correspondence includes the Langlands dual of twisted affine Lie algebras.
However, it is not clear what kind of integrable models correspond to the linear problem for non-simply-laced affine Lie algebras.
(For $B_2^{(1)}$, see \cite{Ito:2016qzt}).
Since the higher-order (pseudo-)ODE associated with the linear problem has a complicated structure, it is very difficult to study its solutions and find the Bethe roots from the analysis of the ODE.
To study the ODE/IM correspondence for affine Lie algebras, it is desirable to formulate the Q-functions based on the linear problem.

In this paper, we will propose a new method to calculate the Bethe roots from the linear problems associated with affine Lie algebras.
For a simply-laced affine Lie algebra, we show that the spectra numerically agree with the result of the NLIEs, which has been constructed for simply-laced Lie algebras \cite{ZinnJustin:1997at,Dunning:2002cu}.
The linear problem for a non-simply-laced affine Lie algebra is defined from the modified affine Toda field equation in a similar way.
A non-simply-laced Lie algebra is obtained by folding the Dynkin diagram of simply-laced one.
We will show that the linear problem, the BAEs and the NLIEs can be also obtained by the folding procedure\footnote{One can also obtain the linear problem for a twisted Lie algebra by different folding procedure \cite{Masoero:2015rcz}.}.
The effective central charge of the corresponding integrable model is calculated by the same procedure, which completes the list of the integrable models corresponding to the linear problems for untwisted affine Lie algebras.
In the present work, we will not discuss the twisted affine Lie algebra \cite{Sun:2012xw,Ito:2015nla,Masoero:2015lga} and the excited states case \cite{Bazhanov:2003ni,Dorey:2006an,Fioravanti:2004cz,Masoero:2018rel,Masoero:2019wqf}, because the Bethe roots become complicated and the NLIEs approach is rather difficult to find them.
These subjects will be discussed in a separate paper.

This paper is organized as follows.
In section \ref{sec:linear_problem_Qfunction}, after introducing basic notions of Lie algebras, we define the linear problem associate with an affine Lie algebra and its asymptotic solutions.
We then define the Q-functions from the linear problem and derive the BAEs from the $\psi$-system.
We also discuss the folding procedure of the linear system.
In section \ref{sec:Q-function}, we introduce the dual linear problem and express the Q-function in terms of the solutions to the dual linear problem and the subdominant solution.
Using this representation, we discuss the asymptotic behaviour of the Q-function.
Next, we explain Cheng's algorithm to obtain the solutions around the origin, which is useful to find the Bethe roots as the zeros of the Q-functions.
Section \ref{sec:NLIE} describes the NLIEs associated with an affine Lie algebra.
We also discuss the folding procedure of the NLIEs for the simply-laced Lie algebra and derive the effective central charge.
In section \ref{sec:comparison}, we compare the Bethe roots obtained from the Q-function in section \ref{sec:Q-function} with those obtained from the NLIEs.
Section \ref{sec:conclusion} is devoted for conclusions and discussions.
In appendix \ref{sec:cartan_representation}, we summarize basic data of simple Lie algebras and their representations which are used in this paper.
In appendix \ref{sec:linear_ops}, we write down the linear system and its dual for affine Lie algebras explicitly.
In appendix \ref{sec:q-funcs}, we summarize the Q-functions for the fundamental representations of a simple Lie algebra.
Finally, in appendix \ref{sec:Cheng_E6}, we apply Cheng's algorithm for the solution of the linear problem and its dual in the case of the affine Lie algebra $E_6^{(1)}$.

\section{Linear problem and Q-function}\label{sec:linear_problem_Qfunction}

In this section, we first summarize the notations of the Lie algebra used in this paper.
We next introduce the linear problem associated with an affine Lie algebra.
Then, we define the Q-function, which connects the subdominant solution at infinity with the solutions around the origin.

\subsection{Lie algebra}\label{sec:Lie}
We begin with some basic definitions of Lie algebras.
Let $\mathfrak{g}$ be a simple Lie algebra, which is generated by $\{ E_{\alpha},H^a\}$ $(\alpha\in\Delta, a=1,\ldots, r)$.
Here $r$ is the rank of $\mathfrak{g}$ and $\Delta$ is the set of roots.
These generators satisfy the commutation relations:
\begin{align}
    &[H^a,H^b]=0,\\
    &[H^a,E_{\alpha}]=\alpha^a\ E_{\alpha},\\
    &[E_{\alpha},E_{\beta}]=\begin{cases} 
                                N_{\alpha,\beta}E_{\alpha+\beta}, & \alpha+\beta\in\Delta,\\
                                \alpha^{\vee}\cdot H, & \alpha+\beta=0,\\
                                0, & \text{otherwise}\, ,
                            \end{cases}
\end{align}
where $\alpha^{\vee}\cdot H = \sum_{a = 1}^{r}\alpha^{\vee a} H^a$ and $N_{\alpha,\beta}$ are the structure constants.
$\alpha^\vee=2\alpha/\alpha^2$ is the co-root of $\alpha$, where we normalize the squared length of the long root to be 2.

Let $\alpha_a$ $(a=1,\dots,r)$ be simple roots of $\mathfrak{g}$.
The Cartan matrix of $\mathfrak{g}$ is defined by $C_{ab}=(\alpha_a\cdot \alpha_b^{\vee})$.
The fundamental weight $\omega_a$ $(a=1,\dots,r)$, which is a dual vector of the simple root $\alpha_a$, is defined by $\omega_a\cdot \alpha_b^{\vee}=\delta_{ab}$.
The associated co-weight $\omega_a^{\vee}$ is defined by $\omega_a^{\vee}=2\omega_a/\alpha_a^2$.
The (co)Weyl vector $\rho$ $(\rho^{\vee})$ is the sum of the (co)fundamental weights: $\rho=\sum_{a=1}^r\omega_a\;(\rho^{\vee}=\sum_{a=1}^r\omega_a^{\vee})$.
The highest root $\theta$ and its co-root $\theta^{\vee}$ are expanded as $\theta=\sum_{a=1}^{r} n_a \alpha_a$ and $\theta^{\vee}=\sum_{a=1}^r n_a^{\vee}\alpha_a^{\vee}$, where $n_a$ ($n_a^{\vee}$) are some positive integers called (co-)labels.
The Dynkin diagrams and the Cartan matrices of simple Lie algebras are presented in the appendix \ref{sec:cartan_representation}.
 
We use $(\rho^{(a)}, V^{(a)})$ ($a=1,\dots,r$) to denote the fundamental representation $\rho^{(a)}$ on the vector space $V^{(a)}$ with the highest weight $\omega_a$.
${\bf e}_i$ ($i=1,\ldots, \mathrm{dim}V^{(a)}$) are the weight vectors of $V^{(a)}$ with the weight $h_i$: $H^b{\bf e}_i=(h_i)^b {\bf e}_i$.

The untwisted affine Lie algebra $\hat{\mathfrak{g}}$ associated with a simple Lie algebra $\mathfrak{g}$ is defined by the extended Dynkin diagram, which is obtained by adding a root $\alpha_0=-\theta+\delta$ \cite{Kac}.
Here $\delta$ is the null element.
We define the generator by $E_{\alpha_0}=E_{-\theta}$.
We also define the label and the co-label $n_0=1$ and $n_0^{\vee}=1$, respectively.
Then the (dual) Coxeter number is defined by $h=\sum_{a=0}^r n_a$ and $h^{\vee}=\sum_{a=0}^{r} n_a^{\vee}$.
Then the affine Lie algebra is generated by $E_{\alpha} \zeta^n$ and $H^a \zeta^n$ ($\alpha\in \Delta$, $a=1,\ldots, r$, $n\in\mathbf{Z}$, $\zeta\in \mathbf{C}$) in addition with a central element.

When the Dynkin diagram of $\mathfrak{g}$ has a symmetry under the interchange of the nodes of the diagram, it induces an automorphism of the Lie algebra.
Using this symmetry, one can relate the simply-laced Lie algebra to non-simply-laced one.
One can also construct the twisted affine Lie algebra by using this automorphism.

Let us illustrate this folding procedure to obtain non-simply-laced Lie algebras, which we will use in this paper.
\paragraph{$\underline{A_{2r-1}\rightarrow C_{r}}$}
Let $\alpha_1,\ldots, \alpha_{2r-1}$ be the simple roots of $A_{2r-1}$ and $\{H^a, E_{\alpha}\}$ its generators.
Then we identify $\alpha_{a}$ with $\alpha_{2r-a}$  ($a=1,\ldots, r$).
We define 
\begin{equation}
    \beta_{a}=\frac{1}{2} (\alpha_a+\alpha_{2r-a}), \quad a=1,\ldots, r-1,\quad \beta_r=\alpha_r,
\end{equation}
which become the simple roots of $C_r$.
The generators $E^{C_r}_{ \beta_a}$ of $C_r$ for simple roots $\beta_a$ are defined from those of $A_{2r-1}$ as
\begin{equation}
    E^{C_r}_{\beta_a}=E_{\alpha_a}+E_{\alpha_{2r-a}}, \quad a=1,\ldots, r-1,\quad E^{C_r}_{ \beta_r}=E_{ \alpha_r}.
\end{equation}
The fundamental representation $V^{(a)}_{A_{2r-1}}$ ($a=1,\ldots, r$) of $A_{2r-1}$ is isomorphic
to $V^{(a)}_{C_r}$: $V^{(a)}_{A_{2r-1}} \simeq V^{(a)}_{C_r}$.
Since $V^{(2r-a)}_{A_{2r-1}}$ is dual to $V^{(a)}_{A_{2r-1}}$, it is isomorphic to $V^{(a)}_{C_r}$.

\paragraph{$\underline{D_{r+1}\rightarrow B_{r}}$}
For $D_{r+1}$ with simple roots $\alpha_1,\ldots, \alpha_{r+1}$, there is a $\mathbf{Z}_2$-symmetry, which interchanges $\alpha_{r}$ with $\alpha_{r+1}$ and other simple roots are unchanged.
We define the simple roots of $B_r$ by
\begin{equation}
    \beta_{a}=\alpha_a, \quad a=1,\ldots, r-1,\quad \beta_r=\frac{1}{2}(\alpha_r+\alpha_{r+1}).
\label{eq:folding_D_to_B}
\end{equation}
The generators of $B_r$ for the simple roots are given by
\begin{equation}
    E^{B_r}_{\beta_a}=E_{\alpha_a},\quad a=1,\ldots, r-1,\quad E^{B_r}_{\beta_r}=E_{\alpha_r}+E_{\alpha_{r+1}}.
\end{equation}

\paragraph{$\underline{D_4\rightarrow G_2}$}
Let $\alpha_1,\alpha_2,\alpha_3, \alpha_4$ be simple roots of $D_4$.
Then we identify $\alpha_1$, $\alpha_3$, $\alpha_4$  by $\mathbf{Z}_3$ -symmetry $(\alpha_1,\alpha_3,\alpha_4)\rightarrow (\alpha_3,\alpha_4,\alpha_1)$ and define
\begin{equation}
    \beta_1=\frac{1}{3}(\alpha_1+\alpha_2+\alpha_3),\quad\beta_2=\alpha_2.
    \label{eq:folding_D4_to_G2}
\end{equation}
$\beta_1,\beta_2$ are the simple roots of $G_2$.
The corresponding generators are given by
\begin{equation}
    E^{G_2}_{\beta_1}=E_{\alpha_1}+E_{\alpha_2}+E_{\alpha_3},\quad E^{G_2}_{\beta_2}=E_{\alpha_2}.
\end{equation}

\paragraph{$\underline{E_6\rightarrow F_4}$}
Let $\alpha_1,\ldots, \alpha_6$ be the simple roots of $E_6$.
The Dynkin diagram of $E_6$ has a $\mathbf{Z}_2$-symmetry which interchanges $\alpha_1$ and $\alpha_5$, $\alpha_2$ and $\alpha_4$. $\alpha_3$ and $\alpha_6$ are unchanged.
Then the simple roots of $F_4$ are defined by
\begin{equation}
    \beta_4=\frac{1}{2}(\alpha_1+\alpha_5),\quad
    \beta_3=\frac{1}{2}(\alpha_2+\alpha_4),\quad
    \beta_2=\alpha_3,\quad \beta_1=\alpha_6.
    \label{eq:folding_E6_to_F4}
\end{equation}
The generators of $F_4$ for the simple roots are defined by
\begin{align}
    E^{F_4}_{\beta_1}&=E_{\alpha_1}+E_{\alpha_5},\quad
    E^{F_4}_{\beta_2}=E_{\alpha_2}+E_{\alpha_4},\quad
    E^{F_4}_{\beta_3}=E_{\alpha_3},\quad
    E^{F_4}_{\beta_4}=E_{\alpha_6}.
    \label{eq:folding_generator_e6tof4}
\end{align}

\subsection{Linear problem and Q-function}
We now define the linear problem for an untwisted affine Lie algebra $\hat{\mathfrak{g}}$.
It is defined by the massless limit of the linear problem of the modified affine Toda field equation \cite{Lukyanov:2010rn,Dorey:2012bx,Ito:2013aea,Adamopoulou:2014fca,kar72958}, which corresponds to the massive integrable model with the BAEs associated to its Langlands dual $\hat{\mathfrak{g}}^{\vee}$. Here $\hat{\mathfrak{g}}$ is a simply-laced or twisted affine Lie algebra.

We start with the linear problem defined on the complex plane. For an affine Lie algebra $\hat{\mathfrak{g}}$ with rank $r$ and its representation $V$, we define
\begin{equation}
    \mathcal{L}_{\hat{\mathfrak{g}}}\qty(x,E,l;\zeta)\ \Psi(x,E,l)=\qty[\dv{x} + A_{\hat{\mathfrak{g}}}]\Psi(x,E,l)=0.
    \label{eq:linear_prob}
\end{equation}
$\Psi$ is $V$-valued function of $x$ and $A_{\hat{\mathfrak{g}}}$ is the $\hat{\mathfrak{g}}$-valued matrix defined by
\begin{align}
    A_{\hat{\mathfrak{g}}} &\coloneqq -\frac{1}{x}\sum_{a = 1}^{r}\ l_a (\alpha^{\vee}_a\cdot  H) + \sum_{a = 1}^{r}\sqrt{n_a^{\vee}}E_{\alpha_a} + \sqrt{n_0^{\vee}}p(x, E)\;\zeta E_{\alpha_0}, \label{eq:potential_op}
\end{align}
where
\begin{align}
    l_a&\coloneqq - \omega_a\cdot g, \qquad p(x,E)\coloneqq x^{h  M}-E\label{eq:def_l_P}.
\end{align}
The parameters $l=(l_1,\dots,l_r)$ are referred as the monodromy parameters.
$g$ is a $r$-dimensional vector satisfying
\begin{equation}
    1+\alpha_a\cdot g>0,\qquad a=0,\dots,r.\label{eq:g_condition}
\end{equation}
These constraints for $g$ come from the condition that the leading term in the solution to the modified affine Toda field equation is logarithmic \cite{Ito:2013aea}.
$\zeta$ is defined to be $+1$ or $-1$, which depends on $\mathfrak{g}$ and $V$.
Concrete expressions of ${\cal L}_{\hat{\mathfrak{g}}}$ are presented in appendix \ref{sec:linear_ops}.

Let $(\rho,V)$ be the representation of $\mathfrak{g}$ with the highest weight $h_1$.
Define the weight vector ${\bf e}_j$ ($j=1,\dots,\mathrm{dim}V$) corresponding to the weight $h_j$,
\begin{equation}
H^a \mathbf{e}_j = (h_j)^a \mathbf{e}_j,
\end{equation}
where ${\bf e}_1$ is the highest weight vector.
$\{ {\bf e}_j\}$ spans an basis in $V$.
For the fundamental representation $V^{(1)}$ of a classical simply-laced Lie algebra $\mathfrak{g}$, this linear system reduces to the (pseudo-)ODE in \cite{Dorey:2006an}.

The simplest example of the linear problem is that of $A_1^{(1)}$ in $V^{(1)}$.
Let $\mathbf{e}_{1} ,\mathbf{e}_{2}$ be the orthonormal basis of $\mathbf{R}^2$.
Then the linear problem  for $\Psi=\psi_1\mathbf{e}_1+\psi_2\mathbf{e}_2$ in $V^{(1)}$ is given by
\begin{equation}
    \qty[\begin{pmatrix} \dv{x}-\frac{l_1}{x} &  0\\ 0 & \dv{x}+\frac{l_1}{x}\end{pmatrix} +  \begin{pmatrix} 0 & 1\\p(x,E) & 0 \end{pmatrix}]\begin{pmatrix} \psi_1\\ \psi_2\end{pmatrix}=0\, .\label{eq:linear_problem_A1}
\end{equation}
Eliminating $\psi_2$, one finds $\psi_1$ satisfies the Schr\"odinger equation with angular momentum and the monomial potential:\cite{Bazhanov:1998wj}
\begin{equation}
    \qty[-\dv{^2}{x^2}+\frac{l_1(l_1-1)}{x^2}+x^{2M}-E]\psi_1=0.
\end{equation}

Under the rotation $(x,E)\to (\omega^k x,\Omega^k E)$ with $\Omega=e^{\frac{2\pi i M}{M+1}}$ and $\omega=e^{\frac{2\pi i }{h(M+1)}}$, the linear differential operator $\mathcal{L}_{\hat{\mathfrak{g}}}$ transforms as \cite{Sun:2012xw}
\begin{equation}
\begin{aligned}
\omega^{k}\mathcal{L}_{\hat{\mathfrak{g}}}(\omega^{k}x,\Omega^k E,l;\zeta)=\omega^{k\rho^{\vee}\cdot H} \mathcal{L}_{\hat{\mathfrak{g}}}(x,E,l;\zeta e^{2\pi ik})
\omega^{-k\rho^{\vee}\cdot H}.
\end{aligned}
\end{equation}
Suppose that $\Psi(x,E)$ is the solution of the linear problem ${\cal L}_{\hat{\mathfrak{g}}}(x,E,l;\zeta)$.
Then we find
\begin{equation}
\Psi_{[k]}(x,E)\coloneqq\omega^{-k\rho^{\vee}\cdot H}\Psi(\omega^k x,\Omega^k E)\label{eq:Symanzik}
\end{equation}
satisfies the linear problem for $\mathcal{L}_{\hat{\mathfrak{g}}}(x,E,l;\zeta)$
evaluated by replacing $E_{\alpha_{0}}\to e^{2\pi ik}E_{\alpha_{0}}$:
\begin{equation}
\mathcal{L}_{\hat{\mathfrak{g}}}(x,E,l;\zeta e^{2\pi ik})
\Psi_{[k]}(x,E)=0.
\end{equation}
We call $\Psi_{[k]}$ in (\ref{eq:Symanzik}) the Symanzik (Sibuya) rotation of $\Psi$\cite{Sibuya}.
We define the representation $V_{[k]}$, where the subscript $[k]$ means that the generator $E_{\alpha_0}$ acts on $V$ as $e^{2\pi i k}E_{\alpha_0}$.
If $\Psi$ is a solution of the linear problem on $V$, then $\Psi_{[k]}$ becomes a solution on $V_{[k]}$.

We discuss the solution of the linear problem at infinity using the WKB approximation.
Consider the gauge transformation: $\Psi(x)\to\Psi^{\prime}(x)=U(x)\Psi(x)$, $\mathcal{L}_{\hat{\mathfrak{g}}}\to\mathcal{L}_{\hat{\mathfrak{g}}}^{\prime}=U(x)\mathcal{L}_{\hat{\mathfrak{g}}}U^{-1}(x)$, where $U(x)=\exp\qty(\log\qty(p(x,E))^{1/h}\rho^{\vee}\cdot H)$.
The linear operator becomes
\begin{equation}
    \mathcal{L}_{\hat{\mathfrak{g}}}^{\prime}=\dv{x}-\frac{1}{x}\sum_{a=1}^rl_a\;\alpha_a\cdot H -\frac{1}{h}\frac{d\log p(x,E)}{dx}\rho^{\vee}\cdot H+p(x,E)^{1/h}\Lambda_+,
\end{equation}
where
\begin{equation}
    \Lambda_+=\sum_{a=1}^r\sqrt{n_a^{\vee}}\ E_{\alpha_a}+\sqrt{n_0^{\vee}}\ \zeta E_{\alpha_0}.
\end{equation}
At large $x$, the $O\qty(1/x)$ terms in $\mathcal{L}_{\hat{\mathfrak{g}}}^{\prime}$ can be ignored.
Defining $\nu_i$ and $\bm{\nu}_i$ ($i=1,\dots,\mathrm{dim}\;V$) as the eigenvalues and the eigenvectors of $\Lambda_+$, the WKB solution $\Psi^{\mathrm{WKB}}(x,E)$ is given by
\begin{equation}
    \Psi^{\mathrm{WKB}}(x,E)=\sum_{i=1}^{\mathrm{dim}V}C_i \exp\qty(-\nu_i\int^x\qty(p(x^{\prime},E))^{1/h}dx^{\prime}-\frac{1}{h}\log(p(x,E))\rho^{\vee}\cdot H)\bm{\nu}_i, \label{eq:WKB_sol}
\end{equation}
where $C_i$ are constants.
We denote $\Psi(x,E)$ the subdominant solution along the positive real axis.
$\Psi(x,E)$ decays fastest in the sector ${\cal S}_0\coloneqq\{ x\in \mathbf{C}\mid \left| \arg x \right| < \frac{\pi}{h(M+1)}\}$ and has the asymptotic behaviour for large $x$:
\begin{align}
	\Psi(x,E,l) &\sim C\exp\qty(-\nu\int^x\qty(p(x^{\prime},E))^{1/h}dx^\prime)\exp\qty(-\frac{1}{h}\log p(x,E)\;\rho^{\vee}\cdot H)\bm{\nu}\label{eq:subdominant}\\
	&\sim C x^{- M\rho^{\vee}\cdot H}\exp\qty(-\frac{\nu}{M+1}\ x^{M+1}){\bm \nu}, \qquad x\to\infty,\label{eq:subdominant_sol}
\end{align}
where $C$ is a constant. $\nu$ is the eigenvalue of generator $\Lambda_+$ with the largest real value and $\bm{\nu}$ the associated eigenvector.
The Symanzik rotation of the solution $\Psi_{[k]}(x,E,l)$ ($k\in \mathbf{Z}$) becomes the subdominant solution in the sector $\mathcal{S}_k$, which is defined by 
\begin{equation}
    \mathcal{S}_k \; : \quad \left| \arg x + \frac{2\pi k}{h(M+1)} \right| < \frac{\pi}{h(M+1)} \,.
\end{equation}
We next discuss the solution of the linear problem around $x=0$.
We find a basis of power series solutions $\mathcal{X}_i(x,E,l)$ ($i=1,\dots,\mathrm{dim}V)$ around $x=0$:
\begin{equation}
    \mathcal{X}_i(x, E,l) = x^{-h_i\cdot g} \mathbf{e}_i + \cdots, \qquad x\to0 \;. \label{eq:small_asymp}
\end{equation}
$\mathcal{X}_i(x, E,l)$ has a monodromy $e^{-2\pi ih_i\cdot g}$ around the origin.
Here $\mathcal{X}_i(x, E,l)$ transforms as
\begin{align}
    \mathcal{X}_{i[k]}(x, E,l)&=\omega^{-kh_i\cdot\qty(\rho^{\vee}+g)}\mathcal{X}_{i}(x, E,l).
\end{align}
Using this basis, the subdominant solution $\Psi(x,E,l)$ can be expanded as
\begin{equation}
  \Psi(x,E,l) = \sum_{i=1}^{\mathrm{dim}V} \mathcal{Q}_i (E,l) \ \mathcal{X}_i (x,E,l) , \label{eq:basis_exp_of_psi}
\end{equation}
which is a generalization of the definition of the Q-function \cite{Bazhanov:1998wj}.
We will identify the coefficients $\mathcal{Q}_i (E,l)$ as the Q-functions of the integrable model characterized by the BAEs.
For this purpose, we focus on the coefficient $\mathcal{Q}_1 (E,l)$ which corresponds to the highest weight state in $V$.
We define
\begin{equation}
    Q(E,l)\coloneqq\mathcal{Q}_1(E,l). \label{eq:Q-function}
\end{equation}
The condition $Q(E,l)=0$ for $E$ will determine the Bethe roots of the integrable model.
This condition corresponds to the special boundary condition for the solution: near the origin, the solution $\Psi$ does not depend on the basis $\mathcal{X}_1$ and decays fastest along the positive real axis.

Let us now consider the linear problem \eqref{eq:linear_prob} for a simple Lie algebra $\mathfrak{g}$, on the fundamental representations $V^{(a)}$ ($a=1,\ldots,r$)  with the highest weight $\omega_a$.
We denote the subdominant solution as $\Psi^{(a)}$ and the basis around the origin as ${\cal X}_i^{(a)}$.
We also denote $Q(E,l)$ in (\ref{eq:Q-function}) as $Q^{(a)}(E,l)$.
These Q-functions are not independent, whose relations are discussed in the following.

We begin with the $A_r^{(1)}$ type linear problem.
The $a$-th fundamental representation $V^{(a)}$ is isomorphic to the $a$-anti-symmetric tensor product of $V^{(1)}$: $V^{(a)}\simeq \wedge^a V^{(1)}$.
The subdominant solution $\Psi^{(a)}$ in $V^{(a)}$ and the Symanzik rotation of $\Psi^{(1)}$'s in $V^{(1)}$ are related by
\begin{equation}
    \Psi^{(a)}=\bigwedge_{b=1}^a \Psi^{(1)}_{\qty[b-\frac{a+1}{2}]},\qquad a=1,\ldots,r.\label{eq:antisymmetric_relation1_A}
\end{equation}
Here the Symanzik rotation is necessary to ensure that the both hand sides satisfy the same linear differential equation and have the same asymptotic behaviour. Substituting the expansion \eqref{eq:basis_exp_of_psi} into \eqref{eq:antisymmetric_relation1_A} and identifying the highest weight vector $\mathcal{X}_1^{(a)}$ with $\mathcal{X}^{(1)}_1\wedge\cdots\wedge\mathcal{X}^{(1)}_a$, one obtains 
\begin{equation}
    Q^{(a)}(E,l) = \sum_{j_0,\ldots,j_{a-1}=1}^{a}\epsilon_{j_0\cdots j_{a-1}}\prod_{k=0}^{a-1} \omega^{-k\lambda^{(1)}_{j_k}}\mathcal{Q}^{(1)}_{j_k\qty[\frac{a-1}{2}-k]} (E,l),\qquad a=1,\ldots,r, \label{eq:A_Qfunc_from_V1}
\end{equation}
where $\lambda^{(a)}_i=\rho^{\vee}\cdot(\omega_a-h_i^{(a)})- h^{(a)}_i\cdot g$ and $\epsilon_{j_0\cdots j_{a-1}}$ is the totally anti-symmetric symbol normalized as $\epsilon_{01\cdots a-1}=1$.
This is nothing but the Wronskian formula which relates Q-function $Q^{(a)}$ to $\mathcal{Q}_i^{(1)}$ \cite{Dorey:2000ma}.

Since $V^{(r+1-a)}$ is dual to $V^{(a)}$, we have also the relation:
\begin{equation}
    \Psi^{(r+1-a)}=\bigwedge_{b=1}^a \Psi^{(r)}_{\qty[b-\frac{a+1}{2}]},\qquad a=2,\ldots,r.\label{eq:antisymmetric_relationr_A}
\end{equation}
From this relation, one can get a formula to calculate Q-function $Q^{(a)}$ from   $\mathcal{Q}_i^{(r)}$.
We can study similar Wronskian type formulas of $Q^{(a)}$ for other affine Lie algebras, which are summarized in appendix \ref{sec:q-funcs}.

The solutions $\Psi^{(a)}$ satisfy another type of relations which is called the $\psi$-system \cite{Sun:2012xw,Ito:2015nla,Masoero:2015lga}.
For the linear problem associated with an affine Lie algebra $\hat{\mathfrak{g}}^{\vee}$, the $\psi$-system leads to the $\hat{\mathfrak{g}}$-type Bethe ansatz equations of the quantum integrable model, where $\hat{\mathfrak{g}}$ is an untwisted affine Lie algebra.
For the representations $V^{(a)}$ of a simply-laced Lie algebra $\mathfrak{g}$, we have the inclusion map $\iota: V^{(a)}\wedge V^{(a)} \rightarrow \otimes_{b=1}^{r} (V^{(b)})^{2\delta_{ab}-C_{ab}}$.
Then, the $\psi$-system \cite{Sun:2012xw,Masoero:2015rcz} for a simply-laced affine Lie algebra $\hat{\mathfrak{g}}$ takes the form\footnote{In \cite{Masoero:2015lga,Masoero:2015rcz}, the $\psi$-system has been proposed in a different form from the present one. However, it is shown to be equivalent to the present one.}
\begin{equation}
  \iota\left( \Psi^{(a)}_{[-1/2]} \wedge \Psi^{(a)}_{[1/2]} \right) = \bigotimes_{b=1}^r \left( \Psi^{(b)} \right)^{2\delta_{ab}-C_{ab}} . \label{eq:psi-system}
\end{equation}
Substituting (\ref{eq:basis_exp_of_psi}) into the $\psi$-system (\ref{eq:psi-system}) and comparing the coefficient of the highest weight state, one finds the relations:
\begin{equation}
    \omega^{-\frac{1}{2}\qty(\lambda_1^{(a)}-\lambda_2^{(a)})}\mathcal{Q}_{1[-1/2]}^{(a)}\mathcal{Q}_{2[1/2]}^{(a)}-\omega^{\frac{1}{2}\qty(\lambda_1^{(a)}-\lambda_2^{(a)})}\mathcal{Q}_{1[1/2]}^{(a)}\mathcal{Q}_{2[-1/2]}^{(a)}=\prod_{b=1}^r\qty[\mathcal{Q}_1^{(b)}]^{2\delta_{ab}-C_{ab}}. \label{eq:relation_of_mathcalQ}
\end{equation}
Evaluating the $1/2$ and $-1/2$ Symanzik rotations of \eqref{eq:relation_of_mathcalQ} at the zeros $E_i^{(a)}$ ($i=0,1,\ldots$) of $Q^{(a)}=\mathcal{Q}_1^{(a)}$ and taking the ratio of them, the authors of \cite{Masoero:2015lga} find the Bethe ansatz equations \cite{Reshetikhin:1987bz,Reshetikhin:1986vd}:
\begin{equation}
    \prod_{b = 1}^{r}\Omega^{C_{ab}\gamma_b/2}\left.\frac{Q_{\qty[C_{ab}/2]}^{(b)}}{Q_{\qty[-C_{ab}/2]}^{(b)}}\right|_{E_i^{(a)}} = -1, \qquad i = 0, 1, \ldots, \label{eq:BAE_for_simply-laced}
\end{equation}
where 
\begin{equation}
\gamma_a \coloneqq \frac{2}{hM}\qty(l_a - \omega_a \cdot \rho^{\vee}), \label{eq:twist_parameter}
\end{equation}
is regarded as the twist parameter in the integrable models (see section \ref{sec:NLIE}).

\subsection{Linear problem for non-simply-laced Lie algebras and folding}
We can study the linear problem and the $\psi$-system for a non-simply-laced affine Lie algebra defined by (\ref{eq:linear_prob}) \cite{Locke, Ito:2016qzt}.
The ODE is also obtained by the folding procedure of simply-laced Lie algebras, which is useful to understand the folding structure of the BAEs and the NLIEs.

Let us discuss the Lie algebra $C_r$, which is obtained by the folding of $A_{2r-1}$. 
The fundamental representation $V^{(a)}_{A_{2r-1}}$ ($a=1,\ldots, r$) of $A_{2r-1}$ is isomorphic to $V^{(a)}_{C_r}$ of $C_r$.
We find the linear problem for $A_{2r-1}^{(1)}$ is reduced to that of $C_r^{(1)}$ by setting the monodromy parameter $l^{A_{2r-1}}$ of $A_{2r-1}^{(1)}$ as  $l_a^{A_{2r-1}^{(1)}}=l_{2r-a}^{A_{2r-1}^{(1)}}$.
The monodromy parameter $l^{C_r^{(1)}}$ of $C_r^{(1)}$ is given by
\begin{equation}
    l_1^{C_r^{(1)}} =l_1^{A_{2r-1}^{(1)}},\qquad l_a^{C_r^{(1)}}= l_a^{A_{2r-1}^{(1)}} -l_{a-1}^{A_{2r-1}^{(1)}}, \quad a=2,\ldots,r.
\end{equation}
Note that setting $l_a^{A_{2r-1}^{(1)}}=l_{2r-a}^{A_{2r-1}^{(1)}}$, the linear problems on $V_{A_{2r-1}}^{(a)}$ and $V_{A_{2r-1}}^{(2r-a)}$ reduce to the same differential system.
The subdominant solution and the basis of the solutions around the origin defined for $A_{2r-1}^{(2)}$ also reduce to those of $C_r^{(1)}$ by this folding procedure.
Then the Q-functions of $A_{2r-1}^{(1)}$ are identified as $Q_{A_{2r-1}^{(1)}}^{(a)}=Q_{A_{2r-1}^{(1)}}^{(2r-a)}$ and the Q-functions of  $C_r^{(1)}$ satisfy
\begin{equation}
   Q^{(a)}_{C_r^{(1)}}= Q^{(a)}_{A_{2r-1}^{(1)}}=\;Q^{(2r-a)}_{A_{2r-1}^{(1)}}, \qquad a=1,\ldots,r .
   \label{eq:q-func_Cr}
\end{equation}
Here we denote Q-functions for $\hat{\mathfrak{g}}$ as $Q_{\hat{\mathfrak{g}}}$.
Applying (\ref{eq:q-func_Cr}) to (\ref{eq:BAE_for_simply-laced}), the BAEs corresponding to $C_r^{(1)}$ type linear problem are given by
\begin{equation}
    \begin{aligned}
        &\prod_{b = 1}^{r}\Omega^{C_{ab}\gamma_b/2}\left.\frac{Q_{\qty[C_{ab}/2]}^{(b)}}{Q_{\qty[-C_{ab}/2]}^{(b)}}\right|_{E_i^{(a)}} = -1, \qquad \text{for}\qquad a=1,\ldots,r-1,\\
        &\Omega^{\gamma_r-\gamma_{r-1}}\left.\frac{\qty[Q_{\qty[-1/2]}^{(r-1)}]^2Q_{\qty[1]}^{(r)}}{\qty[Q_{\qty[1/2]}^{(r-1)}]^2Q_{\qty[-1]}^{(r)}}\right|_{E_i^{(r)}} = -1.
    \end{aligned}\label{eq:BAE_for_C}
\end{equation}
Here $C_{ab}$ is the Cartan matrix of $C_r$ and $\gamma_a$ is defined by (\ref{eq:twist_parameter}).

In a similar way, the BAEs for $B_r^{(1)}$, $F_4^{(1)}$ and
$G^{(1)}_2$ types can be obtained from $D_{r+1}^{(1)}$, $E_{6}^{(1)}$ and $D_{4}^{(1)}$, respectively, which are explained in appendix \ref{sec:q-funcs}.
Here we summarize the BAEs for these algebras:
\begin{description}
    \item[$B_r^{(1)}$:
    ]\mbox{}\\
        \begin{equation}
            \begin{aligned}
            &\prod_{b = 1}^{r}\Omega^{C_{ab}\gamma_b/2}\left.\frac{Q_{\qty[C_{ab}/2]}^{(b)}}{Q_{\qty[-C_{ab}/2]}^{(b)}}\right|_{E_i^{(a)}} = -1, \qquad \text{for}\qquad a=1,\ldots,r-2,r,\\
            &\Omega^{-\frac{1}{2}\gamma_{r-2}+\gamma_{r-1}-\gamma_{r}}\left.\frac{Q_{\qty[-1/2]}^{(r-2)}Q_{\qty[1]}^{(r-1)}\qty[Q_{\qty[-1/2]}^{(r)}]^2}{Q_{\qty[1/2]}^{(r-2)}Q_{\qty[-1]}^{(r-1)}\qty[Q_{\qty[1/2]}^{(r)}]^2}\right|_{E_i^{(r-1)}} = -1.
            \end{aligned}\label{eq:BAE_for_B}
        \end{equation}
    \item[$F_4^{(1)}$:
    ]\mbox{}\\
        \begin{equation}
            \begin{aligned}
            &\prod_{b = 1}^{4}\Omega^{C_{ab}\gamma_b/2}\left.\frac{Q_{\qty[C_{ab}/2]}^{(b)}}{Q_{\qty[-C_{ab}/2]}^{(b)}}\right|_{E_i^{(a)}} = -1, \qquad \text{for}\qquad a=1,2,4,\\
            &\Omega^{-\gamma_{2}+\gamma_{3}-\frac{1}{2}\gamma_{4}}\left.\frac{\qty[Q_{\qty[-1/2]}^{(2)}]^2Q_{\qty[1]}^{(3)}Q_{\qty[-1/2]}^{(4)}}{\qty[Q_{\qty[1/2 ]}^{(2)}]^2Q_{\qty[-1]}^{(3)}Q_{\qty[1/2]}^{(4)}}\right|_{E_i^{(3)}} = -1.
            \end{aligned}\label{eq:BAE_for_F}
        \end{equation}
    \item[$G_2^{(1)}$:
    ]\mbox{}\\
        \begin{equation}
        \Omega^{\gamma_{1}-\frac{1}{2}\gamma_{2}}\left.\frac{Q_{\qty[1]}^{(1)}Q_{\qty[-1/2]}^{(2)}}{Q_{\qty[-1]}^{(1)}Q_{\qty[1/2]}^{(2)}}\right|_{E_i^{(1)}} = -1,\qquad
        \Omega^{-\frac{3}{2}\gamma_{1}+\gamma_{2}}\left.\frac{\qty[Q_{\qty[-1/2]}^{(1)}]^3Q_{\qty[1]}^{(2)}}{\qty[Q_{\qty[1/2]}^{(1)}]^3Q_{\qty[-1]}^{(2)}}\right|_{E_i^{(2)}} = -1. \label{eq:BAE_for_G}
        \end{equation}
\end{description}
Here $C_{ab}$ is the Cartan matrix of the corresponding Lie algebra.

Note that the BAEs (\ref{eq:BAE_for_C})-(\ref{eq:BAE_for_G}) corresponding to the linear problems for the non-simply-laced affine Lie algebras $\hat{\mathfrak g}$ differ from the BAEs for $\hat{\mathfrak g}$, which are related to the linear problem for the Langlands dual $\hat{\mathfrak g}^{\vee}$ \cite{Frenkel:2016gxg}. For a non-simply-laced affine Lie algebra $\hat{\mathfrak g}$, $\hat{\mathfrak g}^{\vee}$ becomes the twisted affine Lie algebra.

\section{Q-function and the dual linear problem}\label{sec:Q-function}
In the previous section, we derived the Bethe ansatz equations for the Q-functions.
An approach to solve them numerically is to convert them into the non-linear integral equation, which will be discussed in the next section.
In this section we explore a method to find the zeros of the Q-function directly from the viewpoint of the linear problem.

For the ABCD type (pseudo-)ODEs, which correspond to $A_r^{(1)},\qty(B_r^{(1)})^{\vee},\qty(C_r^{(1)})^{\vee}$, and $D_r^{(1)}$ type linear problems, the Q-functions for the anti-symmetric tensor products of the first fundamental representation are represented as the Wronskians \cite{Dorey:2000ma,Dorey:2006an}. However, for $D_r^{(1)}$ type, the Q-function corresponding to spinor and conjugate spinor representations, which cannot be expressed in the anti-symmetric tensor product of the first fundamental representation, were missed.
To study the Q-function numerically in an efficient way, it is convenient to introduce the ``adjoint ODE''\cite{Dorey:2006an}, which is satisfied by the dominant cofactor in the Wronskian. However, these ``adjoint ODEs'' were only known for the Q-function of the first fundamental representation of $A_r^{(1)},\qty(B_r^{(1)})^{\vee},\qty(C_r^{(1)})^{\vee}$, and $D_r^{(1)}$.

We will find a general expression of Q-functions for any fundamental representation of untwisted affine Lie algebras by using the asymptotic solution and the solution of the dual linear problems in section \ref{sec:Dual linear problem and Q-functions}. The dual linear problems are defined based on the dual representations. They reduce to the ``adjoint ODEs'' for the first fundamental representations. 
In section \ref{sec:Asymptotic behaviour of Q-function}, we will study the asymptotics of the Q-functions by the WKB approximation.
In section \ref{sec:Cheng's algorithm for linear problems}, we show that Cheng's algorithm, which was applied in \cite{Dorey:2001uw, Dorey:2006an} to obtain the power series solution of the ODE near the origin, can be also applied to the linear problems.
In the last subsection, we explain how the zeros of Q-functions can be obtained by using our new expression.

\subsection{Dual linear problem and Q-functions}\label{sec:Dual linear problem and Q-functions}
Let us consider the linear problem for an affine Lie algebra $\hat{\mathfrak{g}}$ on a representation $V$.
The subdominant solution is expanded in the basis of the solutions around the origin as in (\ref{eq:basis_exp_of_psi}).
The coefficient $\mathcal{Q}_i(E)$ can be obtained as the ratio of the Wronskians:
\begin{equation}
    \mathcal{Q}_{i}(E,l)=
    \text{det} [\mathcal{X}_1,\ldots,\underbrace{\Psi}_{\mathclap{\text{i-th column}}},\ldots,\mathcal{X}_n]
    /\det[\mathcal{X}_1,\ldots,\mathcal{X}_n]. \label{eq:deteminant_def_of_Qs}
\end{equation}
Substituting the WKB expansion (\ref{eq:WKB_sol}) to $\Psi$ and the power series solutions (\ref{eq:small_asymp}) to (\ref{eq:deteminant_def_of_Qs}), one obtains the formula for $\mathcal{Q}_i(E,l)$.
Practically, we need to use finite series approximation of $\mathcal{X}_i$, which allows $x$-dependence on $\mathcal{Q}_i(E,l)$.
For Lie algebras with lower rank, it is found that ${\mathcal Q}_1(E,l)=0$ reproduces the Bethe roots for sufficiently large $x$, where $x$ belongs to both the domains of convergence of $\Psi$ and ${\cal X}_i$.
However, for Lie algebras with higher rank, this formula is difficult to treat numerically due to the large size of the determinant.
In \cite{Dorey:2006an}, where (\ref{eq:deteminant_def_of_Qs}) is represented as the Wronskians of the solutions of the (pseudo-)ODE, it is expanded by the sub-determinants along the column of $\Psi$.
The leading term is shown to be a single sub-determinant which is expressed in terms of ${\cal X}_i$.
Moreover the sub-determinant is shown to obey the adjoint differential equation to the original ODE.
One can use Cheng's algorithm to make a power series expansion of the sub-determinant.
This approach reproduces precise numerical values of the Bethe roots of $Q^{(1)}$, calculated by the NLIEs.

In this paper, we study the matrix version of the adjoint differential equation and a generalization of Cheng's algorithm to the linear differential system.
Consider the dual space $V^*$ of $V$.
There is an inner product $\langle\cdot,\cdot\rangle:V^*\times V\to \mathbf{C}$, where the action of $\mathfrak{g}$ on $V^*$ is defined by
\begin{equation}
    \langle  X^* \psi^*,\chi\rangle = \langle\psi^*,X\chi\rangle,
    \quad \chi\in V,\quad \psi^*\in V^*, \quad X\in \mathfrak{g}.
\end{equation}
Then $-X^*$ satisfy the Lie algebra $\mathfrak{g}$, which defines the dual representation of $\mathfrak{g}$ \cite{Fulton-Harris}.
When $V$ has an $\mathfrak{g}$-invariant quadratic form ${^t \psi} J \chi$ ($\psi,\chi\in V$), where $J$ is an invertible symmetric matrix, then identifying $V^*$ with $V$ via this quadratic form, the dual representation is given by $-J^{-1}{^tX}J$. Here ${^tX}$ denotes the transpose of $X$.
We denote $\bar{\psi}$ for the vector in $V$ corresponding to $\psi^*\in V^*$ by this identification.
Typically, $J$ is the identity matrix, but one can also choose $J$ as the matrix which corresponds to change of the basis due to an automorphism of the Lie algebra.

For the linear differential system (\ref{eq:linear_prob}) and the quadratic form given above, we define the dual linear differential equations by
\begin{equation}
    \mathcal{L}_{\mathfrak{g}}^{\text{dual}}\ \bar{\Psi}(x,E,l) = \qty[\dv{x}+\bar{A}_{\mathfrak{g}}]\bar{\Psi}(x,E,l)=0,\qquad \bar{A}_{\mathfrak{g}}=-J^{-1}\ {^tA_{\mathfrak{g}}}J. \label{eq:dual_linear_problem}
\end{equation}
We follow the same procedure in section \ref{sec:linear_problem_Qfunction} to introduce the subdominant solution $\bar{\Psi}$ and the basis $\bar{\cal X}_i$. 
For a solution $\Psi$ to the linear problem and a solution $\bar{\Psi}$ to the dual linear problem, the inner product $\langle \Psi^*, \Psi\rangle$ is independent of $x$.
In particular, for a basis  $\mathcal{X}_i$ of solutions around the origin, we  define dual basis $\bar{\mathcal{X}}_i$ satisfying
\begin{equation}
    \langle \mathcal{X}^*_i,\mathcal{X}_j\rangle ={}^t\bar{\mathcal{X}}_iJ\mathcal{X}_j= \delta_{ij}. \label{eq:normalization_Xbar}
\end{equation}
Then $\bar{\mathcal{X}}_i$ behaves around the origin as:
\begin{equation}
    \bar{\mathcal{X}}_i(x, E,l) = x^{h_i\cdot g} J^{-1}\mathbf{e}_i + \cdots, \qquad x\to0 \;. \label{eq:small_asymp_dual}
\end{equation}
Using \eqref{eq:normalization_Xbar}, $\mathcal{Q}_{i}(E,l)$ in (\ref{eq:basis_exp_of_psi}) is found to be
\begin{equation}
    \mathcal{Q}_{i}(E,l)=\langle\mathcal{X}^*_i,\Psi\rangle. \label{eq:general_def_of_Qs}
\end{equation}
From this formula, we can see that the Q-function in the tensor product of two representations becomes the product of two Q-functions.
Applying this rule to the anti-symmetric product representation, we obtain the Wronskian formula for the Q-functions as in (\ref{eq:A_Qfunc_from_V1}).
Note that for $A_r^{(1)}$ and $D^{(1)}_{r}$ in the fundamental representation $V^{(1)}$, the dual linear problem is equivalent to the adjoint ODE for the bottom component of $\bar{\Psi}$ \cite{Dorey:2006an}.

Let us check equivalence of two definitions (\ref{eq:deteminant_def_of_Qs}) and (\ref{eq:general_linear_problem}) explicitly by taking $A_1^{(1)}$ as an example.
For the representation $V^{(1)}$ of $A_1$ with the weight vectors $\mathbf{e}_{1,2}\in\mathbf{R}^2$, let $\mathcal{X}_i=\chi_{i,1}{\bf e}_1+\chi_{i,2}{\bf e}_2\;(i=1,2)$ be the basis solution of the linear problem \eqref{eq:linear_problem_A1} and 
$\bar{\mathcal{X}}_i=\bar{\chi}_{i,1}\mathbf{e}_1+\bar{\chi}_{i,2}\mathbf{e}_2\;(i=1,2)$ the basis solution of the dual linear problem which is given by
\begin{equation}
    \qty[\begin{pmatrix} \dv{x}+\frac{l_1}{x} & 0 \\ 0 & \dv{x}-\frac{l_1}{x}\end{pmatrix} -  \begin{pmatrix} 0 & p(x,E)\\1 &0 \end{pmatrix}]\begin{pmatrix} \bar{\psi}_1\\ \bar{\psi}_2\end{pmatrix}=0\, . \label{eq:dual_linear_problem_A1}
\end{equation}
Near origin, the behaviours of the basis and dual basis are
\begin{equation}
\begin{aligned}
    &\mathcal{X}_1\sim x^{l_1}\mathbf{e}_1+\cdots,\qquad \mathcal{X}_2\sim x^{-l_1}\mathbf{e}_2+\cdots,\\
    &\bar{\mathcal{X}}_1\sim x^{-l_1}\mathbf{e}_1+\cdots,\qquad \bar{\mathcal{X}}_2\sim x^{l_1}\mathbf{e}_2+\cdots,\qquad x\to 0.
\end{aligned}\label{eq:dual_small_asym_A1}
\end{equation}
Since the function $\mathcal{X}_1^{\prime}\coloneqq\chi_{2,2}\mathbf{e}_1-\chi_{2,1}\mathbf{e}_2$ has the same asymptotic behaviour of $\bar{\mathcal{X}}_1$ and satisfies the dual linear problem, one finds $\mathcal{X}_1^{\prime}=\bar{\mathcal{X}}_1$.
Similarly, the function $\mathcal{X}_2^{\prime}\coloneqq-\chi_{1,2}\mathbf{e}_1+\chi_{1,1}\mathbf{e}_2$ is equal to $\bar{\mathcal{X}}_2$. Then, one finds the relations:
\begin{equation}
    \bar{\chi}_{1,1}=\chi_{2,2},\qquad\bar{\chi}_{1,2}=-\chi_{2,1},\qquad\bar{\chi}_{2,1}=-\chi_{1,2},\qquad\bar{\chi}_{1,2}=\chi_{1,1}. \label{eq:component_relation_of_basis_A1}
\end{equation}
Noting that $\det[\mathcal{X}_1,\mathcal{X}_2]=1$, one can see the definition \eqref{eq:general_def_of_Qs} is equivalent to the definition of $\mathcal{Q}_i$ in \eqref{eq:deteminant_def_of_Qs}:
\begin{align*}
    \langle\mathcal{X}_i^{\ast},\Psi\rangle&=\bar{\chi}_{i,1}\psi_1+\bar{\chi}_{i,2}\psi_2 \\
    &=(-1)^{i}\qty(\chi_{i,1}\psi_1-\chi_{i,2}\psi_2)=(-1)^i\det[\mathcal{X}_i,\Psi].
\end{align*}

\subsection{Asymptotic behaviour of Q-function}\label{sec:Asymptotic behaviour of Q-function}
We now discuss the asymptotic behaviour of Q-function $Q(E,l)$ for $|E|\gg1$ with ${\rm arg }(-E)<\pi$. 
Setting the initial point of the integration to infinity in \eqref{eq:subdominant}, the subdominant solution $\Psi$ is given by
\begin{equation}
    \Psi(x,E)\sim C \exp\qty(\nu\int_x^{\infty}\qty(p(x^\prime,E)^{1/h}-x^{\prime M})dx^\prime-\nu\frac{x^{M+1}}{M+1}) e^{-1/h\log p(x,E)\rho^{\vee}\cdot H}\bm{\nu}\;,
\end{equation}
where $C$ is a constant.
Choosing $x=0$, the logarithm of the inner product (\ref{eq:general_def_of_Qs}) becomes
\begin{equation}
    \log\;\langle \mathcal{X}^*_1(0,E,l),\Psi(0,E)\rangle
    =\nu\int_0^{\infty}\qty(p(x,E)^{1/h}-x^M)dx+\cdots\ . 
\end{equation}
The integral can be evaluated by using the formula
\begin{equation}
    \kappa(a,b)\coloneqq\int_0^{\infty}\qty((x^a+1)^{1/b}-x^{a/b})dx=\frac{\Gamma(1+1/a)\Gamma(1+1/b)\sin\frac{\pi}{b}}{\Gamma(1+1/a+1/b)\sin\qty(\frac{\pi}{a}+\frac{\pi}{b})}.
    \label{eq:kappa}
\end{equation}
Then one finds that the asymptotic behaviour of $Q(E, l)$ is given by
\begin{equation}
    \log Q(E,l) = \nu\kappa(hM,h)\qty(-E)^{\frac{M+1}{hM}}+\cdots ,\quad  |E|\rightarrow \infty,\quad {\rm arg}(-E)<\pi.
    \label{eq:asym_of_Q}
\end{equation}
Under the assumption of the analyticity of $Q(E,l)$ in the complex $E$-plane and the fact that the exponent of $E$ is less than 1 for $M>1/(h-1)$, the Hadamard factorization theorem says that $Q(E,l)$ factorizes as \cite{Bazhanov:1998wj}
\begin{equation}
    Q(E,l)=Q(0,l)\prod_{i=0}^{\infty}\qty(1-\frac{E}{E_i}), \label{eq:Hadamard}
\end{equation}
where $Q(0,l)$ is a constant and $E_i$ is zeros of $Q(E,l)$.

For the representation $V^{(a)}$, the large $(-E)$ behaviour of the Q-function $Q^{(a)}(E)$ is characterized by the eigenvalue $\nu^{(a)}$ of $\Lambda_+$.
For the linear problems of  simply-laced affine Lie algebra \cite{Masoero:2015lga}, $\nu^{(a)}$ obey the relation
\begin{equation}
    2\nu^{(a)}\cos\frac{\pi}{h} =\sum_{b=1}^r\qty(2\delta_{ab}-C_{ab})\nu^{(b)}, \label{eq:relation_of_nu}
\end{equation}
The vector $(\nu^{(1)},\ldots, \nu^{(r)})$ satisfying \eqref{eq:relation_of_nu} is the Perron-Frobenius (PF) eigenvector of the Cartan matrix \cite{Braden:1989bu,ZinnJustin:1997at,Dunning:2002cu,Dorey:2006an}. 
Table \ref{tab:nu1_of_simply-laced} shows the eigenvalue $\nu^{(1)}$ for simply-laced Lie algebras in the representation $V^{(1)}$.
Table \ref{tab:Perron-Frobenius_eigenvectors} shows the ratio $M_a\coloneqq\nu^{(a)}/\nu^{(1)}$.
These data will be also used in section \ref{sec:NLIE}.
\renewcommand{\arraystretch}{2}
\begin{table}[H]
    \renewcommand{\thetable}{\arabic{table}-a}
    \centering
    \scalebox{0.9}{
    \begin{tabular}{c||c|c|c|c|c}
        $\mathfrak{g}$ & $A_r$ & $D_r$     & $E_6$ & $E_7$ & $E_8$ \\\hline
        $\nu^{(1)}$    & $1$   & $\sqrt{2}$& $\sqrt{2 \sqrt{6} \cos \left(\frac{\pi }{12}\right)}$ & $2\sqrt{2}\cos\qty(\frac{\pi}{18})$ &$\sqrt{512 \sqrt{3} \sin \left(\frac{\pi }{30}\right) \sin \left(\frac{\pi }{5}\right) \cos ^2\left(\frac{2 \pi }{15}\right) \cos ^4\left(\frac{\pi }{5}\right)}$
    \end{tabular}
    }
    \caption{The first component of the PF vector $\nu^{(1)}$ for simply-laced Lie algebras}
    \label{tab:nu1_of_simply-laced}
\end{table}
\renewcommand{\arraystretch}{1}

\renewcommand{\arraystretch}{2}
\begin{table}[H]
    \addtocounter{table}{-1}
    \renewcommand{\thetable}{\arabic{table}-b}
    \centering
    \footnotesize
    \scalebox{0.9}{
    \begin{tabular}{c|c}
        $\mathfrak{g}$ & $M_a$\\
        \hline \hline
        $A_r$ ($h=r+1$) & $M_a = \dfrac{\sin(a\pi/h)}{\sin(\pi/h)}, \quad (a = 1, \dots, r)$\\
        \hline
        $D_r$ ($h=2r-2$) & $M_a = \dfrac{\sin(a\pi/h)}{\sin(\pi/h)}, \quad (a = 1, \dots, r - 2), \qquad M_{r - 1} = M_r = \dfrac{1}{2\sin(\pi/h)}$\\
        \hline
        $E_6$ ($h=12$) & $M_1 = M_5 = 1,  \quad M_2 = M_4 = \dfrac{\sin(2\pi/h)}{\sin(\pi/h)}, \quad M_3 = \dfrac{\sin(3\pi/h)}{\sin(\pi/h)}, \qquad M_6 = \dfrac{\sin(3\pi/h)}{\sin(2\pi/h)}$\\
        \hline
        $E_7$ ($h=18$) & $M_1 = 1, \quad M_2 = \dfrac{\sin(2\pi/h)}{\sin(\pi/h)}, \quad M_3 = \dfrac{\sin(3\pi/h)}{\sin(\pi/h)}, \quad M_4 = \dfrac{\sin(4\pi/h)}{\sin(\pi/h)}$\\
        & $M_5 = \dfrac{\sin(5\pi/h)}{\sin(\pi/h)} - \dfrac{\sin(4\pi/h)}{\sin(2\pi/h)}, \quad M_6 = \dfrac{\sin(6\pi/h)}{\sin(\pi/h)} - \dfrac{\sin(4\pi/h)}{\sin(\pi/h)}, \quad M_7 = \dfrac{\sin(4\pi/h)}{\sin(2\pi/h)}$\\
        \hline
        $E_8$ ($h=30$) & $M_1 = 1, \quad M_2 = \dfrac{\sin(2\pi/h)}{\sin(\pi/h)}, \quad M_3 = \dfrac{\sin(3\pi/h)}{\sin(\pi/h)}, \quad M_4 = \dfrac{\sin(4\pi/h)}{\sin(\pi/h)}, \qquad M_5 = \dfrac{\sin(5\pi/h)}{\sin(\pi/h)}$,\\
        & $M_6 = \dfrac{\sin(6\pi/h)}{\sin(\pi/h)} - \dfrac{\sin(5\pi/h)}{\sin(2\pi/h)}, \quad M_7 = \dfrac{\sin(7\pi/h)}{\sin(\pi/h)} - \dfrac{\sin(5\pi/h)}{\sin(\pi/h)}, \quad M_8 = \dfrac{\sin(5\pi/h)}{\sin(2\pi/h)}$
    \end{tabular}
    }
    \caption{The PF vectors for simply-laced Lie algebras}
    \label{tab:Perron-Frobenius_eigenvectors}
\end{table}
\renewcommand{\arraystretch}{1}

\subsection{Cheng's algorithm for linear problems}\label{sec:Cheng's algorithm for linear problems}
We have seen that the $\mathcal{Q}_{i}(E,l)$ can be determined by the solution of the dual linear problem \eqref{eq:dual_linear_problem} around the origin and the subdominant solution. 
Here we discuss the method to obtain the solution of the linear problem around the origin.
According to \cite{Dorey:2006an}, we call this approach Cheng's algorithm \cite{PhysRev_127_647}, which has been used to obtain power series solutions of the Schr\"odinger equation by iteration. 

Let us consider $n$-dimensional linear problem of the form:
\begin{align}
    {\cal L}\Psi(x)=0, \quad {\cal L}={\cal D}[\bm{q}]+\mathcal{P},
    \label{eq:general_linear_problem}
\end{align}
where
\begin{equation}
    \mathcal{D}[\bm{q}] \coloneqq \mathbf{I}_{n}\dv{x} - \frac{\bm{q}}{x},
    \quad \bm{q}=\mathrm{diag}(q_1,\ldots,q_{n}),
    \label{eq:Cheng_linear_op}
\end{equation}
with $\sum_{i=1}^{n}q_i=0$ and $\mathcal{P}$ is an off-diagonal matrix whose elements are a polynomial in $x$.
We want to find the power series solution of the linear problem \eqref{eq:general_linear_problem} around $x=0$.
We introduce the operator $\mathbf{L}[\bm{q}]$ which acts on the vector $\bm{v}={^t\qty(x^{p_1}, \dots, x^{p_n})}$, whose components are power functions of $x$, as
\begin{equation}
    \mathbf{L}[\bm{q}]\;\bm{v} = {^t\qty(\frac{x^{p_1 + 1}}{p_1 + 1 - q_1},\ldots, \frac{x^{p_{n} + 1}}{p_1 + 1 - q_{n}})}\, .
\end{equation}
$\mathbf{L}[\bm{q}]$ is the inverse operator of $\mathcal{D}[\bm{q}]$ on $\bm{v}$:
\begin{equation}
    \mathbf{L}[\bm{q}]\;\mathcal{D}[\bm{q}]\;\bm{v} = \mathcal{D}[\bm{q}]\;\mathbf{L}[\bm{q}]\;\bm{v} = \bm{v}\, .
     \label{eq:ca_id}
\end{equation}
Using $\mathbf{L}[\bm{q}]$, we construct the following series of the functions of $x$ iteratively:
\begin{equation}
    \mathcal{X}^{m+1}_{i}(x)=\mathcal{X}^{0}_{i}(x)-\mathbf{L}[\bm{q}]\left(\mathcal{P}\mathcal{X}^{m}_{i}(x)\right)\, ,\qquad m=0,1,2,\ldots,
    \label{eq:chengs_sol_iter}
\end{equation}
where
\begin{equation}
    \mathcal{X}^{0}_{i}(x)\coloneqq{^t( 0,\ldots,\underbrace{x^{q_i}}_{\mathclap{\text{i-th component}}},\ldots,0)}.
    \label{eq:chengs_sol_iter1}
\end{equation}
Then the $m\rightarrow \infty$ limit
\begin{align}
    \mathcal{X}_{i}(x)\coloneqq\lim_{m\rightarrow \infty}\mathcal{X}^{m}_{i}(x)
\end{align}
satisfy the linear problem (\ref{eq:general_linear_problem}).
This is proved as follows:
\begin{equation}
    \begin{aligned}
        \mathcal{L}\;\mathcal{X}_{i}(x) &= (\mathcal{D}[\bm{q}] + \mathcal{P})(\mathcal{X}^{0}_{i}(x) - \mathbf{L}[\bm{q}]\qty(\mathcal{P}\mathcal{X}_{i}(x)))\\
        &= \mathcal{P}\qty(\mathcal{X}^{0}_{i}(x) - \mathcal{X}_{i}(x) - \mathbf{L}[\bm{q}](\mathcal{P}\mathcal{X}_{i}(x)))= 0,
    \end{aligned}
    \label{eq:Cheng's_algorithm}
\end{equation}
where we have used ${\cal D}[\bm{q}]\mathcal{X}^{0}_{i}(x)=0$ and (\ref{eq:ca_id}).
We can also check the linear independence of $\left\{ \mathcal{X}_{1}(x),\ldots,\mathcal{X}_{n}(x) \right\}$, which provides the basis of the solution of the linear problem (\ref{eq:general_linear_problem}) around the origin.
Applying this algorithm, we can construct the power series solutions of the linear problem \eqref{eq:linear_prob} and its dual linear problem \eqref{eq:dual_linear_problem} by iteration.

As an example, we demonstrate this procedure for $A_1^{(1)}$.
The matrix form of the linear problem is given by \eqref{eq:linear_problem_A1}. 
The first term of \eqref{eq:linear_problem_A1} in the bracket is $\mathcal{D}[\bm{q}]$ with $\bm{q}=(l_1,-l_1)$ and the second term corresponds to $\mathcal{P}$.
The formal series expansion of the basis $\mathcal{X}_1^{(1)},\mathcal{X}_2^{(1)}$ with the initial condition $\mathcal{X}_1^{(1),0}={^t(x^{l_1},0)}$ and $\mathcal{X}_2^{(1),0}={^t(0,x^{-l_1})}$ are obtained.
If we set $l_1=0,p(x,E)=x^4-E$, we obtain
\begin{equation}
    \mathcal{X}_1^{(1)}={^t\qty(\chi_{1,1},\chi_{1,2})},\quad\quad \mathcal{X}_2^{(1)}={^t\qty(\chi_{2,1},\chi_{1,1})},\label{eq:cheng_sol_A1}
\end{equation}
where $\chi_{1,1},\chi_{1,2},\chi_{2,1}$ are given by
\begin{equation}
    \begin{aligned}
    \chi_{1,1}&=1-\frac{E x^2}{2}+\frac{E^2x^4}{24}+\frac{(24-E^3) x^6}{720}+\cdots,\\
    \chi_{1,2}&=Ex - \frac{E^2 x^3}{6} + \frac{(-24+E^3)x^5}{120}+\cdots, \\
    \chi_{2,1}&=-x+\frac{E x^3}{6}-\frac{E^2 x^5}{120}-\frac{(120-E^3)x^7}{5040}+\cdots.
    \end{aligned}\label{eq:component_cheng_sol_A1}
\end{equation}

The dual linear problem is given by \eqref{eq:dual_linear_problem_A1}, where the first term in the bracket is $\mathcal{D}[\bm{q}]$ with $\bm{q}=(-l_1,l_1)$ and the second term is $\mathcal{P}$. 
Setting the initial conditions $\bar{\mathcal{X}}_1^{(1),0}={^t(x^{-l_1},0)}$ and $\bar{\mathcal{X}}_2^{(1),0}={^t(0,x^{l_1})}$, one obtains the formal series solution
\begin{equation}
    \bar{\mathcal{X}}_1^{(1)}={^t\qty(\chi_{1,1},-\chi_{2,1})},\quad \bar{\mathcal{X}}_2^{(1)}={^t\qty(-\chi_{1,2},\chi_{1,1})}.
    \label{eq:cheng_sol_a1_dual}
\end{equation}
We can also check the normalization \eqref{eq:normalization_Xbar}.
As a non-trivial example, in appendix \ref{sec:Cheng_E6}, we will demonstrate Cheng's algorithm for the $E_6^{(1)}$ linear problem in $V^{(1)}$.

\subsection{Zeros of the Q-function}\label{sec:Zeros of the Q-function}
We now explain a procedure to find the zeros of $\mathcal{Q}_i(E,l)$.
For the linear problem (\ref{eq:linear_prob}) based on $\hat{\mathfrak{g}}$ and the representation $V$ of $\mathfrak{g}$, one can construct $\mathcal{Q}_i(E,l)$ by (\ref{eq:general_def_of_Qs}).
Expanding the dual solution and the subdominant solution in terms of orthonormal weight vectors of $V$ as $\bar{\mathcal{X}}_i=\sum_{i=1}^n\bar{\chi}_{i,j}\mathbf{e}_j$ and $\Psi=\sum_{j=1}^n\psi_j\mathbf{e}_j$, then $\mathcal{Q}_i(E,l)$ becomes
\begin{equation}
    \mathcal{Q}_i(E,l)=\langle \mathcal{X}^*_i,\Psi\rangle=\sum_{j=1}^n \bar{\chi}_{i,j}\ \psi_j\,. \label{eq:component_of_inner_prod}
\end{equation}
As shown previously, this is independent of $x$. 
In order to evaluate the roots of the Q-function numerically, one needs to use the truncated power series solutions.
In this case, it depends on $x$.
We evaluate $\mathcal{Q}_i(E,l)$ approximately at finite but sufficiently large $x$.

To look for the zeros of the Q-function represented by (\ref{eq:component_of_inner_prod}) at finite $x$, we need to consider only the most dominant contribution in the sum.
Among the components of $\Psi$ given by \eqref{eq:subdominant}, the lowest weight component $\psi_n$ corresponds to such a term.
Then, for large but finite $x$, $\mathcal{Q}_i(E,l)$ can be approximated by
\begin{equation}
    \mathcal{Q}_{i}(E,l)\;\sim\;  \bar{\chi}_{i,n}(x_{\text{fixed}},E,l)\ \psi_n(x_{\text{fixed}}) \qquad \text{for}\ x_{\text{fixed}}\gg1. \label{eq:approximation__of_Qi}
\end{equation}
From the solution of the dual linear problems obtained by using Cheng's algorithm, we can determine the zeros of $\mathcal{Q}_i(E,l)$ numerically, which are actually given by the solutions of $\bar{\chi}_{i,n}(x_{\text{fixed}},E,l)=0$.
We are particularly interested in the roots of $Q(E,l)$, which are expected to give the solutions of the BAEs.

Let us explain the above procedure by taking an example of $A_1^{(1)}$ with the potential $p(x,E)=x^4-E$ and the monodromy $l_1=0$. 
The dual linear basis $\{ \bar{\mathcal{X}}_{1}^{(1)},\bar{\mathcal{X}}_{2}^{(1)} \}$ has the component as in \eqref{eq:cheng_sol_a1_dual} and the Q-function is evaluated by the bottom component of $\bar{\mathcal{X}}_{1}^{(1)}$.
Since we can regard $\psi_2^{(1)}(x_{\text{fixed}})$ as a constant at the reference point $x_{\text{fixed}}$, $\mathcal{Q}_i$ can be evaluated by using \eqref{eq:approximation__of_Qi}.
Setting the iteration number of the Cheng's algorithm to be $160$, the series solutions $\bar{\mathcal{X}}^{(1)}_{1}$ and $\bar{\mathcal{X}}^{(1)}_{2}$ satisfy the dual linear problems up to $O\qty(x^{160})$.
At the reference point $x_{\text{fixed}}=5$, one obtains the spectra in table \ref{tab:A1_x^4_(0)} by solving $\mathcal{Q}_1=0$ and $\mathcal{Q}_2=0$. We also show the spectra calculated from thermodynamic Bethe ansatz (TBA) equations in \cite{Dorey:1998pt} for comparison.
\begin{table}[H]
    \centering
    \begin{tabular}{c|cc}
        $i$ & $E_i$ (ODE) & $E_i$ (TBA)\\\hline
        $0$ & $1.06036209048418$    &  $1.06036209048418$ \\
        $1$ & $3.79967302980140$    &  $3.79967302980139$ \\
        $2$ & $7.45569793798635$    &  $7.45569793798672$ \\
        $3$ & $11.6447455113751$    &  $11.6447455113781$ \\
    \end{tabular}
    \caption{Spectra from the ODE and the TBA \cite{Dorey:1998pt} for $A_1^{(1)}$}
    \label{tab:A1_x^4_(0)}
\end{table}
\noindent
One can see the spectra from \eqref{eq:approximation__of_Qi} and the TBA coincides up to $13$ digits.
In section \ref{sec:comparison}, we calculate numerically the spectrum of the linear problems \eqref{eq:linear_prob} for other higher-rank affine Lie algebras, and compare the spectra with those obtained from the NLIEs.

\section{Non-linear integral equations}\label{sec:NLIE}

In this section, we first derive a set of non-linear integral equations (NLIEs) from the Bethe ansatz equations for simply-laced Lie algebras, following \cite{Bazhanov:1996dr,Destri:1992qk,Destri:1994bv,Dorey:1999pv,Dorey:2000ma}.
We then discuss the folding of the NLIEs of simply-laced affine Lie algebras and obtain the NLIEs associated with the folded Dynkin diagrams.
This is equivalent to imposing a condition that makes the solution of BAE symmetric in the simply-laced case.

We start with the Bethe ansatz equations \eqref{eq:BAE_for_simply-laced} for a simply-laced Lie algebra $\mathfrak{g}$ with rank $r$, which is obtained from the linear problem.
In the context of quantum integrable model \cite{Babelon:1982gp,deVega:1986xj,reshetikhin1991algebraic,Reshetikhin:1986vd,1987LMaPh..14..235R,Dunning:2002cu,ZinnJustin:1997at}, the Q-function is the vacuum expectation value of Baxter's Q-operator, $E_i^{(a)}$ zeros of the Q-function $Q^{(a)}(E)$, called the Bethe roots and $\gamma_a$ the twist parameters.
The constant $\Omega$ is defined by $\Omega = \exp(\frac{2\pi i}{h\mu})$, where $ \mu = \frac{1 + M}{hM}$
is the anisotropy parameter.

We introduce the counting functions by
\begin{equation}
    \mathfrak{a}^{(a)}(E) \coloneqq \prod_{b = 1}^{r}\Omega^{-C_{ab}\gamma_b/2}\frac{Q_{[-C_{ab}/2]}^{(b)}(E)}{Q_{[C_{ab}/2]}^{(b)}(E)}, \qquad a = 1, \dots, r.
    \label{eq:counting_function}
\end{equation}
The BAEs \eqref{eq:BAE_for_simply-laced} are equivalent to $\mathfrak{a}^{(a)}(E_k^{(a)}) = -1$.
We regard $\mathfrak{a}^{(a)}(E)$ as a function of $\theta = \mu \log{E}$, which is denoted as $\mathfrak{a}^{(a)}(\theta)$.

When the Q-function is factorized as in (\ref{eq:Hadamard}), we use the residue theorem in the $\theta$-plane to translate the infinite product in ${\mathfrak{a}^{(a)}(\theta)}$ to a contour integral encircling all the zeros.
On the assumption that all of the Bethe roots are real and positive, it is straightforward to transform the Bethe ansatz equations into the non-linear integral equations:
\begin{equation}
    \begin{aligned}
        \ln{\mathfrak{a}^{(a)}(\theta)} = i\pi\Hat{\alpha}_a - i b_0 M_a e^{\theta} &+ \sum_{b = 1}^{r}\int_{\mathcal{C}_1}\dd{\theta'}\varphi_{ab}(\theta - \theta')\ln(1 + \mathfrak{a}^{(b)}(\theta'))\\
        &\quad - \sum_{b = 1}^{r}\int_{\mathcal{C}_2}\dd{\theta'}\varphi_{ab}(\theta - \theta')\ln(1 + \frac{1}{\mathfrak{a}^{(b)}(\theta')})
        \label{eq:NLIE_for_ADE},
    \end{aligned}
\end{equation}
where $b_0$ is a constant.
The contour $\mathcal{C}_1$($\mathcal{C}_2$) run from $-\infty$ to $+\infty$, just below (above) the real $\theta$-axis.
The kernel function is defined by
\begin{equation}
    \varphi_{ab}(\theta) = \int_{-\infty}^{\infty}\frac{\dd{k}}{2\pi}e^{ik\theta}\qty(\delta_{ab} - \frac{\sinh(\mu\pi k)}{\sinh((h\mu - 1)\pi k/h)\cosh(\pi k/h)}C_{ab}^{-1}(k)),
\end{equation}
where $C_{ab}(k)$ is the deformed Cartan matrix:
\begin{equation}
    C_{ab}(k) \coloneqq
    \begin{dcases}
        2, & a = b,\\
        \frac{C_{ab}}{\cosh(\pi k/h)}, & a \neq b,
        \label{eq:deformed_Cartan_matrix}
    \end{dcases}
\end{equation}
and $C_{ab}$ the Cartan matrix of ${\mathfrak g}$.
The constant $\Hat{\alpha}_a$ arises from the phase factors $\Omega^{-C_{ab}\gamma_b}$ in the BAEs and is related to $\gamma_a$ by
\begin{equation}
    \Hat{\alpha}_a \coloneqq (1 - h\mu)^{-1}\;\gamma_a,
\end{equation}
where $\gamma_a$ is defined in \eqref{eq:twist_parameter}.
The constant $M_a$ is derived from the zero modes of the kernel and becomes the Perron-Frobenius vector satisfying \eqref{eq:relation_of_nu}, which is given in table \ref{tab:Perron-Frobenius_eigenvectors}.
Here $M_a$ are normalized such that $M_1=1$.

In the large $E$ (or large $\theta$) limit, setting $l_a$ to be zero, the driving term $-ib_0M_ae^{\theta}$ is dominant and the integration part in the NLIE can be ignored. 
Then the BAEs can be solved approximately as
\begin{equation}
    \begin{aligned}
        E_n^{(a)} &\sim \qty(\frac{\pi}{b_0M_a}\qty(\omega_a \cdot \rho + 2n + 1))^{1/\mu}, \qquad n = 0, 1, 2, \ldots,
    \end{aligned}
    \label{eq:Bethe_root_asymptotics}
\end{equation}
which is valid for large $n$.
We can also recover this asymptotic formula from the WKB solution \cite{Dorey:1999pv}.
The WKB solution \eqref{eq:subdominant_sol} in $V^{(a)}$ defined for large $x$ is analytically continued to the region $x<x_0$ around the turning point $x_0 = E^{\frac{1}{hM}}$ \cite{Landau:1991wop}:
\begin{equation}
    \begin{aligned}
        \Psi^{(a)} &\sim \exp(\nu^{(a)}\cos{\frac{\pi}{h}}\int_{x}^{x_0}\abs{p(x)}^{1/h}\dd{x} - \frac{1}{h}\log{\abs{p(x)}}\rho^{\vee} \cdot H)\\
        &\qquad \times\cos(\nu^{(a)}\sin{\frac{\pi}{h}}\int_{x}^{x_0}\abs{p(x)}^{1/h}\dd{x} - \frac{\pi}{h}\rho^{\vee} \cdot H)\bm{\nu}^{(a)}\, .
    \end{aligned}
\end{equation}
At $x=0$, we impose the boundary condition such that $\Psi=0$.
Focusing on the first component $h_1^{(a)}=\omega_a$, which is the dominant term, we get the quantization condition:
\begin{equation}
    \nu^{(a)}\sin{\frac{\pi}{h}}\int_{0}^{x_0}\abs{p(x)}^{1/h}\dd{x} = \frac{\pi}{2}\qty(\frac{2}{h}\rho^{\vee} \cdot \omega_a + 2n + 1), \qquad n = 0, 1, 2, \ldots .
    \label{eq:quantization_condtion}
\end{equation}
Here the integral can be evaluated by using the formula
\begin{equation}
    \int_{0}^{1}(1 - x^a)^{1/b} \dd{x} = \frac{\sin(\pi/a + \pi/b)}{\sin(\pi/b)}\kappa(a, b),
\end{equation}
where $\kappa(a, b)$ is defined in \eqref{eq:kappa}. 
Identifying
\begin{equation}
    b_0 = 2\sin(\pi\mu)\kappa(hM, h)\nu^{(1)},\\
\end{equation}
the condition (\ref{eq:quantization_condtion}) exactly reproduces the asymptotic behaviour (\ref{eq:Bethe_root_asymptotics}) of the Bethe roots at large $\theta$.
The effective central charge can be calculated using the solution of the NLIEs as \cite{Dorey:1999pv,Dunning:2002cu}
\begin{equation}
    c_{\mathrm{eff}} = 2\times\frac{3}{\pi^2}\sum_{a}^{r}i b_0M_a\qty[\int_{\mathcal{C}_1}\dd{\theta}e^{\theta}\log(1 + \mathfrak{a}^{(a)}(\theta)) - \int_{\mathcal{C}_2}\dd{\theta}e^{\theta}\log(1 + \frac{1}{\mathfrak{a}^{(a)}(\theta)})].
    \label{eq:effective_central_charge}
\end{equation}
Note that this is given by the UV limit of the massive NLIEs \cite{ZinnJustin:1997at} and the factor $2$ comes from the the kink and anti-kink profiles of the counting function.
Evaluating (\ref{eq:effective_central_charge}) by using the $\theta \rightarrow -\infty$ limit of $\mathfrak{a}^{(a)}$:
\begin{equation}
    i\ln{\mathfrak{a}^{(a)}(-\infty)} = -\frac{\pi}{M + 1}\sum_{b = 1}^{r}C_{ab}\Hat{\alpha}_b,
    \label{eq:a_-infty}
\end{equation}
we find \cite{ZinnJustin:1997at,Dunning:2002cu}
\begin{equation}
    c_{\mathrm{eff}}^{\mathrm{UV}} = r - \frac{3}{M + 1}\sum_{a, b = 1}^{r}\Hat{\alpha}_aC_{ab}\Hat{\alpha}_b.
    \label{eq:effc}
\end{equation}
We have also confirmed the formula  \eqref{eq:effc} numerically.

In the following, we derive the NLIEs for the non-simply-laced affine Lie algebra $C_r^{(1)}$, $B_r^{(1)}$, $F_4^{(1)}$ and $G_2^{(1)}$ by using the folding procedure of $A_{2r-1}^{(1)}$, $D_{r+1}^{(1)}$, $E_6^{(1)}$ and $D_4^{(1)}$, respectively.

\paragraph{\underline{$C_{r}^{(1)}$}}
By the folding procedure of $A_{2r-1}$, we obtain $C_r$-type Lie algebra. 
The Perron-Frobenius vector satisfies $M_a=M_{2r-a}$  ($a=1,\ldots, r-1$). 
If we put the conditions $\Hat{\alpha}_a = \Hat{\alpha}_{2r - a}$ ($a=1,\ldots, r-1$) in the NLIEs \eqref{eq:NLIE_for_ADE}, we find that the equations for $\mathfrak{a}^{(a)}$ and $\mathfrak{a}^{(2r - a)}$ are the same.
Then by identifying $\mathfrak{a}^{(a)}$ and $\mathfrak{a}^{(2r - a)}$ in \eqref{eq:NLIE_for_ADE}, we get the set of $r$ integral equations:
\begin{equation}
    \begin{aligned}
        \ln{\widetilde{\mathfrak{a}}^{(a)}(\theta)} = i\pi\Hat{\widetilde{\alpha}}_a - ib_0\widetilde{M}_ae^{\theta} &+ \sum_{b = 1}^{r}\int_{\mathcal{C}_1}\dd{\theta'}\widetilde{\varphi}_{ab}(\theta - \theta')\ln(1 + \widetilde{\mathfrak{a}}^{(b)}(\theta'))\\
        &\quad - \sum_{b = 1}^{r}\int_{\mathcal{C}_2}\dd{\theta'}\widetilde{\varphi}_{ab}(\theta - \theta')\ln(1 + \frac{1}{\widetilde{\mathfrak{a}}^{(b)}(\theta')}),
    \end{aligned}\label{eq:reduced_NLIEs}
\end{equation}
where
\begin{equation}
    \Hat{\widetilde{\alpha}}_a = \Hat{\alpha}_a, \quad \widetilde{M}_a = M_a, \quad \widetilde{\mathfrak{a}}^{(a)} = \mathfrak{a}^{(a)} \qquad a = 1, \dots, r.
    \label{eq:identifying_condition}
\end{equation}
Here the kernel functions are defined by
\begin{equation}
    \widetilde{\varphi}_{ab}(\theta) = \int_{-\infty}^{\infty}\frac{\dd{k}}{2\pi}e^{ik\theta}\qty[\delta_{ab} - \frac{\sinh(\pi\mu k)}{\sinh((h_{C_r}\mu - 1)\pi k/h_{C_r})\cosh(\pi k/h_{C_r})}\qty(C_{C_r}^{-1})_{ab}(k)],
\end{equation}
which is obtained by the folding of the deformed Cartan matrix. The effective central charge (\ref{eq:effc}) becomes
\begin{equation}
    c_{\mathrm{eff}} = 2\qty(\frac{2r - 1}{2} - \frac{3}{M + 1}\sum_{a, b = 1}^{r}\Hat{\widetilde{\alpha}}_a(K_{C_r})_{ab}\Hat{\widetilde{\alpha}}_b),
\end{equation}
where $(K_{C_r})_{ab} \coloneqq (\beta_a^{\vee} \cdot \beta_b^{\vee})$ is the symmetrized Cartan matrix of $C_r$ where $\beta_a^{\vee}$ are the co-simple roots.

\paragraph{\underline{$B_r^{(1)}$}}
The Perron-Frobenius eigenvector of $D_{r+1}$ satisfies $M_r = M_{r + 1}$.
Imposing $\Hat{\alpha}_r = \Hat{\alpha}_{r + 1}$ in the NLIEs \eqref{eq:NLIE_for_ADE}, we find that the equations for $\mathfrak{a}^{(r)}$ and $\mathfrak{a}^{(r + 1)}$ are identical.
Then we get the reduced NLIEs (\ref{eq:reduced_NLIEs}) with \eqref{eq:identifying_condition}.
The kernel function $\widetilde{\varphi}_{ab}$ is given by
\begin{equation}
    \widetilde{\varphi}_{ab}(\theta) = \int_{-\infty}^{\infty}\frac{\dd{k}}{2\pi}e^{ik\theta}\qty[\delta_{ab} - \frac{\sinh(\pi\mu k)}{\sinh((h_{B_r}\mu - 1)\pi k/h_{B_r})\cosh(\pi k/h_{B_r})}(C_{B_r}^{-1})_{ab}(k)].
\end{equation}
The effective central charge becomes
\begin{equation}
    c_{\mathrm{eff}} = 2\qty(\frac{r + 1}{2} - \frac{3}{1 + M}\sum_{a, b = 1}^{r}\Hat{\widetilde{\alpha}}_a(K_{B_r})_{ab}\Hat{\widetilde{\alpha}}_b),
\end{equation}
where $(K_{B_r})_{ab} \coloneqq (\beta_a^{\vee} \cdot \beta_b^{\vee})$ and $\beta_a^{\vee}$ are the co-simple roots of $B_r$ type Lie algebra.

\paragraph{\underline{$F_4^{(1)}$}}
The Perron-Frobenius eigenvector of $E_6$ satisfies $M_1 = M_5$ and $M_2 = M_4$.
Imposing $\Hat{\alpha}_1 = \Hat{\alpha}_5$ and $\Hat{\alpha}_2 = \Hat{\alpha}_4$ in the NLIE \eqref{eq:NLIE_for_ADE}, we find that the equations for $\mathfrak{a}^{(1)}$ and $\mathfrak{a}^{(5)}$, $\mathfrak{a}^{(2)}$ and $\mathfrak{a}^{(4)}$ are identical.
Then we get the reduced NLIEs (\ref{eq:reduced_NLIEs}), where
\begin{equation}
    \begin{aligned}
        &\Hat{\widetilde{\alpha}}_a = \Hat{\alpha}_a, \quad \widetilde{M}_a = M_a, \quad \widetilde{\mathfrak{a}}^{(a)} = \mathfrak{a}^{(a)}, \qquad a = 1, 2, 3,\\
        &\Hat{\widetilde{\alpha}}_4 = \Hat{\alpha}_6, \quad \widetilde{M}_4 = M_6, \quad \widetilde{\mathfrak{a}}^{(4)} = \mathfrak{a}^{(6)}.
    \end{aligned}
\end{equation}
The kernel function $\widetilde{\varphi}_{ab}$ is defined by
\begin{equation}
    \widetilde{\varphi}_{ab}(\theta) = \int_{-\infty}^{\infty}\frac{\dd{k}}{2\pi}e^{ik\theta}\qty(\delta_{ab} - \frac{\sinh(\pi\mu k)}{\sinh((h_{F_4}\mu - 1)\pi k/h_{F_4})\cosh(\pi k/h_{F_4})}(C_{F_4}^{-1})_{ab}(k)),
\end{equation}
where $a,b = 1, \dots, 4$.
The effective central charge becomes
\begin{equation}
    c_{\mathrm{eff}} = 2\qty(3 - \frac{3}{1 + M}\sum_{a, b = 1}^{4}\Hat{\widetilde{\alpha}}_a(K_{F_4})_{ab}\Hat{\widetilde{\alpha}}_b),
\end{equation}
where $(K_{F_4})_{ab} \coloneqq (\beta_a^{\vee} \cdot \beta_b^{\vee})$ and $\beta_a^{\vee}$ are the co-simple roots of $F_4$ type Lie algebra.

\paragraph{\underline{$G_2^{(1)}$}}
The Perron-Frobenius eigenvector of $D_4$ satisfies $M_1 = M_3 = M_4$.
Restricting $\Hat{\alpha}_1 = \Hat{\alpha}_3 = \Hat{\alpha}_4$ in the NLIEs \eqref{eq:NLIE_for_ADE}, we find that the equations for $\mathfrak{a}^{(1)}$, $\mathfrak{a}^{(3)}$ and $\mathfrak{a}^{(4)}$ are the same one.
Then we get the  NLIEs (\ref{eq:reduced_NLIEs}) with \eqref{eq:identifying_condition}.
The kernel function $\widetilde{\varphi}_{ab}$ is defined by
\begin{equation}
    \widetilde{\varphi}_{ab}(\theta) = \int_{-\infty}^{\infty}\frac{\dd{k}}{2\pi}e^{ik\theta}\qty(\delta_{ab} - \frac{\sinh(\pi\mu k)}{\sinh((h_{G_2}\mu - 1)\pi k/h_{G_2})\cosh(\pi k/h_{G_2})}(C_{G_2}^{-1})_{ab}(k)),
\end{equation}
where $a,b = 1, 2$.
The effective central charge becomes
\begin{equation}
    c_{\mathrm{eff}} = 3\qty(\frac{4}{3} - \frac{3}{1 + M}\sum_{a, b = 1}^{2}\Hat{\widetilde{\alpha}}_a(K_{G_2})_{ab}\Hat{\widetilde{\alpha}}_b),
\end{equation}
where $(K_{G_2})_{ab} \coloneqq (\beta_a^{\vee} \cdot \beta_b^{\vee})$ and $\beta_a^{\vee}$ are the co-simple roots of $G_2$ type Lie algebra.

\section{Zeros of the Q-functions}\label{sec:comparison}
In this section, we compare the zeros of the Q-functions derived from the linear problem with the Bethe roots obtained from the NLIEs numerically for a simply-laced affine Lie algebra.
We also compare the spectra from the linear problem for a non-simply-laced affine Lie algebra with those of the NLIEs, which are obtained by folding procedure.
To solve the NLIEs, we use the fast Fourier transformation (FFT) with $2^{20}$ discrete points and the cutoff $16$.

\subsection{\texorpdfstring{$A_r^{(1)}$}{Ar}}
We begin with $A_r^{(1)}$ type affine Lie algebras.
In section \ref{sec:linear_problem_Qfunction}, we have seen that our approach of finding the zeros of the Q-function is in good agreement with those of the integrable model for $A_1^{(1)}$.
As an example, we present the comparison of the zeros of the Q-functions and the spectra from the NLIEs for $A_5^{(1)}$ in table \ref{tab:A5_x^2_(5/12,1/3,0,-1/3,-5/12)}. 
$Q^{(1)}$, $Q^{(2)}$ and $Q^{(3)}$ are calculated for the fundamental representation $V^{(1)}$ and its anti-symmetric products, while $Q^{(4)}$ and $Q^{(5)}$ are from $V^{(5)}$ and $\wedge^2 V^{(5)}$.
For the computation of $Q^{(4)}$, we observe that the results from the $\wedge^2 V^{(5)}$ show better agreement compared with those of $\wedge^4 V^{(1)}$.
In general, higher anti-symmetric products contain more errors than lower ones, which can be seen also in other affine Lie algebras.

\begin{table}[H]
    \centering
    \begin{tabular}{c|c|c|c|c|c}
        & \multicolumn{5}{c}{ODE}\\
        \cline{2-6}
        $i$ & $E_i^{(1)}$ & $E_i^{(2)}$ & $E_i^{(3)}$ & $E_i^{(4)}$ & $E_i^{(5)}$\\
        \hline
        $0$ & $14.24310$ & $10.01615$ & $9.480092$ & $11.33599$ & $17.93235$ \\
        $1$ & $45.54930$ & $24.40178$ & $21.85322$ & $26.44404$ & $50.82952$ \\
        $2$ & $87.09862$ & $43.73395$ & $37.66599$ & $46.11654$ & $93.58282$ \\
        $3$ & $136.7745$ & $66.32806$ & $56.23095$ & $69.05999$ & $144.2729$ \\
        \hline \hline
        & \multicolumn{5}{c}{IM}\\
        \cline{2-6}
        $i$ & $E_i^{(1)}$ & $E_i^{(2)}$ & $E_i^{(3)}$ & $E_i^{(4)}$ & $E_i^{(5)}$\\
        \hline
        $0$ & $14.24299$ & $10.01615$ & $9.480138$ & $11.33594$ & $17.93245$ \\
        $1$ & $45.54960$ & $24.40197$ & $21.85310$ & $26.44387$ & $50.82976$ \\
        $2$ & $87.09861$ & $43.73416$ & $37.66613$ & $46.11712$ & $93.58444$ \\
        $3$ & $136.7736$ & $66.32891$ & $56.23084$ & $69.06014$ & $144.2723$ \\
    \end{tabular}
    \caption{Spectra of $A_5^{(1)}$ with $p(x,E) = x^2 - E$, $l= \qty(5/12, 1/3, 0, -1/3, -5/12)$. We have set the iteration number to be 180 and the reference point $x_{\text{fixed}}=32$.}
    \label{tab:A5_x^2_(5/12,1/3,0,-1/3,-5/12)}
\end{table}

\begin{figure}[H]
    \centering
    \scriptsize
    \begin{tikzpicture}
        \def\nrr{4.6}
        \node[inner sep=0pt] (img) at (0,0) {\includegraphics[width=9cm]{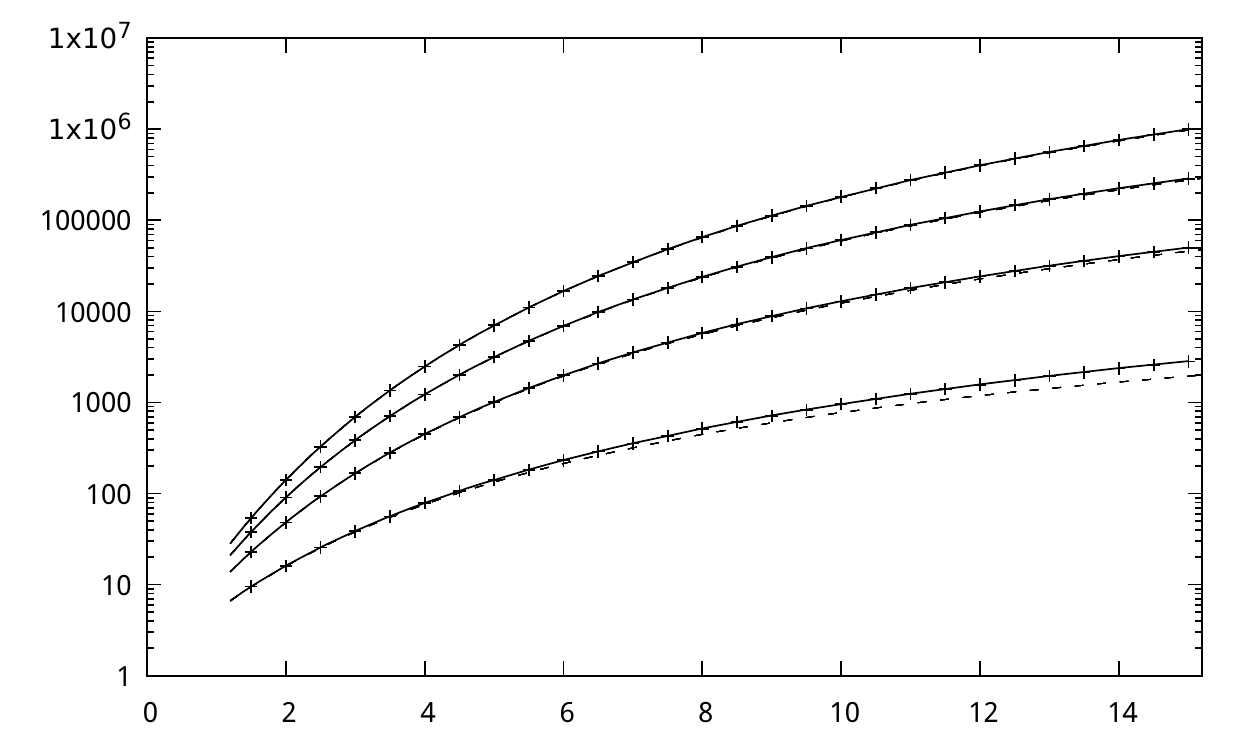}};
        \node (hM) at (4.55,-2.15) {\footnotesize{$hM$}};
        \node (E) at (-3.45,2.8) {\footnotesize{$E$}};
        
        \node (E01) at (\nrr,0.1) {$E_0^{(1)}$};
        \node (E11) at (\nrr,0.95) {$E_1^{(1)}$};
        \node (E21) at (\nrr,1.45) {$E_2^{(1)}$};
        \node (E21) at (\nrr,1.85) {$E_3^{(1)}$};
    \end{tikzpicture}
    \caption{The $hM$-dependence of the spectra $E_n^{(1)} (n = 0, \dots, 3)$ for $A_5^{(1)}$ with $l = (0, 0, 0, 0, 0)$. Solid lines show the spectra from the NLIEs, points the spectra from the ODE, and dashed lines the asymptotic formula (\ref{eq:quantization_condtion}).}
    \label{fig:A5_hM-E_plot}
\end{figure}

In figure \ref{fig:A5_hM-E_plot}, we show the $hM$-dependence of the first four spectra of the Bethe roots.
From table \ref{tab:A5_x^2_(5/12,1/3,0,-1/3,-5/12)} and figure \ref{fig:A5_hM-E_plot}, we can see that the ODE result is in good agreement with that of the NLIEs.

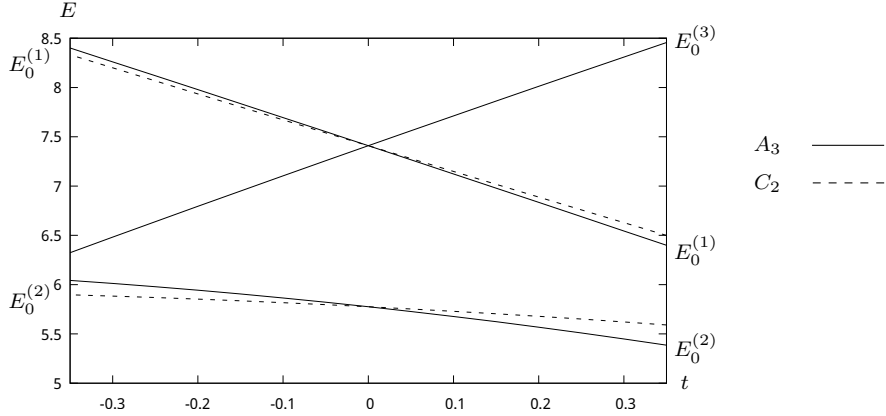
\begin{figure}[H]
    \centering
    \scriptsize
    \begin{tikzpicture}
        \def\nrr{4.55};
        \def\nll{-4.3};
        \def\lgd{\nrr + 1.1};
        \def\lgdlen{1.4};
        
        \node[inner sep=0pt] (img) at (0,0) {\includegraphics[width=9cm]{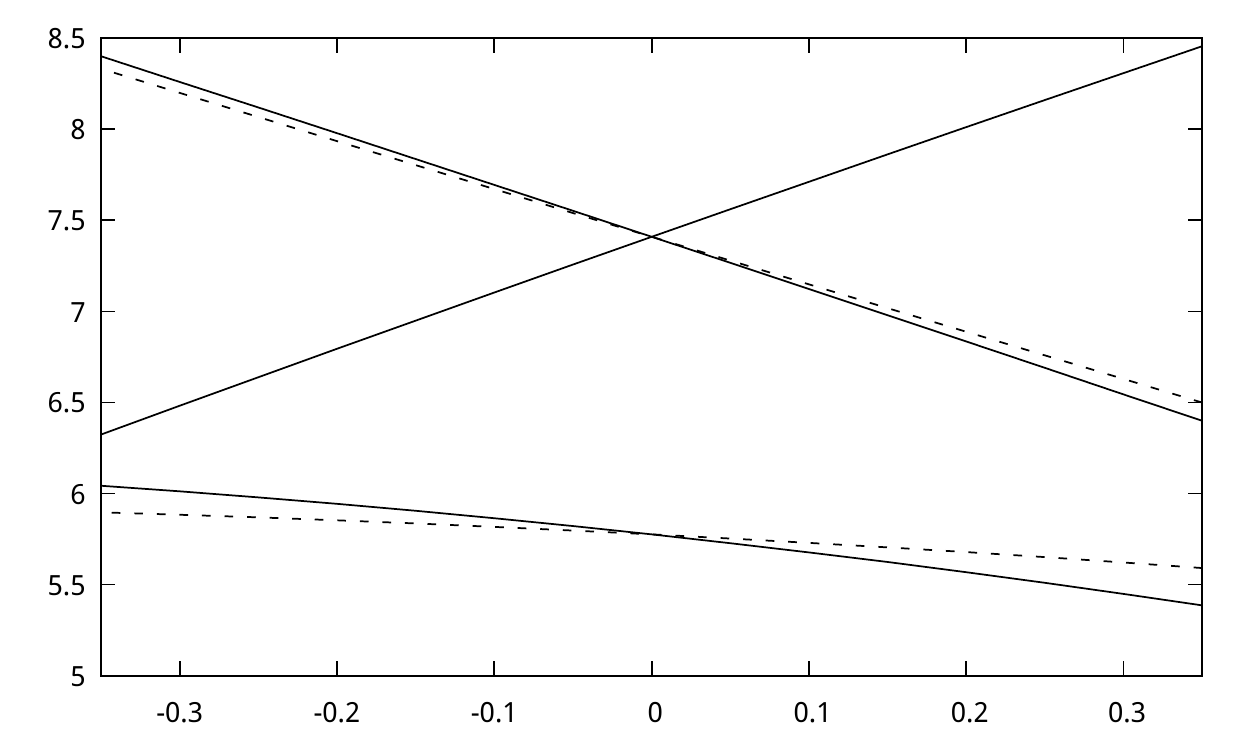}};
        \node (t) at (4.4,-2.15) {$t$};
        \node (E) at (-3.8,2.8) {$E$};
        
        \node (A3E1) at (\nrr,-0.4) {$E_0^{(1)}$};
        \node (A3E2) at (\nrr,-1.7) {$E_0^{(2)}$};
        \node (A3E3) at (\nrr,2.4) {$E_0^{(3)}$};
        
        \node (C2E1) at (\nll,2.1) {$E_0^{(1)}$};
        \node (C2E2) at (\nll,-1.05) {$E_0^{(2)}$};
        
        \node (l1) at (\lgd,1) {$A_3 \quad$};
        \draw [solid] (\lgd + \lgdlen,1) -- (l1);
        \node (l2) at (\lgd,0.5) {$C_2 \quad$};
        \draw [dashed] (\lgd + \lgdlen,0.5) -- (l2);
    \end{tikzpicture}
    \caption{The monodromy dependence of the spectra $E_0^{(a)} (a=1,2,3)$ for $A_3^{(1)}$. Solid lines show the spectra with $\hat{l}=(1,0,-1)$. Dashed lines show the spectra with $\hat{l}=(-1,0,-1)$, which corresponds to those of $C_2^{(1)}$ with $\hat{l}=(-1,0)$.}
    \label{fig:A3_C2_plot}
\end{figure}

We can also compare the spectra for $C_r^{(1)}$ with those of the NLIEs, which is obtained by folding procedure of $A_{2r-1}^{(1)}$.
We have confirmed its correspondence numerically for $r=2$ and $3$.
Based on observation of the numerical agreement between the ODE and the NLIEs, we plot the $l$-dependence of the ground state of the Bethe roots for $A_3^{(1)}$ in figure \ref{fig:A3_C2_plot}.
We change the monodromy parameters along some fixed directions $\hat{l}$: $l=t \hat{l}$, where $t$ is a real parameter. 
For the direction $\hat{l}$ with $\hat{l}_1=\hat{l}_3$, the spectrum reduces to that of $C_2^{(1)}$.
At $t=0$, some spectra coincide due to the symmetry of the Dynkin diagram. 

\subsection{\texorpdfstring{$D_r^{(1)}$}{Dr}}
Next we show some numerical results for $D_r^{(1)}$.
In contract to the analysis based on the pseudo-ODE \cite{Dorey:2006an}, we can calculate zeros of the Q-functions $Q^{(a)}$ ($a=1,\ldots,r$) for all the fundamental representations including the vector, spinor and conjugate-spinor representations. 
Here we compare two spectra for the linear problem for $D_4^{(1)}$ with $l=(1/6, 0, 1/4, -1/4)$ in table \ref{tab:D4_x^2_(1/6, 0, 1/4, -1/4)}, where the ODE results are in good agreement with those of the NLIEs.

\begin{table}[H]
    \centering
    \begin{tabular}{c|c|c|c|c}
        & \multicolumn{4}{c}{ODE}\\
        \cline{2-5}
        $i$ & $E_i^{(1)}$ & $E_i^{(2)}$ & $E_i^{(3)}$ & $E_i^{(4)}$ \\
        \hline
        $0$ & $10.56678$ & $8.077763$ & $10.33052$ & $11.77689$ \\
        $1$ & $29.94890$ & $16.91405$ & $29.62878$ & $31.57108$ \\
        $2$ & $55.28678$ & $28.92213$ & $54.89724$ & $57.25113$ \\
        $3$ & $85.36180$ & $42.73555$ & $84.91285$ & $87.62041$ \\
        \hline \hline
        & \multicolumn{4}{c}{IM}\\
        \cline{2-5}
        $i$ & $E_i^{(1)}$ & $E_i^{(2)}$ & $E_i^{(3)}$ & $E_i^{(4)}$ \\
        \hline
        $0$ & $10.56674$ & $8.077795$ & $10.33047$ & $11.77681$ \\
        $1$ & $29.94883$ & $16.91423$ & $29.62906$ & $31.57127$ \\
        $2$ & $55.28642$ & $28.92151$ & $54.89679$ & $57.25155$ \\
        $3$ & $85.36186$ & $42.73573$ & $84.91172$ & $87.62048$ \\
    \end{tabular}
    \caption{Spectra of $D_4^{(1)}$  with $p(x,E) = x^2 - E$ and $l= \qty(1/6, 0, 1/4, -1/4)$. We set the iteration number to be $240$ and the reference point to be $31$. }
    \label{tab:D4_x^2_(1/6, 0, 1/4, -1/4)}
\end{table}

\begin{figure}[H]
    \centering
    \scriptsize
    \begin{tikzpicture}
        \def\nrr{4.6}
        \node[inner sep=0pt] (img) at (0,0) {\includegraphics[width=9cm]{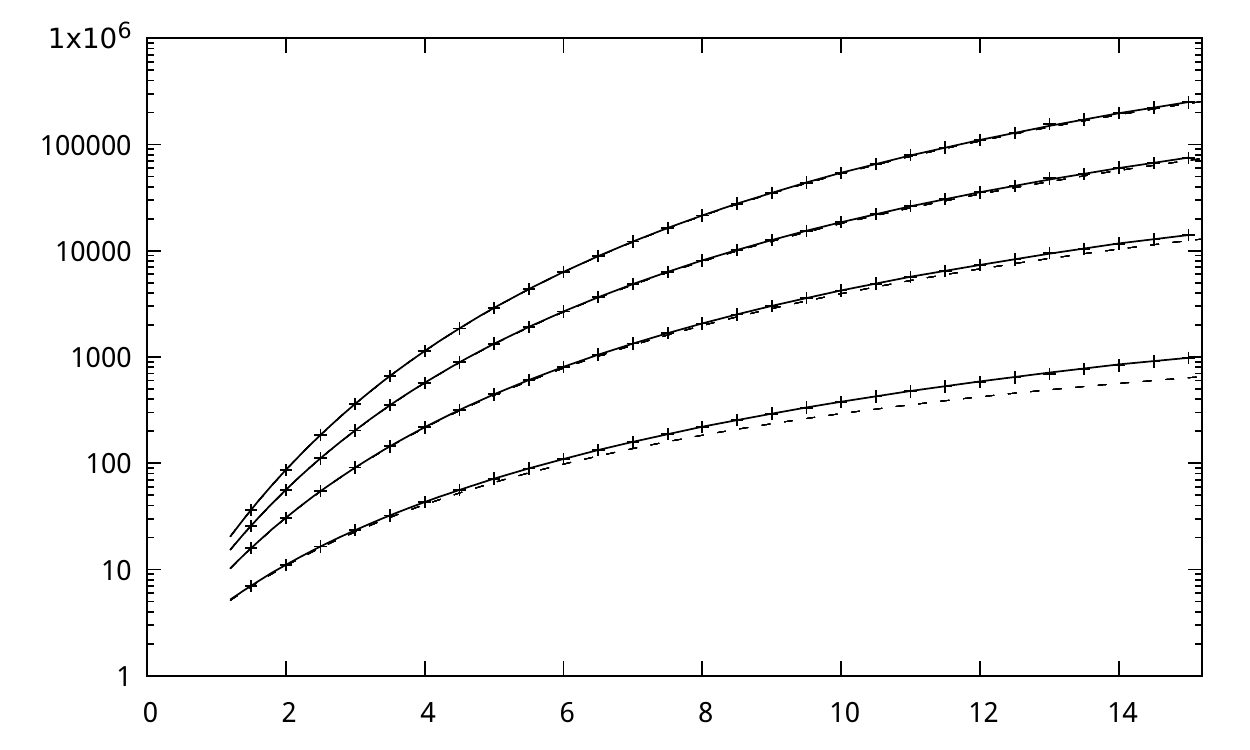}};
        \node (hM) at (4.55,-2.15) {\footnotesize{$hM$}};
        \node (E) at (-3.4,2.8) {\footnotesize{$E$}};
        
        \node (E01) at (\nrr,0.1) {$E_0^{(1)}$};
        \node (E11) at (\nrr,1) {$E_1^{(1)}$};
        \node (E21) at (\nrr,1.55) {$E_2^{(1)}$};
        \node (E21) at (\nrr,2.05) {$E_3^{(1)}$};
    \end{tikzpicture}
    \caption{The $hM$-dependence of the spectra $E_n^{(1)} (n = 0, \dots, 3)$ for $D_4^{(1)}$ with $l=(0,0,0,0)$. Solid lines show the spectra from the NLIEs, points the spectra from the ODE, and dashed lines the asymptotic formula (\ref{eq:quantization_condtion}).}
    \label{fig:D4_hM-E_plot}
\end{figure}
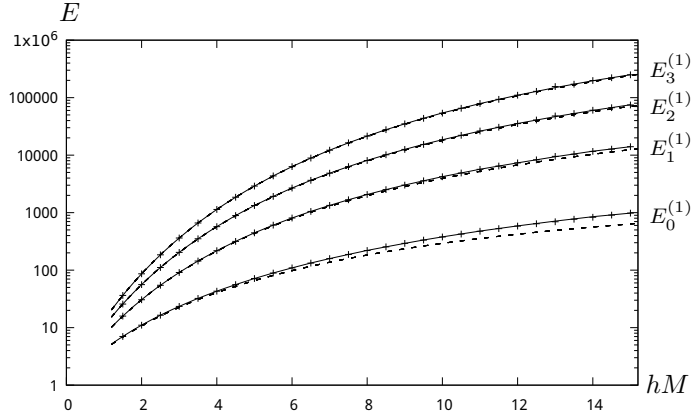

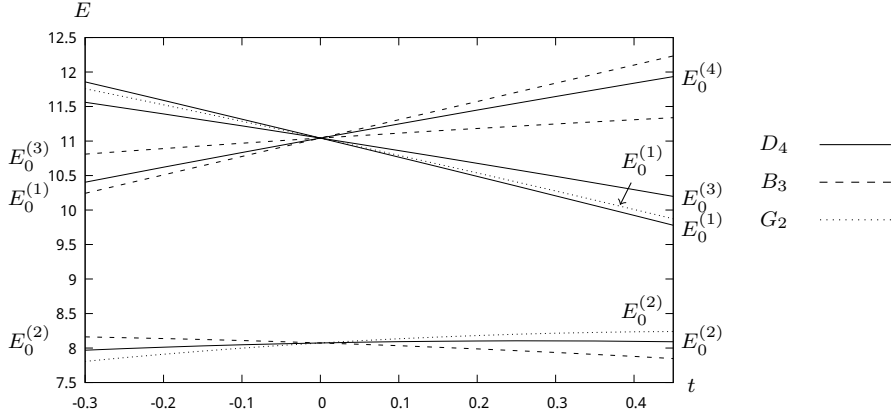
\begin{figure}[H]
    \centering
    \scriptsize
    \begin{tikzpicture}
        \def\nrr{4.55};
        \def\nll{-4.4};
        \def\nrl{3.75};
        \def\lgd{\nrr + 1.1};
        \def\lgdlen{1.4};
        
        \node[inner sep=0pt] (img) at (0,0) {\includegraphics[width=9cm]{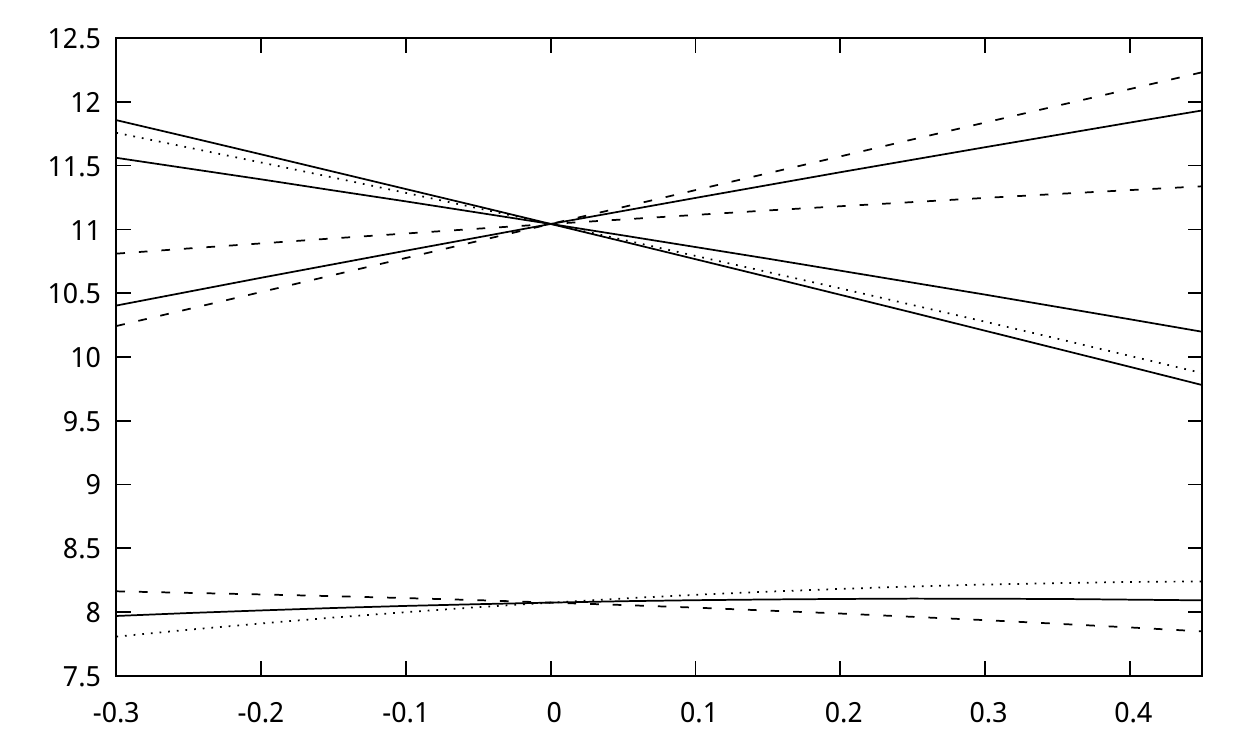}};
        \node (t) at (4.4,-2.2) {$t$};
        \node (E) at (-3.7,2.8) {$E$};

        \node (D4E1) at (\nrr,-0.1) {$E_0^{(1)}$};
        \node (D4E2) at (\nrr,-1.6) {$E_0^{(2)}$};
        \node (D4E3) at (\nrr,0.3) {$E_0^{(3)}$};
        \node (D4E4) at (\nrr,1.9) {$E_0^{(4)}$};

        \node (B3E1) at (\nll,0.3) {$E_0^{(1)}$};
        \node (B3E2) at (\nll,-1.6) {$E_0^{(2)}$};
        \node (B3E3) at (\nll,0.85) {$E_0^{(3)}$};
        
        \node (G2E1) at (\nrl,0.8) {$E_0^{(1)}$};
        \draw [<-,solid] (\nrl - 0.3,0.2) -- (G2E1);
        \node (G2E2) at (\nrl,-1.2) {$E_0^{(2)}$};
        
        \node (l1) at (\lgd,1) {$D_4 \quad$};
        \draw [solid] (\lgd + \lgdlen,1) -- (l1);
        \node (l2) at (\lgd,0.5) {$B_3 \quad$};
        \draw [dashed] (\lgd + \lgdlen,0.5) -- (l2);
        \node (l3) at (\lgd, 0) {$G_2 \quad$};
        \draw [dotted] (\lgd + \lgdlen,0) -- (l3);
    \end{tikzpicture}
    \caption{The monodromy dependence of the spectra $E_0^{(a)} (a=1,\ldots,4)$ for $D_4^{(1)}$. Solid lines show the spectra with $\hat{l}=(1,0,2/3,-2/3)$. Dashed lines show the spectra with $\hat{l}=(-1,0,-1/3,-1/3)$, which corresponds to those of $B_3^{(1)}$ with $\hat{l}=(-1,0,-1/3)$. Dotted lines show the spectra with $\hat{l}=(1,0,1,1)$, which corresponds to those of $G_2^{(1)}$ with $\hat{l}=(1,0)$.}
    \label{fig:D4_B3_G2_plot}
\end{figure}

In figure \ref{fig:D4_B3_G2_plot}, we plot the $l$-dependence of the ground state spectra $E^{(a)}_0$ of $D_4^{(1)}$ for various $l=t \hat{l}$. 
Here $\hat{l}$ with $\hat{l}_3=\hat{l}_4$ corresponds to the spectra for $B_3^{(1)}$ and
$\hat{l}$ with $\hat{l}_1=\hat{l}_3=\hat{l}_4$ corresponds to $G_2^{(1)}$.

\subsection{\texorpdfstring{$E_{6}^{(1)}$, $E_{7}^{(1)}$ and $E_{8}^{(1)}$}{Er}}
We now compare the spectra for $E_{r}^{(1)}$ ($r=6,7,8$) that provide a new and non-trivial test of the ODE/IM correspondence. 

Table \ref{tab:E6_x^2_(5/12,1/3,0,-1/3,-5/12,1/10)} shows the spectra for the $E_6^{(1)}$ linear problem with the potential $p(x,E)=x^2-E$ and the monodromy $l=(5/12,1/3,0,-1/3,-5/12,1/10)$.
$Q^{(1)}$ and $Q^{(2)}$ are calculated from the fundamental representation $V^{(1)}$ and $\wedge^2 V^{(1)}$, while $Q^{(4)}$ and $Q^{(5)}$ are from $V^{(5)}$ and $\wedge^2 V^{(5)}$, and $Q^{(3)}$ and $Q^{(6)}$ are from $V^{(6)}$ and $\wedge^2 V^{(6)}$.
We can see that data of $E_i^{(a)}$ ($a=2,3,4$) show less agreements with the results of the ODE compared with those of $E_i^{(a)}$ ($a=1,5,6$).
This is because these Q-functions are constructed from anti-symmetric tensor products of large dimensional representations.

\begin{table}[H]
    \centering
    \begin{tabular}{c|c|c|c|c|c|c}
        & \multicolumn{6}{c}{ODE}\\
        \cline{2-7}
        $i$ & $E_i^{(1)}$ & $E_i^{(2)}$ & $E_i^{(3)}$ & $E_i^{(4)}$ & $E_i^{(5)}$ & $E_i^{(6)}$\\
        \hline
        $0$ & $26.16492$ & $19.04286$ & $16.95324$ & $19.98072$ & $29.04567$ & $21.54848$\\
        $1$ & $76.14709$ & $37.06396$ & $28.48668$ & $38.50821$ & $80.49209$ & $52.00437$\\
        $2$ & $146.8766$ & $64.52501$ & $44.61201$ & $66.15524$ & $152.5322$ & $93.90413$\\
        $3$ & $236.0037$ & $95.98720$ & $63.83908$ & $97.97408$ & $242.8648$ & $145.7213$\\
        \hline \hline
        & \multicolumn{6}{c}{IM}\\
        \cline{2-7}
        $i$ & $E_i^{(1)}$ & $E_i^{(2)}$ & $E_i^{(3)}$ & $E_i^{(4)}$ & $E_i^{(5)}$ & $E_i^{(6)}$\\
        \hline
        $0$ & $26.16452$ & $19.04232$ & $16.95299$ & $19.98020$ & $29.04519$ & $21.54807$ \\
        $1$ & $76.14715$ & $37.06297$ & $28.48688$ & $38.50782$ & $80.49126$ & $52.00351$ \\
        $2$ & $146.8773$ & $64.52390$ & $44.61186$ & $66.15433$ & $152.5313$ & $93.90137$ \\
        $3$ & $236.0021$ & $95.98496$ & $63.83727$ & $97.97372$ & $242.8597$ & $145.7216$ \\
    \end{tabular}
    \caption{Spectra of $E_6^{(1)}$ with $P = x^2 - E$, $l = \qty(5/12, 1/3, 0, -1/3, -5/12, 1/10)$. We set the iteration number to be $1000$, the reference point to be $x_{\text{fixed}}=58$ for $V^{(1)},V^{(5)}$ and $x_{\text{fixed}}=46$ for $V^{(6)}$.}
    \label{tab:E6_x^2_(5/12,1/3,0,-1/3,-5/12,1/10)}
\end{table}

\begin{figure}[H]
    \centering
    \scriptsize
    \begin{tikzpicture}
        \def\nrr{4.6}
        \node[inner sep=0pt] (img) at (0,0) {\includegraphics[width=9cm]{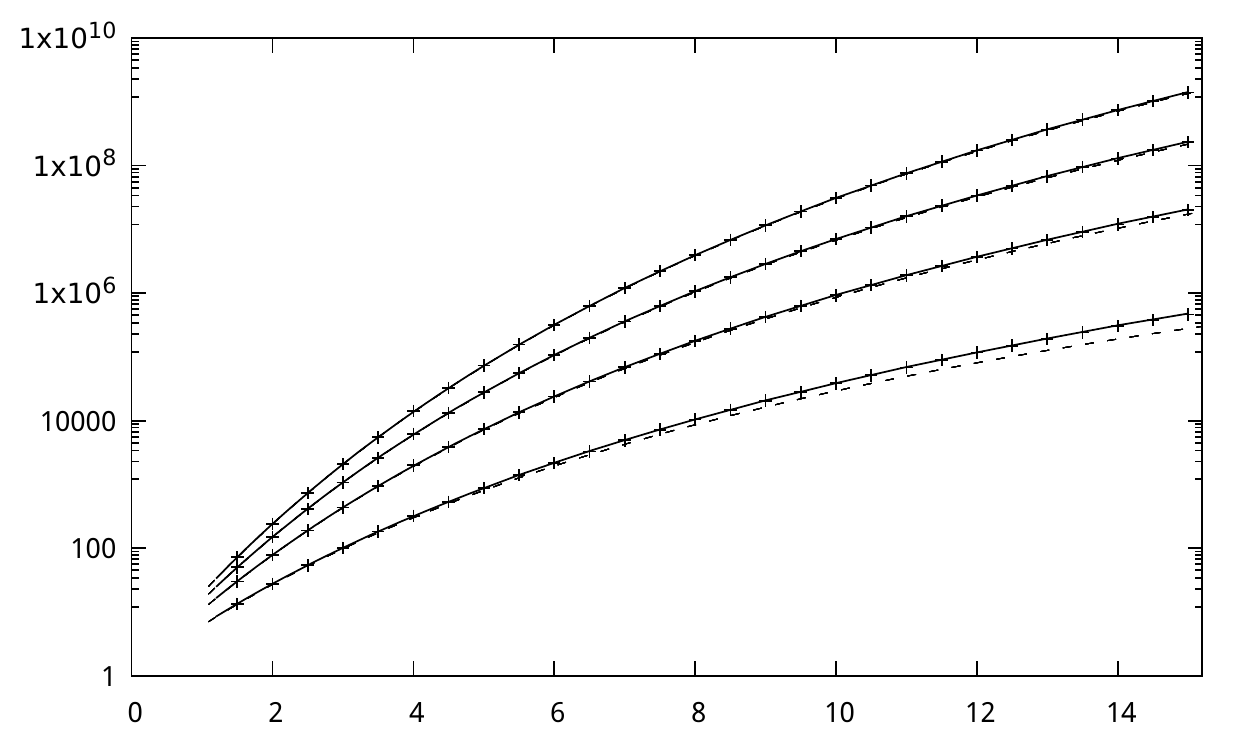}};
        \node (hM) at (4.55,-2.15) {\footnotesize{$hM$}};
        \node (E) at (-3.55,2.8) {\footnotesize{$E$}};
        
        \node (E01) at (\nrr,0.4) {$E_0^{(1)}$};
        \node (E11) at (\nrr,1.2) {$E_1^{(1)}$};
        \node (E21) at (\nrr,1.7) {$E_2^{(1)}$};
        \node (E21) at (\nrr,2) {$E_3^{(1)}$};
    \end{tikzpicture}
    \caption{The $hM$-dependence of the spectra $E_n^{(1)} (n = 0, \dots, 3)$ for $E_6^{(1)}$. Solid lines show the spectra from the NLIEs, points the spectra from the ODE, and dashed lines the asymptotic formula (\ref{eq:quantization_condtion}).}
    \label{fig:E6_hM-E_plot}
\end{figure}

In figure \ref{fig:E6_hM-E_plot}, we show the $hM$-dependence of the spectra $E^{(1)}_n$.
We observe that the ODE results agree with those of the NLIEs numerically.

\begin{figure}[H]
    \centering
    \scriptsize
    \begin{tikzpicture}
        \def\nrr{4.55}
        \def\nll{-4.5};
        \def\lgd{\nrr + 1.1};
        \def\lgdlen{1.4};

        \node[inner sep=0pt] (img) at (0,0) {\includegraphics[width=9cm]{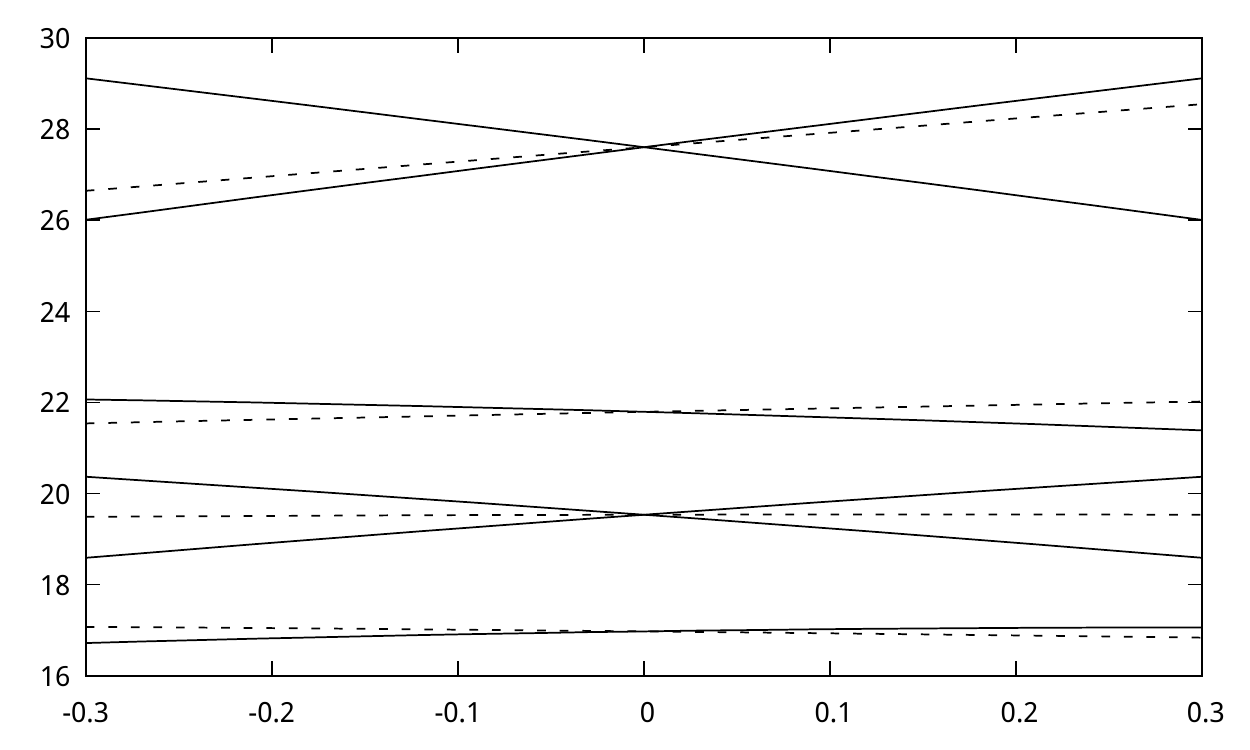}};
        \node (t) at (4.4,-2.15) {$t$};
        \node (E) at (-3.9,2.8) {$E$};
        
        \node (E6E1) at (\nrr,1.1) {$E_0^{(1)}$};
        \node (E6E2) at (\nrr,-1.3) {$E_0^{(2)}$};
        \node (E6E3) at (\nrr,-1.75) {$E_0^{(3)}$};
        \node (E6E4) at (\nrr,-0.75) {$E_0^{(4)}$};
        \node (E6E5) at (\nrr,2.2) {$E_0^{(5)}$};
        \node (E6E6) at (\nrr,-0.35) {$E_0^{(6)}$};
        
        \node (F4E1) at (\nll,1.3) {$E_0^{(1)}$};
        \node (F4E2) at (\nll,-1) {$E_0^{(2)}$};
        \node (F4E3) at (\nll,-1.8) {$E_0^{(3)}$};
        \node (F4E4) at (\nll,-0.4) {$E_0^{(4)}$};
        
        \node (l1) at (\lgd,1) {$E_6 \quad$};
        \draw [solid] (\lgd + \lgdlen,1) -- (l1);
        \node (l2) at (\lgd,0.5) {$F_4 \quad$};
        \draw [dashed] (\lgd + \lgdlen,0.5) -- (l2);
    \end{tikzpicture}
    \caption{The monodromy dependence of the spectra $E_0^{(a)} (a=1,\ldots,6)$ for $E_6^{(1)}$. Solid lines show the spectra with $\hat{l}=(3/2,2,-1/2,2,3/2,1/2)$. Dashed lines show the spectra with $\hat{l}=(-1,-1/3,0,-1/3,-1,-1/2)$, which corresponds to those of $F_4^{(1)}$ with $\hat{l}=(-1,-1/3,0,-1/2)$.}
    \label{fig:E6_F4_plot}
\end{figure}
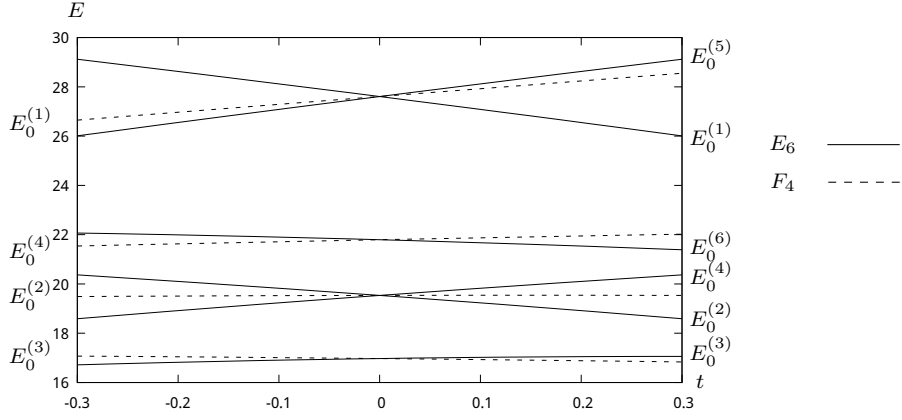

In figure \ref{fig:E6_F4_plot}, we plot the $l$-dependence of $E_0^{(a)}$ for various monodromy parameters $l=t\hat{l}$.
Here for $\hat{l}$, with $\hat{l}_1=\hat{l}_5$ and $\hat{l}_2=\hat{l}_4$, the spectra reduce to those of $F_4^{(1)}$.

Table \ref{tab:E7_x^2_(0,0,0,0,0,0,0)} shows the spectra for $E_7^{(1)}$ with the potential $p(x,E)=x^2-E$ and the monodromy $l=0$. 
$Q^{(a)}$ ($a=1,2,3$) are calculated from the fundamental representation $V^{(1)}$ and its antisymmetric products, while $Q^{(a)}$ ($a=4,5,6$) are from $V^{(6)}$. 
We did not calculate zeros of $Q^{(7)}$ because of our limitation of computational resources.

\begin{table}[H]
    \centering
    \begin{tabular}{c|c|c|c|c|c|c}
        & \multicolumn{6}{c}{ODE}\\
        \cline{2-7}
        $i$ & $E_i^{(1)}$ & $E_i^{(2)}$ & $E_i^{(3)}$ & $E_i^{(4)}$ & $E_i^{(5)}$ & $E_i^{(6)}$\\
        \hline
        $0$ & $46.88833$ & $34.06543$ & $28.51571$ & $26.53575$  & $29.83398$ & $39.17977$ \\
        $1$ & $133.0989$ & $61.91322$ & $46.84773$ & $39.36328$  & $50.29978$ & $98.54030$ \\
        $2$ & $257.3902$ & $109.7830$ & $71.10383$ & $57.00889$  & $82.59258$ & $182.0167$ \\
        $3$ & $416.8349$ & $160.7729$ & $102.4082$ & $77.68440$  & $117.6790$ & $287.6200$ \\
        \hline \hline
        & \multicolumn{6}{c}{IM}\\
        \cline{2-7}
        $i$ & $E_i^{(1)}$ & $E_i^{(2)}$ & $E_i^{(3)}$ & $E_i^{(4)}$ & $E_i^{(5)}$ & $E_i^{(6)}$\\
        \hline
        $0$ & $46.88577$ & $34.06368$ & $28.51386$ & $26.53481$ & $29.83252$ & $39.17691$ \\
        $1$ & $133.0943$ & $61.91087$ & $46.84458$ & $39.36135$ & $50.29832$ & $98.53555$ \\
        $2$ & $257.3906$ & $109.7788$ & $71.10062$ & $57.00469$ & $82.58823$ & $182.0144$ \\
        $3$ & $416.8400$ & $160.7689$ & $102.4038$ & $77.68159$ & $117.6720$ & $287.6197$ \\
    \end{tabular}
    \caption{Spectra of $E_7^{(1)}$ with $P = x^2 - E$, $l = \qty(0, 0, 0, 0, 0, 0, 0)$. We set the iteration number to be $1000$, reference point to be $x_{\text{fixed}}=64$ for $V^{(1)}$ and $x_{\text{fixed}}=55$ for $V^{(6)}$.}
    \label{tab:E7_x^2_(0,0,0,0,0,0,0)}
\end{table}

Finally, table \ref{tab:E8_x^2_(0,0,0,0,0,0,0,0)} shows the spectra for $E_8^{(1)}$ with potential $p(x,E)=x^2-E$ and the monodromy $l=0$.
We calculated $Q^{(a)}$ ($a=1,2,3,4,5$) from the fundamental representation $V^{(1)}$ and its anti-symmetric products.
$Q^{(a)}$ ($a=6,7,8$) could not be calculated because of our limitation of computational resources.

\begin{table}[H]
    \centering
    \begin{tabular}{c|c|c|c|c|c}
        & \multicolumn{5}{c}{ODE}\\
        \cline{2-6}
        $i$ & $E_i^{(1)}$ & $E_i^{(2)}$ & $E_i^{(3)}$ & $E_i^{(4)}$ & $E_i^{(5)}$ \\
        \hline
        $0$ & $79.11456$ & $61.55086$ & $52.70845$ & $48.00023$ & $46.30476$ \\
        $1$ & $206.8568$ & $98.86243$ & $78.43953$ & $67.37008$ & $60.39086$ \\
        $2$ & $390.6975$ & $174.0779$ & $111.3530$ & $90.99633$ & $79.34707$ \\
        $3$ & $627.7404$ & $242.8390$ & $159.3855$ & $119.2882$ & $100.8886$ \\
        \hline \hline
        & \multicolumn{5}{c}{IM}\\
        \cline{2-6}
        $i$ & $E_i^{(1)}$ & $E_i^{(2)}$ & $E_i^{(3)}$ & $E_i^{(4)}$ & $E_i^{(5)}$ \\
        \hline
        $0$ & $79.09889$ & $61.53558$ & $52.71002$ & $48.02433$ & $46.35178$ \\
        $1$ & $206.7926$ & $98.84128$ & $78.38027$ & $67.33502$ & $60.42081$ \\
        $2$ & $390.4583$ & $174.0345$ & $111.3402$ & $90.82841$ & $79.20079$ \\
        $3$ & $627.4773$ & $242.7122$ & $159.5877$ & $119.2778$ & $100.4175$ \\
    \end{tabular}
    \caption{Spectra of $E_8^{(1)}$ with $P = x^2 - E$, $l = \qty(0, 0, 0, 0, 0, 0, 0, 0)$. We set the iteration number $800$ and reference point $x_{\text{fixed}}=67$. }
    \label{tab:E8_x^2_(0,0,0,0,0,0,0,0)}
\end{table}

We can also check the coincidence of the results from the ODE and the NLIEs for various parameter sets. To summarize, we have confirmed that the linear problem (\ref{eq:linear_prob}) for an affine Lie algebra provides the solutions of the Bethe ansatz equations of the corresponding affine Lie algebra.

\section{Conclusions and discussions}\label{sec:conclusion}
In this paper we have studied the ODE/IM correspondence for untwisted affine Lie algebras.
The ODE is defined by the massless limit of the linear system associated with the modified affine Toda field equation.
It is the first-order linear differential system which has two singularities at infinity and the origin.
The Q-function is defined by the inner product of the solutions to the dual linear problem around the origin and the subdominant solution at infinity.
The periodicity of the Q-function can be characterized by the solution around the origin, while the asymptotic behaviour at a large spectral parameter is determined by the subdominant function.
It is interesting to study the present construction of the Q-function in operator formalism and its relation to the CFT approach\cite{Bazhanov:1996dr}.

We also used Cheng's algorithm to obtain the power series solution around the origin.
Together with the behaviour of the subdominant solutions, we have calculated the spectrum of the ODE by finding the zeros of the Q-functions numerically.
These are found to agree with the Bethe roots obtained by the NLIEs.
In particular, we have calculated the Bethe roots for spinor representations of $D_r^{(1)}$.
Moreover, the Bethe roots for $E$-type Lie algebras are calculated.
For non-simply-laced Lie algebras, where the corresponding ODEs are obtained by folding simply-laced Lie algebras, the NLIEs are also derived from the folding procedure.
The effective central charges of the integrable models have been evaluated.

It is interesting to generalize $p(x,E)$, which is a monomial in $x$ in the present paper, to general polynomial \cite{Ito:2018eon,Ito:2019llq}.
In particular, for $p(x,E)=(x^{hM/K}-E)^K$, which is expected to correspond to the GKO (Goddard-Kent-Olive) coset model with level $K$, the BAEs are found in \cite{Dorey:2006an} for classical Lie algebras. Moreover, for twisted affine Lie algebras, the corresponding BAEs have been constructed \cite{Dorey:2006an,Sun:2012xw,Ito:2015nla,Masoero:2015lga}.
However, these cases include the complex Bethe roots, which require further numerical analysis.
It is also interesting to find the inner product representation of the Q-function for massive ODE/IM correspondence\cite{Lukyanov:2010rn,Dorey:2012bx,Ito:2013aea,Adamopoulou:2014fca,kar72958}, where we need to consider holomorphic and anti-holomorphic equations simultaneously.

It is interesting to find other types of functional relations and integral equations, such as T-/Y-system and TBA equations.
In particular, the free energy of the TBA system is related to the minimal surface area and the gluon scattering amplitudes/light-like polygonal Wilson loops via the AdS/CFT correspondence \cite{Alday:2010vh,Hatsuda:2010cc,Dorey:2019ngq, Fioravanti:2020udo}.
The monodromy parameters introduce non-trivial mixing between T- and Y-functions \cite{Maldacena:2010kp,Gao:2013dza}.
The free energy is expected to the form factors of four-dimensional ${\cal N}=4$ gauge theories.
For $B_2^{(1)}$, the corresponding  Y-system is found to be $A_3/\mathbf{Z}_2$-type \cite{Ito:2016qzt}, which gives the same result as the ODE/IM correspondence.

These are applications to strong coupling physics of four-dimensional ${\cal N}=2$ supersymmetric gauge theories through the TBA system \cite{Gaiotto:2009hg,Gaiotto:2014bza} and the quantum SW curve\cite{Ito:2017ypt,Grassi:2018bci}.
The non-linear integral equations would provide a new understanding of the exact WKB method
of the higher-order ODE \cite{berk1982new} and its resurgence structure \cite{Ito:2018eon,Ito:2019llq}.

\subsection*{Acknowledgements}
We would like to thank Simon Ekhammar, Davide Fioravanti, Marco Rossi, Junji Suzuki and Dmytro Volin for useful discussions.
The work of K.I. is supported in part by Grant-in-Aid for Scientific Research 18K03643 and 17H06463 from Japan Society for the Promotion of Science (JSPS).
The work of H.S. is supported by the grant ``Exact Results in Gauge and String Theories'' from the Knut and Alice Wallenberg foundation.

\appendix

\section{Dynkin diagrams, Cartan matrices and Representations}\label{sec:cartan_representation}
In this appendix, we summarize basic data of simple Lie algebras and their representations which are used in this paper.

\paragraph{Dynkin diagram}
The Dynkin diagrams of simple Lie algebras are listed as follows:

\begin{figure}[H]
    \centering
    \begin{tabular}{cc}
        \begin{minipage}[c]{0.48\hsize}
            \begin{tikzpicture}[scale=.4]
                \tikzset{node/.style={draw,circle,thick,inner sep=0pt,minimum size=7pt}}
                \draw (-1,0) node[anchor=east]  {$A_{r}: \quad$};
                \node[node, label=below:$\alpha_1$] (n1) at (0, 0) {};
                \node[node, label=below:$\alpha_2$] (n2) at (2, 0) {};
                \node (nd) at (4, 0) {$\cdots$};
                \node[node, label=below:$\alpha_{r - 1}$] (n3) at (6, 0) {};
                \node[node, label=below:$\alpha_r$] (n4) at (8, 0) {};
                \draw (n1) -- (n2) -- (nd) -- (n3) -- (n4);
            \end{tikzpicture}
        \end{minipage}
        &
        \begin{minipage}[c]{0.48\hsize}
            \begin{tikzpicture}[scale=.4]
                \tikzset{node_w/.style={draw,circle,thick,inner sep=0pt,minimum size=7pt}}
                \tikzset{node_b/.style={draw,circle,thick,fill=black,inner sep=0pt,minimum size=7pt}}
                \draw (-1,0) node[anchor=east]  {$B_{r}: \quad$};
                \node[node_w, label=below:$\alpha_1$] (n1) at (0, 0) {};
                \node[node_w, label=below:$\alpha_2$] (n2) at (2, 0) {};
                \node (nd) at (4, 0) {$\cdots$};
                \node[node_w, label=below:$\alpha_{r - 2}$] (n3) at (6, 0) {};
                \node[node_w, label=below:$\alpha_{r - 1}$] (n4) at (8, 0) {};
                \node[node_b, label=below:$\alpha_r$] (n5) at (10, 0) {};
                \draw[thick] (n1) -- (n2) -- (nd) -- (n3) -- (n4);
                \draw[thick] (n4.north east) -- (n5.north west);
                \draw[thick] (n4.south east) -- (n5.south west);
            \end{tikzpicture}
        \end{minipage}
        \\ \\
        \begin{minipage}[c]{0.48\hsize}
            \begin{tikzpicture}[scale=.4]
                \tikzset{node_w/.style={draw,circle,thick,inner sep=0pt,minimum size=7pt}}
                \tikzset{node_b/.style={draw,circle,thick,fill=black,inner sep=0pt,minimum size=7pt}}
                \draw (-1,0) node[anchor=east]  {$C_{r}: \quad$};
                \node[node_b, label=below:$\alpha_1$] (n1) at (0, 0) {};
                \node[node_b, label=below:$\alpha_2$] (n2) at (2, 0) {};
                \node (nd) at (4, 0) {$\cdots$};
                \node[node_b, label=below:$\alpha_{r - 2}$] (n3) at (6, 0) {};
                \node[node_b, label=below:$\alpha_{r - 1}$] (n4) at (8, 0) {};
                \node[node_w, label=below:$\alpha_r$] (n5) at (10, 0) {};
                \draw[thick] (n1) -- (n2) -- (nd) -- (n3) -- (n4);
                \draw[thick] (n4.north east) -- (n5.north west);
                \draw[thick] (n4.south east) -- (n5.south west);
            \end{tikzpicture}
        \end{minipage}
        &
        \begin{minipage}[c]{0.48\hsize}
            \begin{tikzpicture}[scale=.4]
                \tikzset{node/.style={draw,circle,thick,inner sep=0pt,minimum size=7pt}}
                \draw (-1,0) node[anchor=east]  {$D_{r}: \quad$};
                \node[node, label=below:$\alpha_1$] (n1) at (0, 0) {};
                \node[node, label=below:$\alpha_2$] (n2) at (2, 0) {};
                \node (nd) at (4, 0) {$\cdots$};
                \node[node, label=below:$\alpha_{r - 2}$] (n3) at (6, 0) {};
                \node[node, label=below:$\alpha_{r - 1}$] (n4) at (8, 1.3) {};
                \node[node, label=below:$\alpha_r$] (n5) at (8, -1.3) {};
                \draw[thick] (n1) -- (n2) -- (nd) -- (n3);
                \draw[thick] (n4) -- (n3) -- (n5);
            \end{tikzpicture}
        \end{minipage}
        \\
        \begin{minipage}[c]{0.48\hsize}
            \begin{tikzpicture}[scale=.4]
                \tikzset{node/.style={draw,circle,thick,inner sep=0pt,minimum size=7pt}}
                \draw (-1,0) node[anchor=east]  {$E_6: \quad$};
                \node[node, label=below:$\alpha_1$] (n1) at (0, 0) {};
                \node[node, label=below:$\alpha_2$] (n2) at (2, 0) {};
                \node[node, label=below:$\alpha_3$] (n3) at (4, 0) {};
                \node[node, label=below:$\alpha_4$] (n4) at (6, 0) {};
                \node[node, label=below:$\alpha_5$] (n5) at (8, 0) {};
                \node[node, label=above:$\alpha_6$] (n6) at (4, 2) {};
                \draw[thick] (n1) -- (n2) -- (n3) -- (n4) -- (n5);
                \draw[thick] (n3) -- (n6);
            \end{tikzpicture}
        \end{minipage}
        &
        \begin{minipage}[c]{0.48\hsize}
            \begin{tikzpicture}[scale=.4]
                \tikzset{node/.style={draw,circle,thick,inner sep=0pt,minimum size=7pt}}
                \draw (-1,0) node[anchor=east]  {$E_7: \quad$};
                \node[node, label=below:$\alpha_1$] (n1) at (0, 0) {};
                \node[node, label=below:$\alpha_2$] (n2) at (2, 0) {};
                \node[node, label=below:$\alpha_3$] (n3) at (4, 0) {};
                \node[node, label=below:$\alpha_4$] (n4) at (6, 0) {};
                \node[node, label=below:$\alpha_5$] (n5) at (8, 0) {};
                \node[node, label=below:$\alpha_6$] (n6) at (10, 0) {};
                \node[node, label=above:$\alpha_7$] (n7) at (6, 2) {};
                \draw[thick] (n1) -- (n2) -- (n3) -- (n4) -- (n5) -- (n6);
                \draw[thick] (n4) -- (n7);
            \end{tikzpicture}
        \end{minipage}
        \\ \\
        \begin{minipage}[c]{0.48\hsize}
            \begin{tikzpicture}[scale=.4]
                \tikzset{node/.style={draw,circle,thick,inner sep=0pt,minimum size=7pt}}
                \draw (-1,0) node[anchor=east]  {$E_8: \quad$};
                \node[node, label=below:$\alpha_1$] (n1) at (0, 0) {};
                \node[node, label=below:$\alpha_2$] (n2) at (2, 0) {};
                \node[node, label=below:$\alpha_3$] (n3) at (4, 0) {};
                \node[node, label=below:$\alpha_4$] (n4) at (6, 0) {};
                \node[node, label=below:$\alpha_5$] (n5) at (8, 0) {};
                \node[node, label=below:$\alpha_6$] (n6) at (10, 0) {};
                \node[node, label=below:$\alpha_7$] (n7) at (12, 0) {};
                \node[node, label=above:$\alpha_8$] (n8) at (8, 2) {};
                \draw[thick] (n1) -- (n2) -- (n3) -- (n4) -- (n5) -- (n6) -- (n7);
                \draw[thick] (n5) -- (n8);
            \end{tikzpicture}
        \end{minipage}
        &
        \begin{minipage}[c]{0.48\hsize}
            \begin{tikzpicture}[scale=.4]
                \tikzset{node_w/.style={draw,circle,thick,inner sep=0pt,minimum size=7pt}}
                \tikzset{node_b/.style={draw,circle,thick,fill=black,inner sep=0pt,minimum size=7pt}}
                \draw (-1,0) node[anchor=east]  {$F_4: \quad$};
                \node[node_b, label=below:$\alpha_1$] (n1) at (0, 0) {};
                \node[node_b, label=below:$\alpha_2$] (n2) at (2, 0) {};
                \node[node_w, label=below:$\alpha_3$] (n3) at (4, 0) {};
                \node[node_w, label=below:$\alpha_4$] (n4) at (6, 0) {};
                \draw[thick] (n1) -- (n2);
                \draw[thick] (n2.north east) -- (n3.north west);
                \draw[thick] (n2.south east) -- (n3.south west);
                \draw[thick] (n3) -- (n4);
            \end{tikzpicture}
        \end{minipage}
        \\ \\ \\
        \begin{minipage}[c]{0.48\hsize}
            \begin{tikzpicture}[scale=.4]
                \tikzset{node_w/.style={draw,circle,thick,inner sep=0pt,minimum size=7pt}}
                \tikzset{node_b/.style={draw,circle,thick,fill=black,inner sep=0pt,minimum size=7pt}}
                \draw (-1,0) node[anchor=east]  {$G_2: \quad$};
                \node[node_b, label=below:$\alpha_1$] (n1) at (0, 0) {};
                \node[node_w, label=below:$\alpha_2$] (n2) at (2, 0) {};
                \draw[thick] (n1) -- (n2);
                \draw[thick] (n1.north east) -- (n2.north west);
                \draw[thick] (n1.south east) -- (n2.south west);
            \end{tikzpicture}
        \end{minipage}
    \end{tabular}
    \label{fig:Dynkin-diagram}
\end{figure}

\paragraph{Cartan matrices}
The Cartan matrices for simple Lie algebras are given by
\begin{align*}
        &A_{r \geq 1} : \quad \mqty(2 & -1 &  &  & \\ -1 & 2 & \ddots &  & \\ & \ddots & \ddots & -1 & \\ &  & -1 & 2 & -1 \\ & & & -1 & 2), \qquad
        B_{r \geq 2} : \quad \mqty(2 & -1 & & & \\ -1 & 2 & \ddots & & \\ & \ddots & \ddots & -1 & \\ & & -1 & 2 & -2 \\ & & & -1 & 2 \\),\\
        &C_{r \geq 2} : \quad \mqty(2 & -1 & & & \\ -1 & 2 & \ddots & & \\ & \ddots & \ddots & -1 & \\ & & -1 & 2 & -1 \\ & & & -2 & 2 \\), \qquad
        D_{r \geq 3} : \quad \mqty(2 & -1 & & & & \\ -1 & 2 & \ddots & & & \\ & \ddots & \ddots & -1 & & \\ & & -1 & 2 & -1 & -1 \\ & & & -1 & 2 & \\ & & & -1 & & 2),\\
        &E_6 : \quad \mqty(2 & -1 & & & & \\ -1 & 2 & -1 & & & \\ & -1 & 2 & -1 & & -1 \\ & & -1 & 2 & -1 & \\ & & & -1 & 2 & \\ & & -1 & & & 2), \qquad
        E_7 : \quad \mqty(2 & -1 & & & & & \\ -1 & 2 & -1 & & & & \\ & -1 & 2 & -1 & & & \\ & & -1 & 2 & -1 & & -1 \\ & & & -1 & 2 & -1 & \\ & & & & -1 & 2 & \\ & & & -1 & & & 2),\\
        &E_8 : \quad \mqty(2 & -1 & & & & & & \\ -1 & 2 & -1 & & & & & \\ & -1 & 2 & -1 & & & & \\ & & -1 & 2 & -1 & & & \\ & & & -1 & 2 & -1 & & -1 \\ & & & & -1 & 2 & -1 & \\ & & & & & -1 & 2 & \\ & & & & -1 & & & 2), \qquad
        F_4 : \quad \mqty(2 & -1 & & \\ -1 & 2 & -1 & \\ & -2 & 2 & -1 \\ & & -1 & 2),\\
        &G_2 : \quad \mqty(2 & -1 \\ -3 & 2).
\end{align*}

\paragraph{Coxeter numbers and the highest roots}
The Coxeter numbers and the highest roots are summarized in table \ref{tab:h_theta_for_semisimple_algebra}.

\renewcommand{\arraystretch}{1.5}
\begin{table}[H]
    \centering
    \begin{tabular}{c||c|c}
        $\mathfrak{g}$  & $h$ & $\theta^{\vee}$ \\ \hline
        $A_r$ & $r+1$ & $\alpha_1^{\vee}+\cdots+\alpha_r^{\vee}$ \\
        $B_r$ & $2r$ & $\alpha_1^{\vee}+2\alpha_2^{\vee}+\cdots+2\alpha_{r-1}^{\vee}+\alpha_r^{\vee}$ \\
        $C_r$ & $2r$ & $\alpha_1^{\vee}+\cdots+\alpha_r^{\vee}$ \\
        $D_r$ & $2r-2$ & $\alpha_1^{\vee}+2\alpha_2^{\vee}+\cdots+2\alpha_{r-2}^{\vee}+\alpha_{r-1}^{\vee}+\alpha_r^{\vee}$ \\
        $E_6$ & $12$ & $\alpha_1^{\vee}+2\alpha_2^{\vee}+3\alpha_3^{\vee}+2\alpha_4^{\vee}+\alpha_5^{\vee}+2\alpha_6^{\vee}$ \\
        $E_7$ & $18$ & $\alpha_1^{\vee}+2\alpha_2^{\vee}+3\alpha_3^{\vee}+4\alpha_4^{\vee}+3\alpha_5^{\vee}+2\alpha_6^{\vee}+2\alpha_7^{\vee}$ \\
        $E_8$ & $30$ &  $2\alpha_1^{\vee}+3\alpha_2^{\vee}+4\alpha_3^{\vee}+5\alpha_4^{\vee}+6\alpha_5^{\vee}+4\alpha_6^{\vee}+2\alpha_7^{\vee}+3\alpha_8^{\vee}$ \\
        $F_4$ & $12$ & $\alpha_1^{\vee}+2\alpha_2^{\vee}+3\alpha_3^{\vee}+2\alpha_4^{\vee}$ \\
        $G_2$ & $6$ & $\alpha_1^{\vee}+2\alpha_2^{\vee}$ 
    \end{tabular}
    \caption{Coxeter number $h$ and the co-highest root $\theta^{\vee}$ for simple Lie algebras.}
    \label{tab:h_theta_for_semisimple_algebra}
\end{table}
\renewcommand{\arraystretch}{1}

\subsection{Folding of simply-laced Lie algebras}

Here we describe the folding of Dynkin diagrams of simply-laced Lie algebras of 
$A^{(1)}_{2r-1}$, $D^{(1)}_{r+1}$, $E^{(1)}_6$ and $D^{(1)}_4$.\\

\noindent
$\underline{A_{2r - 1}^{(1)} \rightarrow C_r^{(1)}}$
\begin{figure}[H]
    \centering
    \small
    \begin{tikzpicture}[scale=.47]
        \tikzset{node/.style={draw,circle,thick,inner sep=0pt,minimum size=7pt}}
        \tikzset{node_w/.style={draw,circle,thick,inner sep=0pt,minimum size=7pt}}
        \tikzset{node_b/.style={draw,circle,thick,fill=black,inner sep=0pt,minimum size=7pt}}
        \node[node, label=below:$\alpha_1$] (n1) at (0, 0) {};
        \node[node, label=below:$\alpha_2$] (n2) at (2, 0) {};
        \node (nd) at (4, 0) {$\cdots$};
        \node[node, label=below:$\alpha_{2r - 2}$] (n3) at (6, 0) {};
        \node[node, label=below:$\alpha_{2r - 1}$] (n4) at (8, 0) {};
        \draw (n1) -- (n2) -- (nd) -- (n3) -- (n4);
        \node[node, label=above:$\alpha_0$] (n0) at (4, 2) {};
        \draw (n1) -- (n0) -- (n4);
        \node (g1) at (0, -1.2) {};
        \node (g4) at (8, -1.2) {};
        \node (g2) at (2, -1.2) {};
        \node (g3) at (6, -1.2) {};
        \draw [<->] (g1) to [out=300, in=240] (g4);
        \draw [<->] (g2) to [out=310, in=230] (g3);
        \node (ra) at (11, 0) {$\Longrightarrow$};
        \node[node_b, label=below:$\beta_0$] (n0) at (14, 0) {};
        \node[node_b, label=below:$\beta_1$] (n1) at (16, 0) {};
        \node[node_b, label=below:$\beta_2$] (n2) at (18, 0) {};
        \node (nd) at (20, 0) {$\cdots$};
        \node[node_b, label=below:$\beta_{r - 2}$] (n3) at (22, 0) {};
        \node[node_b, label=below:$\beta_{r - 1}$] (n4) at (24, 0) {};
        \node[node_w, label=below:$\beta_r$] (n5) at (26, 0) {};
        \draw[thick] (n0) -- (n1) -- (n2) -- (nd) -- (n3) -- (n4);
        \draw[thick] (n4.north east) -- (n5.north west);
        \draw[thick] (n4.south east) -- (n5.south west);
    \end{tikzpicture}
\end{figure}

\noindent
$\underline{D_{r + 1}^{(1)} \rightarrow B_r^{(1)}}$

\begin{figure}[H]
    \centering
    \small
    \begin{tikzpicture}[scale=.47]
        \tikzset{node/.style={draw,circle,thick,inner sep=0pt,minimum size=7pt}}
        \tikzset{node_w/.style={draw,circle,thick,inner sep=0pt,minimum size=7pt}}
        \tikzset{node_b/.style={draw,circle,thick,fill=black,inner sep=0pt,minimum size=7pt}}
        \node[node, label=below:$\alpha_0$] (n0) at (0, 0) {};
        \node[node, label=below:$\alpha_1$] (n1) at (2, 0) {};
        \node[node, label=below:$\alpha_2$] (n2) at (4, 0) {};
        \node (nd) at (6, 0) {$\cdots$};
        \node[node, label=below:$\alpha_{r - 1}$] (n3) at (8, 0) {};
        \node[node, label=above:$\alpha_r$] (n4) at (10, 1.3) {};
        \node[node, label=below:$\alpha_{r + 1}$] (n5) at (10, -1.3) {};
        \draw[thick] (n0) -- (n1) -- (n2) -- (nd) -- (n3);
        \draw[thick] (n4) -- (n3) -- (n5);
        \node (g4) at (10.4, 1.3) {};
        \node (g5) at (10.4, -1.3) {};
        \draw [<->] (g4) to [out=330, in=30] (g5);
        \node (ra) at (14, 0) {$\Longrightarrow$};
        \node[node_w, label=below:$\beta_0$] (n0) at (17, 0) {};
        \node[node_w, label=below:$\beta_1$] (n1) at (19, 0) {};
        \node[node_w, label=below:$\beta_2$] (n2) at (21, 0) {};
        \node (nd) at (23, 0) {$\cdots$};
        \node[node_w, label=below:$\beta_{r - 2}$] (n3) at (25, 0) {};
        \node[node_w, label=below:$\beta_{r - 1}$] (n4) at (27, 0) {};
        \node[node_b, label=below:$\beta_r$] (n5) at (29, 0) {};
        \draw[thick] (n0) -- (n1) -- (n2) -- (nd) -- (n3) -- (n4);
        \draw[thick] (n4.north east) -- (n5.north west);
        \draw[thick] (n4.south east) -- (n5.south west);
    \end{tikzpicture}
\end{figure}

\noindent
$\underline{E_6^{(1)} \rightarrow F_4^{(1)}}$

\begin{figure}[H]
    \centering
    \small
    \begin{tikzpicture}[scale=.47]
        \tikzset{node/.style={draw,circle,thick,inner sep=0pt,minimum size=7pt}}
        \tikzset{node_w/.style={draw,circle,thick,inner sep=0pt,minimum size=7pt}}
        \tikzset{node_b/.style={draw,circle,thick,fill=black,inner sep=0pt,minimum size=7pt}}
        \node[node, label=below:$\alpha_1$] (n1) at (0, 0) {};
        \node[node, label=below:$\alpha_2$] (n2) at (2, 0) {};
        \node[node, label=below:$\alpha_3$] (n3) at (4, 0) {};
        \node[node, label=below:$\alpha_4$] (n4) at (6, 0) {};
        \node[node, label=below:$\alpha_5$] (n5) at (8, 0) {};
        \node[node, label=right:$\alpha_6$] (n6) at (4, 2) {};
        \node[node, label=right:$\alpha_0$] (n0) at (4, 4) {};
        \draw[thick] (n1) -- (n2) -- (n3) -- (n4) -- (n5);
        \draw[thick] (n3) -- (n6) -- (n0);
        \node (g1) at (0, -1.2) {};
        \node (g5) at (8, -1.2) {};
        \node (g2) at (2, -1.2) {};
        \node (g4) at (6, -1.2) {};
        \draw [<->] (g1) to [out=300, in=240] (g5);
        \draw [<->] (g2) to [out=310, in=230] (g4);
        \node (ra) at (11, 0) {$\Longrightarrow$};
        \node[node_b, label=below:$\beta_1$] (n1) at (14, 0) {};
        \node[node_b, label=below:$\beta_2$] (n2) at (16, 0) {};
        \node[node_w, label=below:$\beta_3$] (n3) at (18, 0) {};
        \node[node_w, label=below:$\beta_4$] (n4) at (20, 0) {};
        \node[node_w, label=below:$\beta_0$] (n0) at (22, 0) {};
        \draw[thick] (n1) -- (n2);
        \draw[thick] (n2.north east) -- (n3.north west);
        \draw[thick] (n2.south east) -- (n3.south west);
        \draw[thick] (n3) -- (n4) -- (n0);
    \end{tikzpicture}
\end{figure}

\noindent
$\underline{D_4^{(1)} \rightarrow G_2^{(1)}}$

\begin{figure}[H]
    \centering
    \small
    \begin{tikzpicture}[scale=.47]
        \tikzset{node/.style={draw,circle,thick,inner sep=0pt,minimum size=7pt}}
        \tikzset{node_w/.style={draw,circle,thick,inner sep=0pt,minimum size=7pt}}
        \tikzset{node_b/.style={draw,circle,thick,fill=black,inner sep=0pt,minimum size=7pt}}
        \node[node, label=below:$\alpha_0$] (n0) at (0, 0) {};
        \node[node, label=below:$\alpha_1$] (n1) at (2, 0) {};
        \node[node, label=below:$\alpha_2$] (n2) at (4, 0) {};
        \node[node, label=above:$\alpha_3$] (n3) at (5.8, 1.3) {};
        \node[node, label=below:$\alpha_4$] (n4) at (5.8, -1.3) {};
        \draw[thick] (n0) -- (n1) -- (n2);
        \draw[thick] (n3) -- (n2) -- (n4);
        \node (g3r) at (6.2, 1.3) {};
        \node (g4r) at (6.2, -1.3) {};
        \draw [<->] (g3r) to [out=330, in=30] (g4r);
        \node (g1a) at (2, 0.4) {};
        \node (g3l) at (5.2, 1.5) {};
        \draw [<->] (g1a) to [out=60, in=160] (g3l);
        \node (g1b) at (2, -1.2) {};
        \node (g4l) at (5.2, -1.5) {};
        \draw [<->] (g1b) to [out=320, in=200] (g4l);
        \node (ra) at (9, 0) {$\Longrightarrow$};
        \node[node_b, label=below:$\beta_0$] (n0) at (12, 0) {};
        \node[node_b, label=below:$\beta_1$] (n1) at (14, 0) {};
        \node[node_w, label=below:$\beta_2$] (n2) at (16, 0) {};
        \draw[thick] (n0) -- (n1) -- (n2);
        \draw[thick] (n1.north east) -- (n2.north west);
        \draw[thick] (n1.south east) -- (n2.south west);
    \end{tikzpicture}
\end{figure}

\subsection{Representations}

Here we summarize the matrices for the fundamental representation $(\rho^{(a)},V^{(a)})$ of a simple Lie algebra $\mathfrak{g}$, which is used in this paper.
In the following, we show the generators $E_{\alpha_a}$ for the positive simple roots $\alpha_a$ and the extended root $\alpha_0$.
For negative simple roots, the corresponding generators are obtained by taking their transpose and similarity transformation $S$:
\begin{equation}
    E_{-\alpha_a}=S\;{}^t E_{\alpha_a}\;S^{-1}.\label{eq:generators_for_negative_roots}
\end{equation}
The matrix $S$ is an identity matrix except adjoint representation.
The Cartan generators are defined by $\alpha_a^{\vee}\cdot H=[E_{\alpha_a},E_{-\alpha_a}]$.
$e_{a,b}$ denotes the matrix whose $(i,j)$ element is $\delta_{ia}\delta_{bj}$.

\subsubsection{\texorpdfstring{$A_r$}{Ar}}
$(\rho^{(1)},V^{(1)})$ has dimension $(r+1)$.
The generators  are given by
\begin{equation}
    \rho^{(1)}\qty(E_{\alpha_0}) = e_{r+1,1}, \quad \rho^{(1)}\qty(E_{\alpha_a})=e_{a,a+1},\quad a=1,\ldots,r.
\end{equation}
$(\rho^{(r)},V^{(r)})$ is dual to $(\rho^{(1)},V^{(1)})$ and has dimension $(r+1)$.
The generators are given by
\begin{equation}
    \rho^{(r)}\qty(E_{\alpha_0}) = e_{r+1,1}, \quad \rho^{(r)}\qty(E_{\alpha_{a}})=e_{r+1-a,r+2-a},\quad a=1,\ldots,r.
\end{equation}

\subsubsection{\texorpdfstring{$B_r$}{Br}}
The fundamental representation $(\rho^{(1)},V^{(1)})$ has dimension  $2r+1$.
The generators are given by
\begin{equation}
    \begin{aligned}
        \rho^{(1)}\qty(E_{\alpha_0}) = e_{2r,1}+e_{2r+1,2},\quad \rho^{(1)}\qty(E_{\alpha_r})=\sqrt{2}\qty( e_{r,r+1}+e_{r+1,r+2}),\\
        \rho^{(1)}\qty(E_{\alpha_a})=e_{a,a + 1} + e_{2r+1-a,2r+2-a},\quad a=1,\ldots,r-1.
    \end{aligned}
\end{equation}
The fundamental representations $V^{(r)}$ had dimension $2^r$.
Here we show the representation $V^{(r)}$ for $B_{r}$ for $r=2$ and $3$:\\
For $r=2$
\begin{equation}
    \rho^{(2)}\qty(E_{\alpha_1})=e_{2,3},\quad \rho^{(2)}\qty(E_{\alpha_2})=e_{1,2}+e_{3,4},\quad \rho^{(2)}\qty(E_{\alpha_0})=e_{4,1}.
    \label{eq:rep_V2_of_B2}
\end{equation}
For $r=3$
\begin{equation}
    \begin{aligned}
        &\rho^{(3)}\qty(E_{\alpha_1})=e_{3,4}+e_{5,6},\quad \rho^{(3)}\qty(E_{\alpha_2})=e_{2,3}+e_{6,7},\\ &\rho^{(3)}\qty(E_{\alpha_3})=e_{1,2}+e_{3,5}+e_{4,6}+e_{7,8},\quad\rho^{(3)}\qty(E_{\alpha_0})=e_{7,1}+e_{8,2}.
    \end{aligned}\label{eq:rep_V3_of_B3}
\end{equation}

\subsubsection{\texorpdfstring{$C_r$}{Cr}}
The representation $(\rho^{(1)},V^{(1)})$ has dimension $2r$.
The generators are given by
\begin{equation}
    \begin{aligned}
        &\rho^{(1)}\qty(E_{\alpha_0}) = e_{2r,1},\quad \rho^{(1)}\qty(E_{\alpha_r})=e_{r,r+1},\\
        &\rho^{(1)}\qty(E_{\alpha_a}) = e_{a,a+1}+e_{2r-a,2r+1-a},\quad a=1,\ldots,r-1.
    \end{aligned}
\end{equation}

\subsubsection{\texorpdfstring{$D_r$}{Dr}}
The representation $(\rho^{(1)},V^{(1)})$ has dimension $2r$.
The generators are
\begin{equation}
    \begin{aligned}
        &\rho^{(1)}\qty(E_{\alpha_0}) = e_{2r-1,1}+e_{2r,2}, \quad \rho^{(1)}\qty( E_{\alpha_r}) =e_{r-1,r+1} + e_{r,r+2},\\
        &\rho^{(1)}\qty(E_{\alpha_a}) = e_{a,a+1} + e_{2r-a,2r+1-a}, \quad a = 1, \ldots, r-1.
    \end{aligned}\label{eq:rep_V1_of_Dr}
\end{equation}
The fundamental representations $(\rho^{(r-1)},V^{(r-1)})$ and $(\rho^{(r)},V^{(r)})$ have dimension $2^{r-1}$.
Here we show the representation $V^{(r-1)}$ of $D_r$ for $r=3$ and $4$.\\
For $r=3$
\begin{equation}
    \rho^{(2)}\qty(E_{\alpha_1})=e_{2,3},\quad \rho^{(2)}\qty(E_{\alpha_2})=e_{1,2},\quad \rho^{(2)}\qty(E_{\alpha_3})=e_{3,4},\quad 
    \rho^{(2)}\qty(E_{\alpha_0})=e_{4,1}.\label{eq:rep_V2_of_D3}
\end{equation}
For $r=4$
\begin{equation}
    \begin{aligned}
        &\rho^{(3)}\qty(E_{\alpha_1})=e_{3,4}+e_{5,6},\quad \rho^{(3)}\qty(E_{\alpha_2})=e_{2,3}+e_{6,7},\\ 
        &\rho^{(3)}\qty(E_{\alpha_3})=e_{1,2}+e_{7,8},\quad \rho^{(3)}\qty(E_{\alpha_4})=e_{3,5}+e_{4,6},\quad \rho^{(3)}\qty(E_{\alpha_0})=e_{7,1}+e_{8,2}.
    \end{aligned}\label{eq:rep_V3_of_D4}
\end{equation}
Swapping the generators $E_{\alpha_{r-1}}\leftrightarrow E_{\alpha_r}$, we obtain the representation $(\rho^{(r)},V^{(r)})$.

\subsubsection{\texorpdfstring{$E_6$}{E6}}
The fundamental representation $(\rho^{(1)},V^{(1)})$ has dimension $27$.
The generators are given by
\begin{equation}
    \begin{aligned}
        \rho^{(1)}\qty(E_{\alpha_1}) &= e_{1, 2} + e_{12, 15} + e_{14, 17} + e_{16, 19} + e_{18, 21} + e_{20, 22},\\
        \rho^{(1)}\qty(E_{\alpha_2}) &= e_{2, 3} + e_{10, 12} + e_{11, 14} + e_{13, 16} + e_{21, 23} + e_{22, 24},\\
        \rho^{(1)}\qty(E_{\alpha_3}) &= e_{3, 4} + e_{8, 10} + e_{9, 11} + e_{16, 18} + e_{19, 21} + e_{24, 25},\\
        \rho^{(1)}\qty(E_{\alpha_4}) &= e_{4, 5} + e_{6, 8} + e_{11, 13} + e_{14, 16} + e_{17, 19} + e_{25, 26},\\
        \rho^{(1)}\qty(E_{\alpha_5}) &= e_{5, 7} + e_{8, 9} + e_{10, 11} + e_{12, 14} + e_{15, 17} + e_{26, 27},\\
        \rho^{(1)}\qty(E_{\alpha_6}) &= e_{4, 6} + e_{5, 8} + e_{7, 9} + e_{18, 20} + e_{21, 22} + e_{23, 24},\\
        \rho^{(1)}\qty(E_{\alpha_0}) &= e_{20, 1} + e_{22, 2} + e_{24, 3} + e_{25, 4} + e_{26, 5} + e_{27, 7}.
    \end{aligned}
\end{equation}
The representation $(\rho^{(5)},V^{(5)})$ is obtained from $V^{(1)}$ by swapping the  generators $E_{\alpha_1}\leftrightarrow E_{\alpha_5}$ and $E_{\alpha_2}\leftrightarrow E_{\alpha_4}$.
The representation $(\rho^{(6)},V^{(6)})$ is the adjoint representation with  dimension $78$, which is given by
\begin{align}
    \rho^{(6)}\qty(E_{\alpha_1}) =&e_{4, 7}+ e_{6, 9}+ e_{10, 12}+ e_{11, 13}+ e_{14, 17}+ e_{15, 19}+ e_{18, 22}+ e_{20, 24}\notag\\
    &+ e_{25, 29}+ e_{30, 35}- e_{36, 37}+ 2 e_{37, 43}- e_{38, 43}- e_{44, 49}- e_{50, 54}- e_{55, 59}\notag\\
    &- e_{57, 61}- e_{60, 64}- e_{62, 65}- e_{66, 68}- e_{67, 69}- e_{70, 73}- e_{72, 75},\notag\\
    \rho^{(6)}\qty(E_{\alpha_2}) = &e_{3, 4} + e_{5, 6} + e_{8, 11} + e_{12, 16} + e_{17, 21} + e_{19, 23} + e_{22, 26} + e_{24, 28}\notag\\ 
    &+ e_{29, 34} - e_{30, 36} - e_{35, 38} - e_{37, 44} + 2 e_{38, 44} - e_{39, 44} + e_{43, 49} - e_{45, 50}\notag\\
    &- e_{51, 55} - e_{53, 57} - e_{56, 60} - e_{58, 62} - e_{63, 67} - e_{68, 71} - e_{73, 74} - e_{75, 76}, \notag\\
    \rho^{(6)}\qty(E_{\alpha_3}) = &e_{2, 3} + e_{6, 10} + e_{9, 12} + e_{11, 14} + e_{13, 17} + e_{23, 27} - e_{25, 30} + e_{26, 31}+ e_{28, 33}\notag\\ 
    & - e_{29, 35} - e_{34, 39} - e_{38, 45} + 2 e_{39, 45} - e_{40, 45} - e_{42, 45} + e_{44, 50}- e_{46, 51} \notag\\
    & - e_{48, 53}+ e_{49, 54} - e_{52, 56} - e_{62, 66} - e_{65, 68} - e_{67, 70} - e_{69, 73} - e_{76, 77},\notag \\
    \rho^{(6)}\qty(E_{\alpha_4}) = &e_{3, 5} + e_{4, 6} + e_{7, 9} + e_{14, 18} + e_{17, 22} - e_{20, 25} + e_{21, 26} - e_{24, 29}\notag\\
    &+ e_{27, 32} - e_{28, 34} -  e_{33, 40} - e_{39, 46} + 2 e_{40, 46} - e_{41, 46} + e_{45, 51} - e_{47, 52}\\
    &+ e_{50, 55} - e_{53, 58} + e_{54, 59} - e_{57, 62} - e_{61, 65} - e_{70, 72} - e_{73, 75} - e_{74, 76}, \notag\\
    \rho^{(6)}\qty(E_{\alpha_5}) = &e_{5, 8} + e_{6, 11} + e_{9, 13} + e_{10, 14} + e_{12, 17} - e_{15, 20} + e_{16, 21} - e_{19, 24}\notag\\
    &- e_{23, 28} - e_{27, 33} - e_{32, 41} - e_{40, 47} + 2 e_{41, 47} + e_{46, 52} + e_{51, 56} + e_{55, 60}\notag\\
    &- e_{58, 63} + e_{59, 64} - e_{62, 67} - e_{65, 69} - e_{66, 70} - e_{68, 73} - e_{71, 74}, \notag\\
    \rho^{(6)}\qty(E_{\alpha_6}) = &e_{1, 2} - e_{10, 15} - e_{12, 19} + e_{14, 20} - e_{16, 23} + e_{17, 24} - e_{18, 25} + e_{21, 28}\notag\\
    &- e_{22, 29} - e_{26, 34} - e_{31, 42} - e_{39, 48} + 2 e_{42, 48} + e_{45, 53} + e_{50, 57} - e_{51, 58}\notag\\
    &+ e_{54, 61} - e_{55, 62} + e_{56, 63} - e_{59, 65} + e_{60, 67} + e_{64, 69} - e_{77, 78},\notag\\
    \rho^{(6)}\qty(E_{\alpha_0}) = &+e_{42,1}+e_{48,2}-e_{59,3}+e_{60,4}+e_{62,5}-e_{63,6}-e_{61,7}+e_{64,8}+e_{65,9}\notag\\
    &-e_{66,10}+e_{67,11}-e_{68,12}+e_{69,13}-e_{70,14}-e_{71,15}+e_{73,16}+e_{65,9}-e_{66,10}\notag\\
    &+e_{67,11}-e_{68,12}+e_{69,13}-e_{70,14}-e_{71,15}+e_{73,16}-e_{74,17}+e_{72,18}-e_{75,19}\notag\\
    &+e_{76,20}-e_{77,31}-e_{78,37}-2 e_{78,38}-3 e_{78,39}-2 e_{78,40}-e_{78,41}-2 e_{78,42}.\notag
\end{align}
$S$ in \eqref{eq:generators_for_negative_roots} is defined by
\begin{equation}
    S=\mathbf{1}_{36} \oplus C_{E_6} \oplus \mathbf{1}_{36},
    \label{eq:basis_transf_mat_E6V6}
\end{equation}
where $C_{E_6}$ is the Cartan matrix of $E_6$.

\subsubsection{\texorpdfstring{$E_7$}{E7}}
The fundamental representation $(\rho^{(1)},V^{(1)})$ has dimension 56.
The generators are
\begin{align}
    \rho^{(1)}\qty(E_{\alpha_1})=&e_{1, 2} + e_{16, 19} + e_{18, 22} + e_{21, 25} + e_{23, 28} +  e_{26, 31}\notag\\
    &+ e_{27, 30} + e_{29, 33} + e_{32, 36} + e_{35, 39} +  e_{38, 42} + e_{55, 56}, \notag\\
    \rho^{(1)}\qty(E_{\alpha_2})=&e_{2, 3} + e_{13, 16} + e_{15, 18} + e_{17, 21} + e_{20, 23} + e_{24, 27}\notag\\
    &+ e_{31, 34} + e_{33, 37} + e_{36, 40} + e_{39, 43} +  e_{42, 45} + e_{54, 55}, \notag\\
    \rho^{(1)}\qty(E_{\alpha_3})=&e_{3, 4} + e_{11, 13} + e_{12, 15} + e_{14, 17} + e_{23, 26} +  e_{27, 29}\notag\\
    &+ e_{28, 31} + e_{30, 33} + e_{40, 41} + e_{43, 44} + e_{45, 47} + e_{53, 54}, \notag\\
    \rho^{(1)}\qty(E_{\alpha_4})=&e_{4, 5} + e_{9, 11} + e_{10, 12} + e_{17, 20} + e_{21, 23}+ e_{25, 28}\notag\\
    & + e_{29, 32} + e_{33, 36} + e_{37, 40} + e_{44, 46} + e_{47, 49} + e_{52, 53},\\ 
    \rho^{(1)}\qty(E_{\alpha_5})=&e_{5, 6} + e_{7, 9} + e_{12, 14} + e_{15, 17} + e_{18, 21}+ e_{22, 25}\notag\\ 
    & + e_{32, 35} + e_{36, 39} + e_{40, 43} + e_{41, 44} + e_{49, 51} + e_{50, 52}, \notag\\
    \rho^{(1)}\qty(E_{\alpha_6})=&e_{6, 8} + e_{9, 10} + e_{11, 12} + e_{13, 15} + e_{16, 18}+e_{19, 22}\notag\\
    &+ e_{35, 38} + e_{39, 42} + e_{43, 45} + e_{44, 47} + e_{46, 49} + e_{48, 50},\notag\\ 
    \rho^{(1)}\qty(E_{\alpha_7})=&e_{5, 7} + e_{6, 9} + e_{8, 10} + e_{20, 24} + e_{23, 27} + e_{28, 30}\notag\\
    &+ e_{26, 29} + e_{31, 33} + e_{34, 37} + e_{46, 48} +  e_{49, 50} + e_{51, 52},\notag\\
    \rho^{(1)}\qty(E_{\alpha_0})=&e_{38,1}+e_{42,2}+e_{45,3}+e_{47,4}+e_{49,5}+e_{51,6}\notag\\
    &+e_{50,7}+e_{52,9}+e_{53,11}+e_{54,13}+e_{55,16}+e_{56,19}.\notag
\end{align}
For adjoint representation $(\rho^{(6)},V^{(6)})$, which has dimension $133$, the generators are given by
\begin{align}
    \rho^{(6)}\qty(E_{\alpha_1})=& e_{7,9}+ e_{10, 12} + e_{14, 16} + e_{17, 20} + e_{18, 21} + e_{19, 24} + e_{22, 26} + e_{23, 28}\notag\\
    & + e_{27, 31} e_{30, 34} + e_{33, 36} - e_{42, 44} - e_{43, 46} - e_{45, 49} - e_{48, 53} - e_{52, 59} \notag\\
    &- e_{58, 69} - e_{68, 76}+ 2 e_{69, 76} + e_{75, 82} + e_{81, 86} + e_{85, 89} + e_{88, 91}  \notag\\
    &+ e_{90, 92} - e_{98, 101} - e_{100, 104}- e_{103, 107}- e_{106, 111} - e_{108, 112}  \notag\\
    &-e_{110, 115} - e_{113, 116} - e_{114, 117} - e_{118, 120} - e_{122, 124} - e_{125, 127}, \notag\\
    \rho^{(6)}\qty(E_{\alpha_2})=& e_{5, 7} + e_{8, 10} + e_{11, 14} + e_{13, 17} + e_{15, 19} + e_{21, 25} + e_{26, 29} + e_{28, 32} \notag\\
    & + e_{31, 35}+ e_{34, 37}+ e_{36, 39} - e_{44, 47} - e_{46, 50} - e_{49, 54} + e_{52, 58} - e_{53, 60}\notag\\
    & - e_{59, 68} - e_{67, 75} + 2 e_{68, 75}- e_{69, 75} + e_{74, 81} - e_{76, 82} + e_{80, 85} + e_{84, 88} \notag\\
    &+ e_{87, 90} - e_{95, 98} - e_{97, 100} - e_{99, 103}- e_{102, 106} - e_{105, 108} - e_{109, 113} \notag\\
    & - e_{115, 119} - e_{117, 121} - e_{120, 123}  - e_{124, 126} - e_{127, 129},\notag\\
    \rho^{(6)}\qty(E_{\alpha_3})=& e_{4, 5} + e_{6, 8} + e_{14, 18} + e_{16, 21} + e_{17, 22} + e_{19, 23} + e_{20, 26} + e_{24, 28}  \notag\\
    &+ e_{35, 38} + e_{37, 40}+ e_{39, 41}- e_{47, 51} + e_{48, 52} - e_{50, 55} + e_{53, 59} - e_{54, 61} \notag\\
    & - e_{60, 67} - e_{66, 74}+ 2 e_{67, 74}- e_{68, 74} + e_{73, 80} - e_{75, 81} + e_{79, 84} - e_{82, 86} \notag\\
    & + e_{83, 87} - e_{93, 95} - e_{94, 97}- e_{96, 99}- e_{106, 110} - e_{108, 114} -e_{111, 115}  \notag\\
    &- e_{112, 117} - e_{113, 118} - e_{116, 120}- e_{126, 128} - e_{129, 130},\notag\\
    \rho^{(6)}\qty(E_{\alpha_4})=& e_{3, 4} + e_{8, 11} + e_{10, 14} + e_{12, 16} + e_{22, 27} + e_{23, 30} + e_{26, 31} + e_{28, 34} + e_{29, 35}\notag\\
    & + e_{32, 37}+ e_{41, 57} + e_{45, 48} + e_{49, 53} - e_{51, 56} + e_{54, 60} - e_{55, 62} - e_{61, 66}  \notag\\
    & - e_{65, 73}+ 2 e_{66, 73}- e_{67, 73}- e_{70, 73} + e_{72, 79} - e_{74, 80} - e_{77, 93} + e_{78, 83}   \notag\\
    & - e_{81, 85} - e_{86, 89}- e_{97, 102}- e_{99, 105}- e_{100, 106}-e_{103, 108} - e_{104, 111}  \notag\\
    & - e_{107, 112} - e_{118, 122} - e_{120, 124} - e_{123, 126} - e_{130, 131},\\
    \rho^{(6)}\qty(E_{\alpha_5})=& e_{2, 3} + e_{11, 13} + e_{14, 17} + e_{16, 20} + e_{18, 22} + e_{21, 26} + e_{25, 29} + e_{30, 33} \notag\\
    &+ e_{34, 36}  + e_{37, 39}+ e_{40, 41} + e_{43, 45} + e_{46, 49} + e_{50, 54} + e_{55, 61} - e_{56, 63} \notag\\
    &- e_{62, 65} - e_{64, 72} + 2 e_{65, 72}- e_{66, 72} + e_{71, 78} - e_{73, 79} - e_{80, 84} - e_{85, 88}\notag\\
    & - e_{89, 91} - e_{93, 94} - e_{95, 97} - e_{98, 100}- e_{101, 104} - e_{105, 109} - e_{108, 113}   \notag\\
    & - e_{112, 116} - e_{114, 118} - e_{117, 120}- e_{121, 123} - e_{131, 132},\notag\\
    \rho^{(6)}\qty(E_{\alpha_6})=& e_{1, 2} + e_{13, 15} + e_{17, 19} + e_{20, 24} + e_{22, 23} + e_{26, 28} + e_{27, 30} + e_{29, 32} \notag\\
    & + e_{31, 34}+ e_{35, 37}+ e_{38, 40} + e_{42, 43} + e_{44, 46} + e_{47, 50} + e_{51, 55} + e_{56, 62}\notag\\
    &  - e_{63, 64}+ 2 e_{64, 71}- e_{65, 71}- e_{72, 78} - e_{79, 83} - e_{84, 87}- e_{88, 90}  \notag\\
    &  - e_{91, 92}- e_{94, 96}- e_{97, 99} - e_{100, 103}- e_{102, 105} - e_{104, 107} - e_{106, 108} \notag\\
    &  - e_{110, 114} - e_{111, 112}- e_{115, 117} - e_{119, 121} - e_{132, 133},\notag\\
    \rho^{(1)}\qty(E_{\alpha_7})=&e_{4, 6} + e_{5, 8} + e_{7, 10} + e_{9, 12} + e_{27, 42} + e_{30, 43} - e_{31, 44} + e_{33, 45} - e_{34, 46}\notag\\
    &+ e_{35, 47} - e_{36, 49} + e_{37, 50} - e_{38, 51} + e_{39, 54} - e_{40, 55} - e_{41, 61} - e_{57, 70} \notag\\
    &- e_{66, 77}+ 2 e_{70, 77} + e_{73, 93} + e_{79, 94} - e_{80, 95} + e_{83, 96} - e_{84, 97} + e_{85, 98}\notag\\
    &- e_{87, 99} + e_{88, 100}- e_{89, 101} + e_{90, 103} - e_{91, 104} - e_{92, 107} - e_{122, 125} \notag\\
    &- e_{124, 127} - e_{126, 129} - e_{128, 130},\notag\\
    \rho^{(6)}\qty(E_{\alpha_0})=&-e_{64, 1} - e_{71, 2} - e_{78, 3} - e_{83, 4} - e_{87, 5} - e_{90, 7} - e_{92, 9} - e_{96, 6} \notag\\
    & + e_{99, 8} - e_{103, 10}- e_{105, 11} + e_{107, 12} + e_{108, 14} + e_{109, 13} - e_{112, 16} \notag\\
    & - e_{113, 17} - e_{114, 18} + e_{116, 20}+ e_{117, 21}+ e_{118, 22} - e_{120, 26} - e_{121, 25} \notag\\
    & - e_{122, 27} + e_{123, 29} + e_{124, 31}+ e_{125, 42} - e_{126, 35}+ e_{127, 44}+ e_{128, 38}  \notag\\
    & + e_{129, 47} + e_{130, 51} + e_{131, 56} + e_{132, 63}+ 2 e_{133, 64} + 3 e_{133, 65} \notag\\
    &+ 4 e_{133, 66} + 3 e_{133, 67} + 2 e_{133, 68} + e_{133, 69} + 2 e_{133, 70}.\notag
\end{align}
$S$ in \eqref{eq:generators_for_negative_roots} is given by
\begin{equation}
    S=\mathbf{1}_{63} \oplus JC_{E_7}J^{-1} \oplus \mathbf{1}_{63}. \label{eq:basis_transf_mat_E7V6}
\end{equation}
Here $C_{E_7}$ is the Cartan matrix of $E_7$ and $J$ is defined by
\begin{equation}
    J=\sum_{i=1}^7e_{i,8-i}.
\end{equation}

\subsubsection{\texorpdfstring{$E_8$}{E8}}
The fundamental representation $(\rho^{(1)},V^{(1)})$ is the adjoint representation with dimension $248$, which is given by
\begin{align}
    \rho^{(1)}\qty(E_{\alpha_1})=&e_{1,2}+e_{18,21}+e_{22,25}+e_{26,29}+e_{30,33}+e_{34,37}+e_{35,38}+e_{39,42}+e_{40,43}\notag\\
    &+e_{44,48}+e_{45,49}+e_{50,54}+e_{51,55}+e_{56,60}+e_{57,61}+e_{58,62}+e_{63,67}\notag\\
    &+e_{64,68}+e_{69,74}+e_{70,75}+e_{76,81}+e_{77,82}+e_{83,89}+e_{84,90}+e_{91,97}\notag\\
    &+e_{98,104}+e_{105,111}+e_{112,119}-e_{120,121}+2 e_{121,129}-e_{122,129}-e_{130,137}\notag\\
    &-e_{138,144}-e_{145,151}-e_{152,158}-e_{159,165}-e_{160,166}-e_{167,172}-e_{168,173}\notag\\
    &-e_{174,179}-e_{175,180}-e_{181,185}-e_{182,186}-e_{187,191}-e_{188,192}-e_{189,193}\notag\\
    &-e_{194,198}-e_{195,199}-e_{200,204}-e_{201,205}-e_{206,209}-e_{207,210}-e_{211,214}\notag\\
    &-e_{212,215}-e_{216,219}-e_{220,223}-e_{224,227}-e_{228,231}-e_{247,248},\notag\\
    \rho^{(1)}\qty(E_{\alpha_2})=&e_{2,3}+e_{15,18}+e_{19,22}+e_{23,26}+e_{27,30}+e_{31,34}+e_{32,35}+e_{36,40}+e_{41,45}\notag\\
    &+e_{42,47}+e_{46,51}+e_{48,53}+e_{52,58}+e_{54,59}+e_{60,65}+e_{61,66}+e_{67,72}\notag\\
    &+e_{68,73}+e_{74,79}+e_{75,80}+e_{81,87}+e_{82,88}+e_{89,95}+e_{90,96}+e_{97,103}\notag\\
    &+e_{104,110}+e_{111,118}-e_{112,120}-e_{119,122}-e_{121,130}+2 e_{122,130}-e_{123,130}\notag\\
    &+e_{129,137}-e_{131,138}-e_{139,145}-e_{146,152}-e_{153,159}-e_{154,160}-e_{161,167}\notag\\
    &-e_{162,168}-e_{169,174}-e_{170,175}-e_{176,181}-e_{177,182}-e_{183,188}-e_{184,189}\notag\\
    &-e_{190,195}-e_{191,197}-e_{196,201}-e_{198,203}-e_{202,207}-e_{204,208}-e_{209,213}\notag\\
    &-e_{214,217}-e_{215,218}-e_{219,222}-e_{223,226}-e_{227,230}-e_{231,234}-e_{246,247},\notag\\
    \rho^{(1)}\qty(E_{\alpha_3})=&e_{3,4}+e_{13,15}+e_{16,19}+e_{20,23}+e_{24,27}+e_{28,32}+e_{34,39}+e_{37,42}+e_{40,44}\notag\\
    &+e_{43,48}+e_{45,50}+e_{49,54}+e_{51,57}+e_{55,61}+e_{58,64}+e_{62,68}+e_{65,71}\notag\\
    &+e_{72,78}+e_{79,85}+e_{80,86}+e_{87,93}+e_{88,94}+e_{95,101}+e_{96,102}+e_{103,109}\notag\\
    &-e_{105,112}+e_{110,117}-e_{111,119}-e_{118,123}-e_{122,131}+e_{103,109}-e_{105,112}\notag\\
    &+e_{110,117}-e_{111,119}-e_{118,123}-e_{122,131}+2 e_{123,131}-e_{124,131}+e_{130,138}\notag\\
    &-e_{132,139}+e_{137,144}-e_{140,146}-e_{147,153}-e_{148,154}-e_{155,161}-e_{156,162}\notag\\
    &-e_{163,169}-e_{164,170}-e_{171,177}-e_{178,184}-e_{181,187}-e_{185,191}-e_{188,194}\notag\\
    &-e_{192,198}-e_{195,200}-e_{199,204}-e_{201,206}-e_{205,209}-e_{207,212}-e_{210,215}\notag\\
    &-e_{217,221}-e_{222,225}-e_{226,229}-e_{230,233}-e_{234,236}-e_{245,246},\notag\\
    \rho^{(1)}\qty(E_{\alpha_4})=&e_{4,5}+e_{11,13}+e_{14,16}+e_{17,20}+e_{27,31}+e_{30,34}+e_{32,36}+e_{33,37}+e_{35,40}\notag\\
    &+e_{38,43}+e_{50,56}+e_{54,60}+e_{57,63}+e_{59,65}+e_{61,67}+e_{64,70}+e_{66,72}+e_{68,75}\notag\\
    &+e_{73,80}+e_{85,92}+e_{93,99}+e_{94,100}-e_{98,105}+e_{101,107}+e_{102,108}-e_{104,111}\notag\\
    &+e_{109,116}-e_{110,118}-e_{117,124}-e_{123,132}+2 e_{124,132}-e_{125,132}+e_{131,139}\notag\\
    &-e_{133,140}+e_{138,145}-e_{141,147}-e_{142,148}+e_{144,151}-e_{149,155}-e_{150,156}\notag\\
    &-e_{157,164}-e_{169,176}-e_{174,181}-e_{177,183}-e_{179,185}-e_{182,188}-e_{184,190}\notag\\
    &-e_{186,192}-e_{189,195}-e_{193,199}-e_{206,211}-e_{209,214}-e_{212,216}-e_{213,217}\notag\\
    &-e_{215,219}-e_{218,222}-e_{229,232}-e_{233,235}-e_{236,238}-e_{244,245},\notag\\
    \rho^{(1)}\qty(E_{\alpha_5})=&e_{5,6}+e_{10,11}+e_{12,14}+e_{20,24}+e_{23,27}+e_{26,30}+e_{29,33}+e_{36,41}+e_{40,45}\notag\\
    &+e_{43,49}+e_{44,50}+e_{48,54}+e_{53,59}+e_{63,69}+e_{67,74}+e_{70,76}+e_{72,79}+e_{75,81}\notag\\
    &+e_{78,85}+e_{80,87}+e_{86,93}-e_{91,98}-e_{97,104}+e_{100,106}-e_{103,110}+e_{107,113}\notag\\
    &+e_{108,115}-e_{109,117}-e_{116,125}-e_{124,133}+2 e_{125,133}-e_{126,133}-e_{128,133}\notag\\
    &+e_{132,140}-e_{134,141}-e_{136,142}+e_{139,146}-e_{143,149}+e_{145,152}+e_{151,158}\\
    &-e_{156,163}-e_{162,169}-e_{164,171}-e_{168,174}-e_{170,177}-e_{173,179}-e_{175,182}\notag\\
    &-e_{180,186}-e_{190,196}-e_{195,201}-e_{199,205}-e_{200,206}-e_{204,209}-e_{208,213}\notag\\
    &-e_{216,220}-e_{219,223}-e_{222,226}-e_{225,229}-e_{235,237}-e_{238,239}-e_{243,244},\notag\\
    \rho^{(1)}\qty(E_{\alpha_6})=&e_{6,7}+e_{8,10}+e_{14,17}+e_{16,20}+e_{19,23}+e_{22,26}+e_{25,29}+e_{41,46}+e_{45,51}\notag\\
    &+e_{49,55}+e_{50,57}+e_{54,61}+e_{56,63}+e_{59,66}+e_{60,67}+e_{65,72}+e_{71,78}+e_{76,83}\notag\\
    &+e_{81,89}-e_{84,91}+e_{87,95}-e_{90,97}+e_{93,101}-e_{96,103}+e_{99,107}-e_{102,109}\notag\\
    &+e_{106,114}-e_{108,116}-e_{115,126}-e_{125,134}+2 e_{126,134}-e_{127,134}+e_{133,141}\notag\\
    &-e_{135,143}+e_{140,147}-e_{142,150}+e_{146,153}-e_{148,156}+e_{152,159}-e_{154,162}\notag\\
    &+e_{158,165}-e_{160,168}-e_{166,173}-e_{171,178}-e_{177,184}-e_{182,189}-e_{183,190}\notag\\
    &-e_{186,193}-e_{188,195}-e_{192,199}-e_{194,200}-e_{198,204}-e_{203,208}-e_{220,224}\notag\\
    &-e_{223,227}-e_{226,230}-e_{229,233}-e_{232,235}-e_{239,241}-e_{242,243},\notag\\
    \rho^{(1)}\qty(E_{\alpha_7})=&e_{7,9}+e_{10,12}+e_{11,14}+e_{13,16}+e_{15,19}+e_{18,22}+e_{21,25}+e_{46,52}+e_{51,58}\notag\\
    &+e_{55,62}+e_{57,64}+e_{61,68}+e_{63,70}+e_{66,73}+e_{67,75}+e_{69,76}+e_{72,80}\notag\\
    &+e_{74,81}-e_{77,84}+e_{78,86}+e_{79,87}-e_{82,90}+e_{85,93}-e_{88,96}+e_{92,99}\notag\\
    &-e_{94,102}-e_{100,108}-e_{106,115}-e_{114,127}-e_{126,135}+2 e_{127,135}+e_{134,143}\notag\\
    &+e_{141,149}+e_{147,155}-e_{150,157}+e_{153,161}-e_{156,164}+e_{159,167}-e_{162,170}\notag\\
    &-e_{163,171}+e_{165,172}-e_{168,175}-e_{169,177}-e_{173,180}-e_{174,182}-e_{176,183}\notag\\
    &-e_{179,186}-e_{181,188}-e_{185,192}-e_{187,194}-e_{191,198}-e_{197,203}-e_{224,228}\notag\\
    &-e_{227,231}-e_{230,234}-e_{233,236}-e_{235,238}-e_{237,239}-e_{240,242},\notag\\
    \rho^{(1)}\qty(E_{\alpha_8})=&e_{6,8}+e_{7,10}+e_{9,12}+e_{24,28}+e_{27,32}+e_{30,35}+e_{31,36}+e_{33,38}+e_{34,40}\notag\\
    &+e_{37,43}+e_{39,44}+e_{42,48}+e_{47,53}-e_{69,77}-e_{74,82}+e_{76,84}-e_{79,88}\notag\\
    &+e_{81,90}-e_{83,91}-e_{85,94}+e_{87,96}-e_{89,97}-e_{92,100}+e_{93,102}-e_{95,103}\notag\\
    &+e_{99,108}-e_{101,109}-e_{107,116}-e_{113,128}-e_{125,136}+2 e_{128,136}+e_{133,142}\notag\\
    &+e_{140,148}-e_{141,150}+e_{146,154}-e_{147,156}+e_{149,157}+e_{152,160}-e_{153,162}\notag\\
    &+e_{155,164}+e_{158,166}-e_{159,168}+e_{161,170}-e_{165,173}+e_{167,175}+e_{172,180}\notag\\
    &-e_{196,202}-e_{201,207}-e_{205,210}-e_{206,212}-e_{209,215}-e_{211,216}-e_{213,218}\notag\\
    &-e_{214,219}-e_{217,222}-e_{221,225}-e_{237,240}-e_{239,242}-e_{241,243},\notag\\
    \rho^{(1)}\qty(E_{\alpha_0})=&-e_{121,1}-e_{129,2}-e_{137,3}-e_{144,4}-e_{151,5}-e_{158,6}-e_{165,7}-e_{166,8}\notag\\
    &-e_{172,9}+e_{173,10}-e_{179,11}-e_{180,12}+e_{185,13}+e_{186,14}-e_{191,15}\notag\\
    &-e_{192,16}-e_{193,17}+e_{197,18}+e_{198,19}+e_{199,20}-e_{203,22}-e_{204,23}\notag\\
    &-e_{205,24}+e_{208,26}+e_{209,27}+e_{210,28}-e_{213,30}-e_{214,31}-e_{215,32}\notag\\
    &+e_{217,34}+e_{218,35}+e_{219,36}-e_{221,39}-e_{222,40}-e_{223,41}+e_{225,44}\notag\\
    &+e_{226,45}+e_{227,46}-e_{229,50}-e_{230,51}-e_{231,52}+e_{232,56}+e_{233,57}\notag\\
    &+e_{234,58}-e_{235,63}-e_{236,64}+e_{237,69}+e_{238,70}-e_{239,76}+e_{240,77}\notag\\
    &+e_{241,83}+e_{242,84}+e_{243,91}+e_{244,98}+e_{245,105}+e_{246,112}\notag\\
    &+e_{247,120}+2 e_{248,121}+3 e_{248,122}+4 e_{248,123}+5 e_{248,124}\notag\\
    &+6 e_{248,125}+4 e_{248,126}+2 e_{248,127}+3 e_{248,128}.\notag
\end{align}
$S$ in \eqref{eq:generators_for_negative_roots} is given by
\begin{equation}
    S=\mathbf{1}_{120} \oplus C_{E_8} \oplus \mathbf{1}_{120}, \label{eq:basis_transf_mat_E8V1}
\end{equation}
where $C_{E_8}$ is the Cartan matrix of $E_8$.

\subsubsection{\texorpdfstring{$F_4^{(1)}$}{F4}}
The fundamental representation $(\rho^{(1)},V^{(1)})$, which has dimension $26$,  is
\begin{equation}
    \begin{aligned}
        \rho^{(1)}\qty(E_{\alpha_1}) =& e_{1, 2} + e_{5, 7} + e_{8, 9} + e_{10, 11} + 2^{-1/2} e_{12, 13} \\
        &+ \qty(3/2)^{1/2}e_{12, 14}+ 2^{-1/2}e_{13, 16} - \qty(3/2)^{1/2}e_{14, 16}\\
        &+ e_{15, 18} + e_{17, 20} +e_{19,21} +  e_{25, 26},\\
        \rho^{(1)}\qty(E_{\alpha_2}) =& e_{2, 3} + e_{4, 5} + e_{6, 8} + e_{10, 12} + 2^{1/2} e_{11, 13} \\
        &+ 2^{1/2} e_{13, 15}+ e_{16, 18} + e_{20, 22} + e_{21, 23}  + e_{24, 25},\\
        \rho^{(1)}\qty(E_{\alpha_3}) =& e_{3, 4} + e_{8, 10} + e_{9, 11} + e_{15, 17} + e_{18, 20} + e_{23, 24}, \\
        \rho^{(1)}\qty(E_{\alpha_4}) =& e_{4, 6} + e_{5, 8} + e_{7, 9} + e_{17, 19} + e_{20, 21} + e_{22, 23},\\
        \rho^{(1)}\qty(E_{\alpha_0}) =& e_{19, 1} + e_{21, 2} + e_{23, 3} + e_{24, 4} + e_{25, 5} + e_{26, 7}.
    \end{aligned}
\end{equation}
The  representation $(\rho^{(4)},V^{(4)})$, which is the adjoint representation with dimension $56$, is given by
\begin{equation}
\begin{aligned}
    \rho^{(4)}\qty(E_{\alpha_1}) =& -e_{4, 5} - e_{6, 7} - 2 e_{7, 9} - e_{8, 10} - 2 e_{10, 12} - e_{11, 14} - e_{13, 17} - 2 e_{14, 15}\\ 
    &- e_{16, 19} - e_{18, 22} - 2 e_{21, 28} - 2^{-1} e_{27, 32} + e_{28, 32} + e_{31, 35} + e_{34, 37} + e_{36, 40}\\
    &+ e_{38, 39} + 2 e_{39, 42} + e_{41, 43} + 2 e_{43, 45} + e_{44, 46} + 2 e_{46, 47} + e_{48, 49},\\
    \rho^{(4)}\qty(E_{\alpha_2}) = & -e_{3, 4} - 2 e_{4, 6} - e_{5, 7} - e_{10, 13} - e_{12, 17} - e_{14, 16} - e_{15, 19} - 2 e_{17, 20}\\
    &+ e_{18, 21} - 2 e_{19, 23} - 2 e_{22, 27} - e_{26, 31} + e_{27, 31} - 2^{-1} e_{28, 31} + e_{30, 34} - e_{32, 35}\\ 
    &+ e_{33, 36} + 2 e_{34, 38} + 2 e_{36, 41} + e_{37, 39} + e_{40, 43} + e_{46, 48} + e_{47, 49} + 2 e_{49, 50},\\
    \rho^{(4)}\qty(E_{\alpha_3}) = & -e_{2, 3} - e_{6, 8} - e_{7, 10} - e_{9, 12} + e_{16, 18} + e_{19, 22} - e_{20, 24} - e_{23, 26}\\
    & -e_{25, 30} + 2 e_{26, 30} - e_{27, 30} + e_{29, 33} - e_{31, 34} - e_{35, 37} + e_{41, 44} \\
    &+ e_{43, 46}+ e_{45, 47} + e_{50, 51},\\
    \rho^{(4)}\qty(E_{\alpha_4}) = &-e_{1, 2} + e_{8, 11} + e_{10, 14} + e_{12, 15} + e_{13, 16} + e_{17, 19} + e_{20, 23} - e_{24, 25}+ 2 e_{25, 29} \\
    &- e_{26, 29} - e_{30, 33} - e_{34, 36} - e_{37, 40} - e_{38, 41} - e_{39, 43} - e_{42, 45} + e_{51, 52},\\
    \rho^{(4)}\qty(E_{\alpha_0}) = &2e_{25,1} +3e_{26,1}+4e_{27,1}+2e_{28,1}-e_{29,2}+e_{33,3}-e_{36,4}+e_{40,5}+e_{41,6}\\
    &-e_{43,7}-e_{44,8}+e_{45,9}+e_{46,10}-e_{47,12}-e_{48,13}+e_{49,17}-e_{50,20}\\
    &+e_{51,24}-e_{52,25}.
\end{aligned}
\end{equation}
$S$ in \eqref{eq:generators_for_negative_roots} is given by
\begin{equation}
     S=P_{24}\oplus JK_{F_4}J^{-1} \oplus P_{24}^{\prime}.\label{eq:basis_transf_mat_F4V4}
\end{equation}
Here $\qty(K_{F_4})_{ab}=\alpha_a^{\vee}\cdot\alpha_b^{\vee}$ is the symmetrized Cartan matrix. $P_{24}$ and $P_{24}^{\prime}$ are $24$ dimensional diagonal matrices, which have the components $P_{24}=\mathrm{diag}\qty(p_1,\ldots,p_{24})$ and $P_{24}^{\prime}=\mathrm{diag}\qty(p_{24},\ldots,p_{1})$ with
\begin{equation}
     p_i=\begin{cases} 2, & i\in\qty{4, 5, 7, 10, 13, 14, 16, 17, 18, 19, 21, 22}, \\ 1, & \text{otherwise}. \end{cases}
\end{equation}
$J$ is defined by
\begin{equation}
    J=\sum_{i=1}^4e_{i,5-i}.
\end{equation}

\subsubsection{\texorpdfstring{$G_2$}{G2}}
The fundamental representation $(\rho^{(1)},V^{(1)})$, which has dimension $7$, is given by
\begin{equation}
    \begin{aligned}
        &\rho^{(1)}\qty(E_{\alpha_0}) = e_{6,1}+e_{7,2}, \quad \rho^{(1)}\qty(E_{\alpha_2}) =e_{2,3}+e_{5,6},\\
        &\rho^{(1)}\qty(E_{\alpha_1})=e_{1,2}+e_{6,7}+\sqrt{2}\qty(e_{3,4}+e_{4,5}).
    \end{aligned}
\end{equation}

\section{\texorpdfstring{$\mathcal{L}_{\hat{\mathfrak{g}}}$}{Lg} and \texorpdfstring{$\mathcal{L}_{\hat{\mathfrak{g}}}^{\text{dual}}$}{Lgdual} for affine Lie algebras}\label{sec:linear_ops}
In this appendix, we summarize the differential operators for the (dual) linear problems associated with affine Lie algebras, which are used in this work.
We also explain the folding structure of the linear operators for non-simply-laced Lie algebras.

\subsection{simply-laced affine Lie algebras}
We begin with simply-laced affine Lie algebras.

\subsubsection{\texorpdfstring{$A_r^{(1)}$}{Ar}}
The  $\mathcal{L}_{A_r^{(1)}}$ for the representation $(\rho^{(a)},V^{(a)})$ is defined by
\begin{equation}
    \mathcal{L}_{A_{r}^{(1)}}=\dv{x}-\frac{1}{x}\sum_{a=1}^r l_a\  \rho^{(a)}(\alpha_a^{\vee}\cdot H)+\sum_{a=1}^r \rho^{(a)}(E_{\alpha_a})+ p(x,E)\;\zeta \rho^{(a)}(E_{\alpha_0}). \label{eq:linear_op_Ar}
\end{equation}
Here $\zeta=1$ for $V^{(1)}$ and $V^{(r)}$. 
The differential operator $\mathcal{L}_{A_r^{(1)}}^{\text{dual}}$ for the dual representation
is given by
\begin{equation}
    \mathcal{L}_{A_r^{(1)}}^{\text{dual}}=\mathbf{1}_{r+1}\dv{x} + \frac{1}{x} \sum_{a=1}^r l_a\;\rho^{(1)}\qty(\alpha_{a}^{\vee}\cdot H)- \sum_{a=1}^r \rho^{(1)}\qty(E_{-\alpha_{a}})-p(x,E)\;\rho^{(1)}\;\qty(E_{-\alpha_0}). \label{eq:dual_linear_op_Ar_V1}
\end{equation}
$\mathcal{L}_{A_r^{(1)}}^{\text{dual}}$ for $V^{(r)}$ is obtained by swapping the generators $E_{\alpha_a}\leftrightarrow E_{\alpha_{r+1-a}}\;(a=1,\ldots,r)$ and $\alpha_{a}^{\vee}\cdot H\leftrightarrow\alpha_{r+1-a}^{\vee}\cdot H\;(a=1,\ldots,r)$ in \eqref{eq:dual_linear_op_Ar_V1}.

\subsubsection{\texorpdfstring{$D_r^{(1)}$}{Dr}}
The differential operator $\mathcal{L}_{D_r^{(1)}}$ is
\begin{equation}
    \mathcal{L}_{D_r^{(1)}} = \dv{x} -\frac{1}{x} \sum_{a = 1}^{r}l_a\;\alpha_a^{\vee} \cdot H + E_{\alpha_1} + \sum_{a = 2}^{r - 2}\sqrt{2}E_{\alpha_a} + E_{\alpha_{r - 1}} + E_{\alpha_r} + p(x, E)\;\zeta E_{\alpha_0}, \label{eq:linear_op_Dr}
\end{equation}
where $\zeta=1$ for $V^{(1)}$ and $\zeta=(-1)^r$ for $V^{(a)}$ ($a=r-1,r$) for the representations \eqref{eq:rep_V2_of_D3} and \eqref{eq:rep_V3_of_D4}.
The operator $\mathcal{L}_{D_r^{(1)}}^{\text{dual}}$ in $V^{(1)}$ is given by
\begin{equation}
    \begin{aligned}
        \mathcal{L}_{D_r}^{\text{dual}}=&\mathbf{1}_{2r}\dv{x} - \frac{1}{x}\sum_{a=1}^r l_a\; \rho^{(1)}\qty(\alpha_{a}^{\vee}\cdot H)-\rho^{(1)}\qty(E_{-\alpha_1})- \sum_{a=1}^r \sqrt{2} \rho^{(1)}\qty(E_{-\alpha_{a}})\\
        &\qquad\qquad\qquad-\rho^{(1)}\qty(E_{-\alpha_{r-1}})-\rho^{(1)}\qty(E_{-\alpha_{r}})-p(x,E)\;\zeta\rho^{(1)}\qty(E_{-\alpha_0}).
    \end{aligned}\label{eq:dual_linear_op_Dr}
\end{equation}
$\mathcal{L}_{D_{r}^{(1)}}^{\text{dual}}$ in $V^{(r-1)}$ for \eqref{eq:rep_V2_of_D3} and \eqref{eq:rep_V3_of_D4} is
\begin{equation}
    \begin{aligned}
        \mathcal{L}_{D_r}^{\text{dual}}=&\mathbf{1}_{2^{r-1}}\dv{x} - \frac{1}{x}\sum_{a=1}^r l_a\; \rho^{(r-1)}\qty(\alpha_{a}^{\vee}\cdot H)-\rho^{(r-1)}\qty(E_{-\alpha_1})- \sum_{a=1}^r \sqrt{2} \rho^{(r-1)}\qty(E_{-\alpha_{a}})\\
        &\qquad\qquad-\rho^{(r-1)}\qty(E_{-\alpha_{r-1}})-\rho^{(r-1)}\qty(E_{-\alpha_{r}})-(-1)^{r}p(x,E)\;\rho^{(r-1)}\qty(E_{-\alpha_0}).
    \end{aligned}\label{eq:dual_linear_op_Dr_Vr-1}
\end{equation}
$\mathcal{L}_{D_{r}^{(1)}}^{\text{dual}}$ in $V^{(r)}$ is obtained by swapping the generators $E_{\alpha_{r-1}}\leftrightarrow E_{\alpha_r}$ and $\alpha_{r-1}^{\vee}\cdot H\leftrightarrow\alpha_{r}^{\vee}\cdot H$ in \eqref{eq:dual_linear_op_Dr_Vr-1}.

\subsubsection{\texorpdfstring{$E_6^{(1)}$}{E6}}
The differential operator $\mathcal{L}_{E_6^{(1)}}$ is
\begin{equation}
    \begin{aligned}
        \mathcal{L}_{E_6^{(1)}} &= \dv{x} - \frac{1}{x}\sum_{a = 1}^{6}l_a\;\alpha_a^{\vee} \cdot H \\
        &+ E_{\alpha_1}+ \sqrt{2} E_{\alpha_2} + \sqrt{3} E_{\alpha_3} + \sqrt{2} E_{\alpha_4} + E_{\alpha_{5}} + \sqrt{2} E_{\alpha_6} + p(x, E)\;\zeta E_{\alpha_0}, 
    \end{aligned}\label{eq:linear_op_E6}
\end{equation}
where $\zeta=1$ for $V^{(1)},V^{(5)}$ and $\zeta=-1$ for $V^{(6)}$.
The operator $\mathcal{L}_{E_6^{(1)}}^{\text{dual}}$ in $V^{(1)}$ is given by 
\begin{equation}
    \begin{aligned}
        \mathcal{L}_{E_6^{(1)}}^{\text{dual}} =& \mathbf{1}_{27}\dv{x} + \frac{1}{x}\sum_{a = 1}^{6}l_{a}\;\rho^{(b)}\qty(-\alpha_a^{\vee} \cdot H)- \rho^{(b)}\qty(E_{-\alpha_1})- \sqrt{2} \rho^{(b)}\qty(E_{-\alpha_2})- \rho^{(b)}\qty(E_{-\alpha_3})\\
        &\qquad \quad - \sqrt{2} \rho^{(b)}\qty(E_{-\alpha_4})+ \rho^{(b)}\qty(E_{-\alpha_{5}})-\sqrt{2} \rho^{(b)}\qty(E_{-\alpha_6}) - p(x, E)\; \rho^{(b)}\qty(E_{-\alpha_0}).
    \end{aligned} \label{eq:dual_linear_op_E6_V1}
\end{equation}
$\mathcal{L}_{E_6^{(1)}}^{\text{dual}}$ in $V^{(5)}$ is obtained by swapping the generators $E_{\alpha_1}\leftrightarrow E_{\alpha_5},E_{\alpha_2}\leftrightarrow E_{\alpha_4},\alpha_{1}^{\vee}\cdot H\leftrightarrow\alpha_{5}^{\vee}\cdot H$ and $\alpha_{2}^{\vee}\cdot H\leftrightarrow\alpha_{4}^{\vee}\cdot H$ in \eqref{eq:dual_linear_op_E6_V1}.
The operator $\mathcal{L}_{E_6^{(1)}}^{\text{dual}}$ in $V^{(6)}$ is 
\begin{equation}
    \begin{aligned}
        \mathcal{L}_{E_6^{(1)}}^{\text{dual}} =& \mathbf{1}_{78}\dv{x} + \frac{1}{x}\sum_{a = 1}^{6}l_{a}\;S^{-1}\rho^{(6)}\qty(-\alpha_a^{\vee} \cdot H)S- S^{-1}\rho^{(6)}\qty(E_{-\alpha_1})S\\
        &\quad- \sqrt{2} S^{-1}\rho^{(6)}\qty(E_{-\alpha_2})S- S^{-1}\rho^{(6)}\qty(E_{-\alpha_3})S- \sqrt{2} S^{-1}\rho^{(6)}\qty(E_{-\alpha_4})S\\
        &\qquad + S^{-1}\rho^{(6)}\qty(E_{-\alpha_{5}})S-\sqrt{2} S^{-1}\rho^{(6)}\qty(E_{-\alpha_6})S + p(x, E)\; S^{-1}\rho^{(6)}\qty(E_{-\alpha_0})S,
    \end{aligned} \label{eq:dual_linear_op_E6_V6}
\end{equation}
where $S$ is defined in \eqref{eq:basis_transf_mat_E6V6}.

\subsubsection{\texorpdfstring{$E_7^{(1)}$}{E7}}
The differential operator $\mathcal{L}_{E_7^{(1)}}$ is
\begin{equation}
    \begin{aligned}
        \mathcal{L}_{E_7^{(1)}} =& \dv{x} -\frac{1}{x} \sum_{a = 1}^{7}l_a\;\alpha_a^{\vee} \cdot H + E_{\alpha_1} + \sqrt{2} E_{\alpha_2} + \sqrt{3} E_{\alpha_3} \\
        & + \sqrt{4} E_{\alpha_4} + \sqrt{3} E_{\alpha_5} + \sqrt{2} E_{\alpha_6} + \sqrt{2} E_{\alpha_7} + p(x, E)\;\zeta E_{\alpha_0}, \label{eq:linear_op_E7}
    \end{aligned}
\end{equation}
where $\zeta=-1$ for $V^{(1)}$ and $\zeta=1$ for $V^{(6)}$.
The differential operator $\mathcal{L}_{E_7^{(1)}}^{\text{dual}}$ in $V^{(1)}$ is given by
\begin{equation}
    \begin{aligned}
        \mathcal{L}_{E_7^{(1)}}^{\text{dual}}=& \mathbf{1}_{56}\dv{x} + \frac{1}{x}\sum_{a = 1}^{7}l_{a}\;\rho^{(1)}\qty(-\alpha_a^{\vee} \cdot H)- \rho^{(1)}\qty(E_{-\alpha_1})- \sqrt{2}\rho^{(1)}\qty(E_{-\alpha_2})\\
        &\quad\qquad - \sqrt{3}\rho^{(1)}\qty(E_{-\alpha_3})- \sqrt{4} \rho^{(1)}\qty(E_{-\alpha_4})- \sqrt{3}\rho^{(1)}\qty(E_{-\alpha_5})\\
        &\qquad\qquad - \sqrt{2} \rho^{(1)}\qty(E_{-\alpha_6}) - \sqrt{2}\rho^{(1)}\qty(E_{-\alpha_{7}}) + p(x, E)\;\rho^{(1)}\qty(E_{-\alpha_0}).
    \end{aligned}\label{eq:dual_linear_op_E7_V1}
\end{equation}
The differential operator $\mathcal{L}_{E_7^{(1)}}^{\text{dual}}$ in $V^{(6)}$ is 
\begin{equation}
    \begin{aligned}
        \mathcal{L}_{E_7^{(1)}}^{\text{dual}}=& \mathbf{1}_{78}\dv{x} + \frac{1}{x}\sum_{a = 1}^{7}l_{a}\;S^{-1}\rho^{(6)}\qty(-\alpha_a^{\vee} \cdot H)S\\
        &- S^{-1}\rho^{(6)}\qty(E_{-\alpha_1})S- \sqrt{2}S^{-1}\rho^{(6)}\qty(E_{-\alpha_2})S- \sqrt{3}S^{-1}\rho^{(6)}\qty(E_{-\alpha_3})S\\
        &\qquad - \sqrt{4} S^{-1}\rho^{(6)}\qty(E_{-\alpha_4})S- \sqrt{3}S^{-1}\rho^{(6)}\qty(E_{-\alpha_5})S- \sqrt{2} S^{-1}\rho^{(6)}\qty(E_{-\alpha_6})S\\
        &\qquad\qquad  - \sqrt{2}S^{-1}\rho^{(6)}\qty(E_{-\alpha_{7}})S + p(x, E)\;S^{-1}\rho^{(6)}\qty(E_{-\alpha_0})S,
    \end{aligned}\label{eq:dual_linear_op_E7_V6}
\end{equation}
where $S$ is defined in \eqref{eq:basis_transf_mat_E7V6}.

\subsubsection{\texorpdfstring{$E_8^{(1)}$}{E8}}
The differential operator $\mathcal{L}_{E_8^{(1)}}$ is
\begin{equation}
    \begin{aligned}
        \mathcal{L}_{E_8^{(1)}} =& \dv{x} - \frac{1}{x}\sum_{a = 1}^{8}l_a\;\alpha_a^{\vee} \cdot H + \sqrt{2}E_{\alpha_1} + \sqrt{3} E_{\alpha_2} + \sqrt{4} E_{\alpha_3} \\
        & + \sqrt{5} E_{\alpha_4} + \sqrt{6} E_{\alpha_5} + \sqrt{4} E_{\alpha_6} + \sqrt{2} E_{\alpha_7} + \sqrt{3} E_{\alpha_8} + p(x, E)\;\zeta E_{\alpha_0}, \label{eq:linear_op_E8}
    \end{aligned}
\end{equation}
where $\zeta=-1$ for $V^{(1)}$. $\mathcal{L}_{E_8^{(1)}}^{\text{dual}}$ in $V^{(1)}$ is given by
\begin{equation}
    \begin{aligned}
        \mathcal{L}_{E_8^{(1)}}^{\text{dual}} =& \mathbf{1}_{248}\dv{x} + \frac{1}{x}\sum_{a = 1}^{8}l_a\;S^{-1}\rho^{(1)}\qty(\alpha_a^{\vee} \cdot H)S\\
        &- \sqrt{2}S^{-1}\rho^{(1)}\qty(E_{\alpha_1})S - \sqrt{3} S^{-1}\rho^{(1)}\qty(E_{\alpha_2})S - \sqrt{4} S^{-1}\rho^{(1)}\qty(E_{\alpha_3})S \\
        &\quad - \sqrt{5} S^{-1}\rho^{(1)}\qty(E_{\alpha_4})S - \sqrt{6} S^{-1}\rho^{(1)}\qty(E_{\alpha_5})S - \sqrt{4} S^{-1}\rho^{(1)}\qty(E_{\alpha_6})S\\
        &\quad\quad- \sqrt{2} S^{-1}\rho^{(1)}\qty(E_{\alpha_7})S - \sqrt{3} S^{-1}\rho^{(1)}\qty(E_{\alpha_8})S + p(x, E)\; S^{-1}\rho^{(1)}\qty(E_{\alpha_0})S, \label{eq:dual_linear_op_E8_V1}
    \end{aligned}
\end{equation}
where $S$ is defined in \eqref{eq:basis_transf_mat_E8V1}.

\subsection{Non-simply-laced affine Lie algebras}

\subsubsection{\texorpdfstring{$B_r^{(1)}$}{Br}}
The differential operator $\mathcal{L}_{B_r^{(1)}}$ is
\begin{equation}
    \mathcal{L}_{B_{r}^{(1)}}=\dv{x}-\frac{1}{x}\sum_{a=1}^r l_a\;\alpha_a^{\vee}\cdot H + E_{\alpha_1}+\sum_{a=2}^{r-1}\sqrt{2} E_{\alpha_a} +E_{\alpha_r} + p(x,E)\; \zeta E_{\alpha_0},\label{eq:linear_op_Br}
\end{equation}
where $\zeta=1$ for $V^{(1)}$ and $\zeta=(-1)^{r+1}$ for $V^{(r)}$ for the representations \eqref{eq:rep_V2_of_B2} and \eqref{eq:rep_V3_of_B3}.
The differential operator $\mathcal{L}_{B_r^{(1)}}^{\text{dual}}$ in $V^{(1)}$ is given by
\begin{equation}
    \begin{aligned}
        \mathcal{L}_{B_r}^{\text{dual}}=&\mathbf{1}_{2r+1}\dv{x} +\frac{1}{x} \sum_{a=1}^r l_a\;\rho^{(1)}\qty(\alpha_a^{\vee}\cdot H)\\
        &- \rho^{(1)}\qty(E_{-\alpha_1}) - \sum_{a=2}^{r-1}\sqrt{2} \rho^{(1)}\qty(E_{-\alpha_a}) - \rho^{(1)}\qty(E_{-\alpha_r}) - p(x,E) \;\rho^{(1)}\qty(E_{-\alpha_0}).
    \end{aligned} \label{eq:dual_linear_op_Br_V1}
\end{equation}
$\mathcal{L}_{B_{r}^{(1)}}^{\text{dual}}$ in $V^{(r)}$ for \eqref{eq:rep_V2_of_B2} and \eqref{eq:rep_V3_of_B3} is
\begin{equation}
    \begin{aligned}
        \mathcal{L}_{B_r}^{\text{dual}}=&\mathbf{1}_{2^r}\dv{x} + \frac{1}{x}\sum_{a=1}^r l_a\;\rho^{(r)}\qty(\alpha_a^{\vee}\cdot H)- \rho^{(r)}\qty(E_{-\alpha_1})\\
        &\qquad - \sum_{a=2}^{r-1}\sqrt{2} \rho^{(r)}\qty(E_{-\alpha_a}) - \rho^{(r)}\qty(E_{-\alpha_r}) - (-1)^{r+1} p(x,E) \;\rho^{(r)}\qty(E_{-\alpha_0}). 
    \end{aligned} \label{eq:dual_linear_op_Br_Vr}
\end{equation}

\subsubsection{\texorpdfstring{$C_r^{(1)}$}{Cr}}
The differential operator $\mathcal{L}_{C_r^{(1)}}$ is 
\begin{equation}
    \mathcal{L}_{C_{r}^{(1)}}=\dv{x}-\frac{1}{x}\sum_{a=1}^r l_a\;\alpha_a^{\vee}\cdot H + \sum_{a=1}^{r} E_{\alpha_a} + p(x,E)\;\zeta E_{\alpha_0}, \label{eq:linear_op_Cr}
\end{equation}
where $\zeta=1$ for $V^{(1)}$.
The differential operator $\mathcal{L}_{C_r^{(1)}}^{\text{dual}}$ in $V^{(1)}$ is
\begin{equation}
    \mathcal{L}_{C_{r}^{(1)}}^{\text{dual}}=\mathbf{1}_{2r}\dv{x}+\frac{1}{x}\sum_{a=1}^r l_a\;\rho^{(1)}\qty(\alpha_a^{\vee}\cdot H) - \sum_{a=1}^{r} \rho^{(1)}\qty(E_{-\alpha_a}) - p(x,E)\;\rho^{(1)}\qty(E_{-\alpha_0}). \label{eq:dual_linear_op_Cr}
\end{equation}

\subsubsection{\texorpdfstring{$F_4^{(1)}$}{F4}}
The differential operator $\mathcal{L}_{F_4^{(1)}}$ is
\begin{equation}
    \begin{aligned}
        \mathcal{L}_{F_4^{(1)}} = \dv{x} - \frac{1}{x}\sum_{a = 1}^{4}l_a\;\alpha_a^{\vee} \cdot H + E_{\alpha_1} + \sqrt{2} E_{\alpha_2} + \sqrt{3} E_{\alpha_3} + \sqrt{2} E_{\alpha_4} + p(x, E)\;\zeta E_{\alpha_0}, 
    \end{aligned}\label{eq:linear_op_F4}
\end{equation}
where $\zeta=1$ for $V^{(1)}$ and $\zeta=-1$ for $V^{(4)}$.
The differential operator $\mathcal{L}_{F_4^{(1)}}^{\text{dual}}$ in $V^{(1)}$ is 
\begin{equation}
    \begin{aligned}
        \mathcal{L}_{F_4^{(1)}}^{\text{dual}}=&\mathbf{1}_{26}\dv{x} +\frac{1}{x} \sum_{a=1}^4 l_a \rho^{(1)}\qty(\alpha_1^{\vee}\cdot H)+\rho^{(1)}\qty(E_{-\alpha_1})+\sqrt{2}\rho^{(1)}\qty(E_{-\alpha_2})\\
        &\qquad\qquad+\sqrt{3}\rho^{(1)}\qty(E_{-\alpha_3})+\sqrt{2}\rho^{(1)}\qty(E_{-\alpha_4})+p(x,E)\;\rho^{(1)}\qty(E_{-\alpha_0}).
    \end{aligned}\label{eq:dual_linear_op_F4_V1}
\end{equation}
The differential operator $\mathcal{L}_{F_4^{(1)}}^{\text{dual}}$ in $V^{(4)}$ is given by
\begin{equation}
    \begin{aligned}
        \mathcal{L}_{F_4^{(1)}}^{\text{dual}}=&\mathbf{1}_{52}\dv{x} +\frac{1}{x} \sum_{a=1}^4 l_a S^{-1}\rho^{(4)}\qty(\alpha_1^{\vee}\cdot H)S\\
        &\qquad+S^{-1}\rho^{(4)}\qty(E_{-\alpha_1})S+\sqrt{2}S^{-1}\rho^{(4)}\qty(E_{-\alpha_2})S+\sqrt{3}S^{-1}\rho^{(4)}\qty(E_{-\alpha_3})S\\
        &\qquad\qquad+\sqrt{2}S^{-1}\rho^{(4)}\qty(E_{-\alpha_4})S+p(x,E)\;S^{-1}\rho^{(4)}\qty(E_{-\alpha_0})S,
    \end{aligned}\label{eq:dual_linear_op_F4_V4}
\end{equation}
where $S$ is defined in \eqref{eq:basis_transf_mat_F4V4}.

\subsubsection{\texorpdfstring{$G_2^{(1)}$}{G2}}
The differential operator $\mathcal{L}_{G_2^{(1)}}$ is
\begin{equation}
    \mathcal{L}_{G_2^{(1)}} = \dv{x} - \frac{1}{x} \sum_{a = 1}^{2}l_a\;\qty(\alpha_a^{\vee} \cdot H) + E_{\alpha_1} + \sqrt{2} E_{\alpha_2} + p(x, E)\;\zeta E_{\alpha_0}, \label{eq:linear_op_G2}
\end{equation}
where $\zeta=1$ for $V^{(1)}$.
The differential operator $\mathcal{L}_{G_2^{(1)}}^{\text{dual}}$ in $V^{(1)}$ is given by
\begin{equation}
    \begin{aligned}
        \mathcal{L}_{G_2^{(1)}}^{\text{dual}} &= \mathbf{1}_{7}\dv{x} -\frac{1}{x} \sum_{a = 1}^{2}l_a\;\qty(\alpha_a^{\vee} \cdot H)\\
        &\qquad- \rho^{(1)}\qty(E_{-\alpha_1}) - \sqrt{2}\rho^{(1)}(E_{-\alpha_2}) - p(x,E)\;\rho^{(1)}\qty(E_{-\alpha_0}).
    \end{aligned}\label{eq:dual_linear_op_G2}
\end{equation}

\subsection{Folding and the linear problems for non-simply-laced affine Lie algebras}
In this part of the appendix, we discuss the relation between the linear problem for non-simply-laced Lie algebra and the linear problem obtained by folding of a simply-laced affine Lie algebra based on the matrix representations $V^{(a)}$.
Here we focus on the foldings: $D^{(1)}_{r+1}\rightarrow B_r^{(1)}$, $E_6^{(1)}\rightarrow F_4^{(1)}$ and $D_4^{(3)}\rightarrow G_2^{(1)}$, where the dimension of the representation changes.

\subsubsection{\texorpdfstring{$D_{r+1}^{(1)} \to B_{r}^{(1)}$}{DB}}
We discuss the folding of the representation $V_{D_{r+1}}^{(1)}$ of $D_{r+1}$, which has dimension $2r+2$.
Let $\mathbf{e}_1,\ldots,\mathbf{e}_{2r+2}\in\mathbf{R}^{2r+2}$ be the weight vectors of  $V_{D_{r+1}}^{(1)}$.
Folding $D_{r+1}$ to $B_{r}$ by identifying the simple roots as in \eqref{eq:folding_D_to_B}, 
$V_{D_{r+1}}^{(1)}$ is decomposed into the representations of $B_r$ as
\begin{equation}
    V_{D_{r+1}}^{(1)}=V_{B_r}^{(1)}\oplus\mathbf{C}, \label{eq:decomposition_D_to_B_V1}
\end{equation}
where $V_{B_r}^{(1)}$ is spanned by $\{ \mathbf{e}_1,\ldots,\mathbf{e}_{r},2^{-1/2}\qty(\mathbf{e}_{r+1}+\mathbf{e}_{r+2}),\mathbf{e}_{r+3},\ldots,\mathbf{e}_{2r+2}\}$ and $\mathbf{C}$ is spanned by $2^{-1/2}\qty(\mathbf{e}_{r+1}-\mathbf{e}_{r+2})$.

Now we apply this decomposition to linear problems.
Let the solution of the linear problem on $V_{D_{r+1}}^{(1)}$ be $\Psi=\sum_{i=1}^{2r+2}\psi_i \mathbf{e}_i$.
Choosing the monodromy parameters $l$ of $D_{r+1}^{(1)}$ as $l_{r}=l_{r+1}$, one can see that the linear combination $\psi_{r+1}-\psi_{r+2}$, which corresponds to the basis $2^{-1/2}\qty(\mathbf{e}_{r+1}-\mathbf{e}_{r+2})$, satisfies 
\begin{equation}
    \dv{x}\qty(\psi_{r+1}-\psi_{r+2})=0.
\end{equation}
Hence $\psi_{r+1}-\psi_{r+2}$ becomes constant.
Then, let us define a vector-valued function $\widetilde{\Psi}=\sum_{i=1}^{2r+1}\widetilde{\psi}_i\widetilde{\mathbf{e}}_i$ where $\qty{\widetilde{\mathbf{e}}_i}_{i=1}^{2r+1}$ are the orthonormal basis of $\mathbf{R}^{2r+1}$ and $\widetilde{\psi}_i$ is defined as
\begin{equation}
    \widetilde{\psi}_{i}=\begin{cases}
        \psi_{i}, & i=1,\ldots,r,\\
        \frac{1}{\sqrt{2}}\qty(\psi_{r+1}+\psi_{r+2}), & i=r+1,\\
        \psi_{i+1}, & i=r+2,\ldots,2r+1.
    \end{cases}
    \label{eq:ident_of_component_D_B}
\end{equation}
Introducing the monodromy parameters $\widetilde{l}_i$ of $B_{r}^{(1)}$ as
\begin{equation}
    \widetilde{l}_i\coloneqq l_i,\quad i=1,\ldots,r-1,\qquad \widetilde{l}_r\coloneqq l_r=l_{r+1},\label{eq:ident_of_monodromy_D_B}
\end{equation}
one finds that $\widetilde{\Psi}$ solves the linear problem on $V_{B_r}^{(1)}$. 
For other representations of $D_{r+1}$, one can see
\begin{equation}
    V_{D_{r+1}}^{(r)}=V^{(r+1)}_{D_{r+1}}=V^{(r)}_{B_r} \label{eq:decomposition_D_to_B_Vr}
\end{equation}
after the folding procedure.
Therefore the linear problems on $V_{D_{r+1}}^{(r)}$ and $V_{D_{r+1}}^{(r+1)}$ are equivalent to that on $V^{(r)}_{B_r}$ under the condition $l_{r}=l_{r+1}$.

\subsubsection{\texorpdfstring{$E_6^{(1)}\to F_4^{(1)}$}{EF}}
We next discuss the folding of the representation $V_{E_6}^{(1)}$ of $E_6$, whose matrix representation is given in appendix \ref{sec:cartan_representation}.
Let $\mathbf{e}_1,\ldots,\mathbf{e}_{27}\in\mathbf{R}^{27}$ be the weight vectors in $V_{E_6}^{(1)}$.
Folding the $E_6$ to $F_4$ by identifying the roots as in \eqref{eq:folding_E6_to_F4}, the weight vectors decompose into the weight vectors of $F_4$ in $V^{(1)}_{F_4}$ and that of the singlet representation ${\bf C}$ of $F_4$:
\begin{equation}
    V^{(1)}_{E_6}=V^{(1)}_{F_4}\oplus\mathbf{C}, \label{eq:decomposition_E_to_F_V1}
\end{equation}
where $V^{(1)}_{F_4}$ is spanned by 
\begin{equation}
    \qty{\widetilde{\mathbf{e}}_i}_{i=1}^{26}=\{ \mathbf{e}_1,\ldots,\mathbf{e}_{12},2^{-1/2}(\mathbf{e}_{13}+\mathbf{e}_{14}),6^{-1/2}(\mathbf{e}_{13}-\mathbf{e}_{14}-2\mathbf{e}_{15}),\mathbf{e}_{16},\ldots,\mathbf{e}_{27}\}
\end{equation}
and $\mathbf{C}$ is spanned by $3^{-1/2}\qty(\mathbf{e}_{13}-\mathbf{e}_{14}+\mathbf{e}_{15})$.
Let the solution of the linear problem for $E_{6}^{(1)}$ in $V_{E_6}^{(1)}$ be $\Psi=\sum_{i=1}^{27}\psi_i\mathbf{e}_i$. 
Choosing the monodromy parameters in $l$ of $E_6^{(1)}$ as $l_{1}=l_{5},l_{2}=l_{4}$, one can see that the linear combination $\psi_{13}-\psi_{14}+\psi_{15}$, which corresponds to the singlet representation, becomes constant.
Let us define a vector-valued function $\widetilde{\Psi}=\sum_{i=1}^{26}\widetilde{\psi}_i\widetilde{\mathbf{e}}_i$ with
\begin{equation}
    \begin{aligned}
        \widetilde{\psi}_{i}&=\begin{cases}
            \psi_{i}, & i=1,\ldots,12,\\
            \frac{1}{\sqrt{2}}\qty(\psi_{13}+\psi_{14}), & i=13,\\
            \frac{1}{\sqrt{6}}\qty(\psi_{13}-\psi_{14}-2\psi_{15}), & i=14,\\
            \psi_{i+1}, & i=15,\ldots,26.
        \end{cases}
    \end{aligned} \label{eq:ident_of_component_E_F}
\end{equation}
Denoting the monodromy parameters $\tilde{l}_i$ as
\begin{equation}
    \widetilde{l}_1\coloneqq l_1=l_5, \quad \widetilde{l}_2\coloneqq l_2=l_4, \quad \widetilde{l}_3 \coloneqq l_3,\quad \widetilde{l}_4\coloneqq l_6, \label{eq:ident_of_monodromy_E_F}
\end{equation}
one can see $\widetilde{\Psi}$ solves the linear problem for $F_4^{(1)}$ in $V_{F_4}^{(1)}$.
For the representation $V^{(6)}_{E_6}$, it is decomposed by $F_4$ as
\begin{equation}
V^{(6)}_{E_6}=V^{(4)}_{F_4}\oplus V^{(1)}_{F_4}. \label{eq:decomposition_E_to_F_V6}
\end{equation}
In this decomposition, the weight vectors $\mathbf{e}_1,\mathbf{e}_2$ of $V^{(1)}_{E_6}$ belong to $V^{(4)}_{F_4}$.
Then, the linear problem in $V^{(6)}_{E_6}$ including the weight vector $\mathbf{e}_1$ reduces to that in $V_{F_4}^{(4)}$ under \eqref{eq:ident_of_monodromy_E_F}.

\subsubsection{\texorpdfstring{$D_4^{(1)}\to G_2^{(1)}$}{DG}}
We discuss the folding of the representation $V_{D_4}^{(1)}$ of $D_4$.
Let $\mathbf{e}_1,\ldots,\mathbf{e}_8\in\mathbf{R}^8$ be the weight vectors of $V_{D_4}^{(1)}$. 
Folding the $D_4$ to $G_2$ by identifying the roots as in \eqref{eq:folding_D4_to_G2}, $V_{D_4}^{(1)}$ decomposes into $V^{(1)}_{G_2}$ and the singlet representation:
\begin{equation}
    V^{(1)}_{D_4}=V^{(1)}_{G_2}\oplus\mathbf{C}, \label{eq:decomposition_D_to_G}
\end{equation}
where $V^{(1)}_{G_2}$ is spanned by $\qty{\widetilde{\mathbf{e}}_i}_{i=1}^{7}=\{ \mathbf{e}_1,\mathbf{e}_2,\mathbf{e}_3,2^{-1/2}\qty(\mathbf{e}_4+\mathbf{e}_5),\mathbf{e}_6,\mathbf{e}_7,\mathbf{e}_8\}$ and $\mathbf{C}$ is spanned by $2^{-1/2}\qty(\mathbf{e}_4-\mathbf{e}_5)$.

Let the solution of the linear problem in $V^{(1)}_{D_4}$ be $\Psi=\sum_{i=1}^{8}\psi_i\mathbf{e}_i$. 
Choosing the monodromy parameter $l$ of $D_4^{(1)}$ as $l_3=l_4$, one can see that the linear combination $\psi_4-\psi_5$, which corresponds to the weight vectors $2^{-1/2}\qty(\mathbf{e}_4-\mathbf{e}_5)$, becomes constant.
Then, we introduce a vector-valued function $\widetilde{\Psi}=\sum_{i=1}^{7}\widetilde{\psi}_i\widetilde{\mathbf{e}}_i$, where $\widetilde{\psi}_i$ is defined by
\begin{equation}
    \widetilde{\psi}_{i}=\begin{cases}
        \psi_{i}, & i=1,2,3,\\
        \frac{1}{\sqrt{2}}\qty(\psi_{4}+\psi_{5}), & i=4,\\
        \psi_{i+1}, & i=5,6,7.
    \end{cases}
\end{equation}
Setting the monodromy parameters $\widetilde{l}_i$ as
\begin{equation} 
    \widetilde{l}_1\coloneqq l_1=l_3=l_4,\quad \widetilde{l}_2\coloneqq l_2, \label{eq:ident_of_monodromy_D_G}
\end{equation}
one can see $\tilde{\Psi}$ solves the linear problem for $G_2^{(1)}$ in $V_{G_2}^{(1)}$.

\section{Wronskian type formulas for the Q-functions}\label{sec:q-funcs}
In this appendix, we summarize the Wronskian type formulas of the Q-functions obtained from the anti-symmetric products of a fundamental representation, which are used to compute the Bethe roots in the section \ref{sec:comparison}.

\paragraph{\underline{$A_r^{(1)}$}}
From $V^{(a)}=\wedge^a V^{(1)}$, we have obtained \eqref{eq:A_Qfunc_from_V1}.
In a similar way, from  $V^{(a)}=\wedge^{r+1-a} V^{(r)}$, we find
\begin{equation}
    Q^{(a)}=\sum_{j_0,\ldots,j_{r-a}=1}^{r+1-a}\epsilon_{j_0\cdots j_{r-a}}\prod_{k=0}^{r-a} \Omega^{-k\lambda^{(r)}_{j_k}/h M}\mathcal{Q}^{(r)}_{j_k\qty[\frac{r-a}{2}-k]}(E),\ \ \ a=2,\ldots,r. \label{eq:A_Qfunc_from_Vr}
\end{equation}

\paragraph{\underline{$D_r^{(1)}$}}
From $V^{(a)}=\wedge^a V^{(1)}$ $(a=1,\ldots,r-2)$, we obtain
\begin{equation}
    Q^{(a)}(E) = \sum_{j_0,\ldots,j_{a-1}=1}^{a}\epsilon_{j_0\cdots j_{a-1}}\prod_{k=0}^{a-1} \Omega^{-k\lambda^{(1)}_{j_k}/h M}\mathcal{Q}^{(1)}_{j_k\qty[\frac{r-a}{2}-k]}(E),\ \ \ a=2,\ldots,r-2. \label{eq:D_Qfunc_from_V1}
\end{equation}

\paragraph{\underline{$E_6^{(1)}$}}
From $\wedge^2 V^{(1)}=V^{(2)}$, $\wedge^{2}V^{(5)}=V^{(4)}$ and $\wedge^2 V^{(6)}=V^{(3)}\oplus V^{(6)}$, one finds
\begin{equation}
    \begin{aligned}
        &Q^{(2)}(E) = \sum_{j_0,j_1=1}^{2}\epsilon_{j_0j_1}\prod_{k=0}^{1} \Omega^{-k\lambda^{(1)}_{j_k}/h M}\mathcal{Q}^{(1)}_{j_k\qty[\frac{1}{2}-k]}(E),\\
        &Q^{(4)}(E)=\sum_{j_0,j_2=1}^{2}\epsilon_{j_0j_1}\prod_{k=0}^{1} \Omega^{-k\lambda^{(5)}_{j_k}/12M}\mathcal{Q}^{(5)}_{j_k\qty[\frac{1}{2}-k]}(E),\\
        &Q^{(3)}(E)=\sum_{j_0,j_2=1}^{2}\epsilon_{j_0j_1}\prod_{k=0}^{1} \Omega^{-k\lambda^{(6)}_{j_k}/12M}\mathcal{Q}^{(6)}_{j_k\qty[\frac{1}{2}-k]}(E).
    \end{aligned}\label{eq:E6_Qfunc_from_V1V5V6}
\end{equation}

\paragraph{\underline{$E_7^{(1)}$}}
We have the following relations for anti-symmetric products:
\begin{equation}
    \begin{aligned}
        \wedge^2 V^{(1)}&=V^{(2)}\oplus\mathbf{C},\qquad \wedge^3 V^{(1)}=V^{(3)}\oplus V^{(1)},\qquad \wedge^4 V^{(1)}=V^{(4)}\oplus \cdots,\\
        \wedge^2 V^{(6)}&=V^{(5)}\oplus V^{(6)},\quad \wedge^3 V^{(6)}=V^{(4)}\oplus \cdots ,
    \end{aligned}
\end{equation}
where $\cdots$ means other representations.
We then find
\begin{equation}
    Q^{(a)}(E) =
    \begin{dcases}
        \sum_{j_0,\ldots,j_{a-1}=1}^{a}\epsilon_{j_0\cdots j_{a-1}}\prod_{k=0}^{a-1} \Omega^{-k\lambda^{(1)}_{j_k}/h M}\mathcal{Q}^{(1)}_{j_k\qty[\frac{a-1}{2}-k]}(E), & a=2,3,\\
        \sum_{j_0,\ldots,j_{6-a}=1}^{7-a}\epsilon_{j_0\cdots j_{6-a}}\prod_{k=0}^{6-a} \Omega^{-k\lambda^{(6)}_{j_k}/h M}\mathcal{Q}^{(6)}_{j_k\qty[\frac{6-a}{2}-k]}(E), & a=5,6.
    \end{dcases}
\end{equation}

\paragraph{\underline{$E_8^{(1)}$}}
We have the following relations for anti-symmetric products:
\begin{equation}
        \wedge^2 V^{(1)}=V^{(2)},\qquad \wedge^a V^{(1)}=V^{(a)}\oplus \cdots,\quad  a=3,4,5.
\end{equation}
where $\cdots$ means other representations.
Then we have
\begin{equation}
    Q^{(a)}(E) = \sum_{j_0,\ldots,j_{a-1}=1}^{a}\epsilon_{j_0\cdots j_{a-1}}\prod_{k=0}^{a-1} \Omega^{-k\lambda^{(1)}_{j_k}/h M}\mathcal{Q}^{(1)}_{j_k\qty[\frac{a-1}{2}-k]}(E),\qquad a=2,\ldots,5.
\end{equation}

\paragraph{\underline{$B_r^{(1)}$}}
From $\wedge^a V^{(1)}=V^{(a)}\;(a=1,\ldots,r-1)$, we find
\begin{equation}
    Q^{(a)}(E) = \sum_{j_0,\ldots,j_{a-1}=1}^{a}\epsilon_{j_0\cdots j_{a-1}}\prod_{k=0}^{a-1} \Omega^{-k\lambda^{(1)}_{j_k}/h M}\mathcal{Q}^{(1)}_{j_k\qty[\frac{a-1}{2}-k]}(E),\ \ \ a=2,\ldots,r-1. \label{eq:calc_of_qfunc_B}
\end{equation}

\paragraph{\underline{$C_r^{(1)}$}}
From $\wedge^a V^{(1)}=V^{(a)}\oplus\mathbf{C}\wedge^{(a-2)}V^{(1)}$, we obtain
\begin{equation}
    Q^{(a)}(E) = \sum_{j_0,\ldots,j_{a-1}=1}^{a}\epsilon_{j_0\cdots j_{a-1}}\prod_{k=0}^{a-1} \Omega^{-k\lambda^{(1)}_{j_k}/h M}\mathcal{Q}^{(1)}_{j_k\qty[\frac{a-1}{2}-k]}(E),\ \ \ a=2,\ldots,r.
\end{equation}

\paragraph{\underline{$F_4^{(1)}$}}
From $\wedge^2 V^{(1)}=V^{(2)}\oplus V^{(4)}$ and $ \wedge^2 V^{(4)}=V^{(3)}\oplus V^{(4)}$, we get
\begin{equation}
\begin{aligned}
    &Q^{(2)}(E)=\sum_{j_0,j_1=1}^{2}\epsilon_{j_0j_1}\prod_{k=0}^{1} \Omega^{-k\lambda^{(1)}_{j_k}/12M}\mathcal{Q}^{(1)}_{j_k\qty[\frac{1}{2}-k]}(E),\\
    &Q^{(3)}(E)=\sum_{j_0,j_1=1}^{2}\epsilon_{j_0j_1}\prod_{k=0}^{1} \Omega^{-k\lambda^{(4)}_{j_k}/12M}\mathcal{Q}^{(4)}_{j_k\qty[\frac{1}{2}-k]}(E).
    \label{eq:F4_Qfunc_from_V1V4}
\end{aligned}
\end{equation}

\paragraph{\underline{$G_2^{(1)}$}}
From $\wedge^2 V^{(1)}=V^{(2)}\oplus V^{(1)}$, we get
\begin{equation}
    Q^{(2)}(E)=\sum_{j_0,j_1=1}^{2}\epsilon_{j_0j_1}\prod_{k=0}^{1} \Omega^{-k\lambda^{(1)}_{j_k}/6M}\mathcal{Q}^{(1)}_{j_k\qty[\frac{1}{2}-k]}(E). \label{eq:G2_Qfunc_from_V1}
\end{equation}

\subsection{Folding and Q-functions}\label{sec:folding_q}
In the appendix \ref{sec:linear_ops}, we have shown that the linear problem for non-simply-laced affine Lie algebra follows from that of simply-laced Lie algebra with specific monodromy parameters.
The subdominant solutions and the solutions around the origin can be also obtained from the folding procedure.
We then find the relation between the Q-functions.
Since we have explained the relations between $C_r^{(1)}$ and $A_{2r-1}^{(1)}$ in section \ref{sec:linear_problem_Qfunction}, we summarize the results in the cases of $B_r^{(1)},F_4^{(1)}$, and $G_2^{(1)}$:

\paragraph{$B_r^{(1)}$}
\begin{equation}
    \begin{aligned}
        Q^{(a)}_{B_{r}^{(1)}}=Q^{(a)}_{D_{r+1}^{(1)}}, \qquad a=1,\ldots,r-1,\qquad Q^{(r)}_{B_{r}^{(1)}}=Q^{(r)}_{D_{r+1}^{(1)}}=Q^{(r+1)}_{D_{r+1}^{(1)}},
    \end{aligned}
\end{equation}

\paragraph{$F_4^{(1)}$}
\begin{equation}
    \begin{aligned}
        &Q^{(1)}_{F_{4}^{(1)}}=Q^{(1)}_{E_{6}^{(1)}}=Q^{(5)}_{E_{6}^{(1)}},\qquad Q^{(3)}_{F_{4}^{(1)}}=Q^{(3)}_{E_{6}^{(1)}},\\ &Q^{(2)}_{F_{4}^{(1)}}=Q^{(2)}_{E_{6}^{(1)}}=Q^{(4)}_{E_{6}^{(1)}},\qquad Q^{(4)}_{F_{4}^{(1)}}=Q^{(6)}_{E_{6}^{(1)}},
    \end{aligned}
\end{equation}

\paragraph{$G_2^{(1)}$}
\begin{equation}
    \begin{aligned}
        Q^{(1)}_{G_{2}^{(1)}}=Q^{(1)}_{D_{4}^{(1)}}=Q^{(3)}_{D_{4}^{(1)}}=Q^{(4)}_{D_{4}^{(1)}},\qquad
        &Q^{(2)}_{G_{2}^{(1)}}=Q^{(2)}_{D_{4}^{(1)}}.
    \end{aligned}
\end{equation}
These relations can be also checked numerically in section \ref{sec:comparison}.

\section{Cheng's algorithm for \texorpdfstring{$E_6^{(1)}$}{E6}}\label{sec:Cheng_E6}
In this section, we demonstrate the Cheng's algorithm for the $E_6^{(1)}$ linear problem in $V^{(1)}$.
The linear operators $\mathcal{D}[\bm{q}]$ and $\mathcal{P}$ become:
\begin{align}
    &\mathcal{D}[\bm{q}]=\bm{1}_{27}\dv{x}-\frac{\bm{q}}{x},\qquad \bm{q}=\sum_{a=1}^6 l_a\;\rho^{(1)}\qty(\alpha_a\cdot H),\\
    &\mathcal{P}=\rho^{(1)}\qty(E_{\alpha_1})+ \sqrt{2} \rho^{(1)}\qty(E_{\alpha_2}) + \sqrt{3} \rho^{(1)}\qty(E_{\alpha_3})+ \sqrt{2} \rho^{(1)}\qty(E_{\alpha_4})\notag\\
    &\qquad\qquad\qquad + \rho^{(1)}\qty(E_{\alpha_{5}}) + \sqrt{2} \rho^{(1)}\qty(E_{\alpha_6}) + p(x, E)\;\rho^{(1)}\qty(E_{\alpha_0}).
\end{align}
Then, using the inverse operator $\mathbf{L}[\bm{q}]$, one can calculate power series solutions $\mathcal{X}_{i}$ ($i=1,\ldots,27$).
As an example, for the potential $p(x,E)=x^2-E$ and the monodromy parameters $l=(5/12, 1/3, 0, -1/3, -5/12, 1/10)$ as in the section \ref{sec:comparison}, the top component of $\mathcal{X}_1$ is
\begin{align}
    \chi_{1,1}=&x^{5/12}-\frac{160451840000 E }{137931559732275309 \sqrt{3}}x^{149/12}+\frac{437678816000 }{30622269497100148413 \sqrt{3}}x^{173/12}\notag\\
    &\qquad +\frac{54927641304141824000000000 E^2 }{12325053095061152120611792093400953456769103}x^{293/12}+\cdots.
\end{align}
On the other hand, the the bottom component of $\bar{\mathcal{X}}_1$ becomes
\begin{align}
    \bar{\chi}_{1,27}=&x^{187/12}-\frac{9697109248000 E}{5766433072467466952630259 \sqrt{3}}x^{331/12}\notag\\
    &\qquad\qquad\qquad\qquad+\frac{5142734920000000 }{12093036229962779542892824383 \sqrt{3}}x^{355/12}+\cdots.
\end{align}

\end{document}